\documentclass[preprint, 
amsmath, amssymb, superscriptaddress, 
nofootinbib, 
preprintnumbers, keywords]{revtex4}

\usepackage{here}
\usepackage{bm}
\usepackage{booktabs}
\usepackage{mathrsfs}
\usepackage{array}

\usepackage{mediabb}
\usepackage{color}
\usepackage{comment}

\newcommand{\AC}[1]{SACOT-$\chi$}
\newcommand{\ZMC}[1]{ZM-$\chi$}
\newcommand{\MZC}[1]{M0-$\chi$}
\newcommand{\SubC}[1]{sub-$\chi$}
\newcommand{\ZMx}[1]{ZM-$x$}
\newcommand{\MZx}[1]{M0-$x$}
\newcommand{\Subx}[1]{sub-$x$}

\newcommand{\mAC}{{\rm SACOT\mathchar`-\chi}}
\newcommand{\mZMC}{{\rm ZM\mathchar`-\chi}}
\newcommand{\mMZC}{{\rm M0\mathchar`-\chi}}
\newcommand{\mSubC}{{\rm sub\mathchar`-\chi}}
\newcommand{\mZMx}{{\rm ZM\mathchar`-}x}
\newcommand{\mMZx}{{\rm M0\mathchar`-}x}

\newcommand{\mM}{{\rm M}}
\newcommand{\mZM}{{\rm ZM}}
\newcommand{\mMZ}{{\rm M0}}

\usepackage{fancybox} 

\usepackage{ulem}
\usepackage{multirow}

\begin{document}
\title{
Charged lepton flavor violation associated with heavy quark 
production in deep inelastic lepton-nucleon scattering via 
scalar exchange
}  

\author{Yuichiro Kiyo}
\email[E-mail: ]{ykiyo@juntendo.ac.jp}
\affiliation{Department of Physics, Juntendo University, 
Inzai, 270-1695 Japan}

\author{Michihisa Takeuchi}
\email[E-mail: ]{m.takeuchi@het.phys.sci.osaka-u.ac.jp}
\affiliation{Department of Physics, Osaka University, Osaka 560-0043, Japan}

\author{Yuichi Uesaka}
\email[E-mail: ]{uesaka@ip.kyusan-u.ac.jp}
\affiliation{Faculty of Science and Engineering, Kyushu Sangyo University, 2-3-1 Matsukadai, Higashi-ku, Fukuoka 813-8503, Japan}

\author{Masato Yamanaka}
\email[E-mail: ]{yamanaka@osaka-cu.ac.jp}
\affiliation{Department of Mathematics and Physics, 
Osaka City University, Osaka 558-8585, Japan}
\affiliation{Nambu Yoichiro Institute of Theoretical and 
Experimental Physics (NITEP), Osaka City University, 
Osaka 558-8585, Japan
}

\preprint{OCU-PHYS 545, NITEP 90, OU-HET-1102}

\date{\today}

\begin{abstract}
We study charged lepton flavor violation (CLFV) associated with heavy quark pair production in lepton-nucleon deep-inelastic scattering $\ell_i N \to \ell_j q\bar{q} X$.
Here $\ell_i$ and $\ell_j$ denote the initial and final leptons; $N$ and $X$ are respectively the initial nucleon and arbitrary final hadronic system.
We employ a model Lagrangian in which a scalar and pseudoscalar mediator  generates the CLFV.
We derive heavy quark structure functions for scalar and pseudoscalar currents and compute momentum distributions of the final lepton for the process.
Our focus is on the heavy quark mass effects in the final lepton momentum distribution.
We clarify the necessity of inclusion of the heavy quark mass
to obtain reliable theory predictions for the CLFV signal searches in the deep-inelastic scattering.

\end{abstract}

\maketitle

\section{Introduction}
\label{sec:Intro}

A variety of new physics models provides contributions
 to the Charged Lepton Flavor Violating (CLFV) observables 
 by extra degrees of freedom, for instance
extensions of particle contents, additional space dimensions, etc.
(See Refs.\cite{Kuno:1999jp,Raidal:2008jk} and references therein
 for reviews on this subject.)
It is often the case that the CLFV mediators couple with not only leptons but also quarks. In that case, noticeable processes which involve hadronic interactions, e.g., $\mu \to e$ conversion in nuclei, $\tau \to \ell_i \pi \pi$ ($\ell_i = e, \mu $), 
different flavor di-lepton $\ell_i \ell_j$ production at hadron collider experiments, would be expected.
No signal for such types of CLFV processes is discovered so far although a lot of effort to search for them has been devoted in the various experiments.
Though these results are translated into the stringent limits on the CLFV interactions, the limits are mainly on the interactions related to light flavor quarks.
This motivates us to revisit the scenarios 
where the CLFV mediators dominantly couples with heavy quarks. 
Such scenarios are motivated theoretically as well as 
experimentally, e.g. extra-dimension models\,\cite{Huber:2003tu, 
Moreau:2006np, Agashe:2006iy, Davidson:2007si}, 
two Higgs doublet models\,\cite{Kanemura:2005hr, 
Davidson:2010xv, Crivellin:2013wna, Botella:2015hoa}, 
leptoquark models\,\cite{Crivellin:2020mjs}, models with flavor 
symmetry~\cite{Tsumura:2009yf}, and the next to minimal 
flavor violation scenarios\,\cite{Agashe:2005hk}. 

In such scenarios, CLFV processes via deep-inelastic scattering (DIS) 
$\ell_i N\to \ell_j X$, where $N$ is a nucleon, $\ell_i$ and $\ell_j$ 
are respectively the initial and final leptons, offer a good prospect 
for CLFV searches. Such processes can be probed at fixed target 
experiments and lepton-hadron colliders.
In both experiments, 
the problems due to pile-up and QCD background are better controlled than in the environments of the hadron-hadron colliders.
In this article, we will focus on the study at the fixed target experiments. 
The event rates in such experiments increase with the beam energy, beam intensity and target density. 
Typical beam energy $E_\ell$ in the next-generation experiments is up to $10~{\rm TeV}$, which corresponds to $\sqrt{s} \lesssim 100$\,GeV.
Although it seems not high, it is sufficient to open 
production thresholds for the CLFV processes.
Therefore it is expected to observe enough signals 
in the CLFV searches at the fixed target experiments.

The DIS processes are studied with a variety of theoretical motivations
in the context of the CLFV~\cite{Gninenko:2001id, Sher:2003vi, 
Kanemura:2004jt, Gonderinger:2010yn, Bolanos:2012zd, 
Liao:2015hqu, Abada:2016vzu, Takeuchi:2017btl, 
Gninenko:2018num, Antusch:2020fyz, Husek:2020fru, Cirigliano:2021img,Antusch:2020vul}, 
as well as a probe for the Standard Model (SM) and new physics  
\cite{Cakir:2009xi, Han:2009pe, Liang:2010gm, Blaksley:2011ey, 
Biswal:2012mp, Dutta:2013mva, Li:2017kfk, Curtin:2017bxr, 
Li:2018wut, Azuelos:2019bwg}. 
The HERA experiment searched for the CLFV DIS processes and put the bound 
on the related parameters~\cite{Aktas:2007ji, Aaron:2011zz}. 
Searches for the CLFV DIS processes are proposed at the upcoming experiments, 
and shown to reach higher sensitivities than the current 
bounds by a few orders of magnitude~\cite{Accardi:2012qut, 
AbelleiraFernandez:2012cc}. 

In this article, we study the CLFV DIS processes, $\ell_i N \to \ell_j X_H$, 
in the scenario where a (pseudo-)scalar CLFV mediator 
dominantly couples with heavy flavor quarks, like the SM Higgs boson. 
Here $X_H$ denotes an inclusive hadronic final state involving heavy 
quarks.
It is worth investigating the CLFV DIS processes associated with heavy flavor quarks, since the CLFV operators involving heavy flavor quarks are usually difficult to probe directly in the  low-energy flavor experiments.
Experimental signals for the processes are characterized as the existence of
a heavy charged lepton $\ell_j$ and heavy quarks in the final state.
Such signals seem distinctive, but there is always a competition between the signals and the background. Thus, precise understanding of the backgrounds and also the accurate theory prediction for the signal processes would be required. In this article, we focus on the latter point.
One of differences between heavy quarks and light quarks is the mass effect, 
and it is important to understand how the heavy quark mass 
affects the CLFV DIS observables. 
At first glance, it looks simple and straightforward, 
but it turns out rather complicated when the issue is related to the problem of the large logarithmic resummation in the perturbative QCD.
A resolution to the problem was given in a series of seminal papers by 
ACOT \cite{Aivazis:1993kh, Aivazis:1993pi} at the leading order (LO), and the result has been extended
to include higher-order effect in a consistent manner 
\cite{Buza:1996wv,Cacciari:1998it, Collins:1998rz,Forte:2010ta}. 
In the present article, we apply the 
ACOT method to the CLFV DIS involving heavy quarks. 
Although we work at LO in QCD strong coupling expansion, 
we include some of the important effects of heavy quark mass 
which were obtained in the studies of QCD structure functions in the literatures 
\cite{Kramer:2000hn,Tung:2001mv, Kretzer:2003it,Stavreva:2012bs}. We aim for the construction of 
heavy quark structure functions of (pseudo-)scalar exchange
coming from CLFV interactions. With the constructed heavy quark structure functions we analyze some distributions 
of the final lepton momentum to investigate how the heavy quark mass effect modifies the 
CLFV DIS observables. We focus on the analysis of the CLFV DIS processes 
associated with heavy quark production in the present article,
and a comprehensive phenomenological study taking other modes 
will be reported in a separate paper.

The present article is organized in the  following way.
In section II.A we will describe the model Lagrangian of
the CLFV (pseudo-)scalar mediator which strongly 
couples with heavy quarks. With the model Lagrangian, 
we will compute the cross section for CLFV DIS associated with heavy quark pair production. We shall introduce a structure function of (pseudo-)scalar current.
In section II.B we will  show the cross section formula and 
momentum distribution of the final lepton, where 
an inverse moment of the structure function shall be 
introduced. 
In section II.C we will compute the heavy quark contribution to the structure function at the leading order in $\alpha_s$ with massive quark.
In sections II.D and E, the SACOT scheme and threshold 
improvement of the heavy quark structure function are 
discussed. The issue here is the unification of heavy 
quark mass effect and large logarithm resummation 
in a consistent manner.
In section III we will perform the numerical analysis of the 
structure functions for the production of bottom and charm quarks.
In section IV we will present the numerical results of the 
CLFV cross section associated with heavy quark production.

\section{CLFV DIS and Heavy quark production}

\subsection{CLFV DIS via scalar or pseudoscalar current}
We start with an interaction Lagrangian for a neutral scalar or 
pseudoscalar field $\phi\in\{S, P\}$ coupled with charged leptons 
$\ell_i, \ell_j$ and heavy flavor 
quarks $q=c,b,t$;
\begin{equation}
\begin{split}
	\mathcal{L}_\phi
	&= 
	- \sum_{ij} \left( 
	\rho_{ij}^\phi \bar{\ell}_j P_L \ell_i \phi 
	+ h.c. 
	\right)
	-\sum_{q}\rho_{qq}^\phi \bar{q} \, \Gamma^\phi q \phi, 
\label{Eq:Lscalar}
\end{split}
\end{equation}
where $i, j$ run over flavor indices of charged leptons, $q$ runs over 
heavy flavor quarks, and $P_L=(1-\gamma_5)/2$.
The vertex factors
$\Gamma^{S}=1$ and $\Gamma^{P}= i\gamma_5$ 
are the matrices in Dirac-space 
respectively for scalar and pseudoscalar cases.
The lepton and quark fields in the interaction Lagrangians are mass eigenstates.
We assume that the CLFV mediators interact with quarks 
through flavor diagonal couplings $\rho_{qq}^{\phi}$, while in the lepton sector the off-diagonal coupling $\rho_{ij}^\phi (i\neq j)$ induces 
the CLFV.

\begin{figure}[htb!]
\begin{center}
\includegraphics[width=6cm]{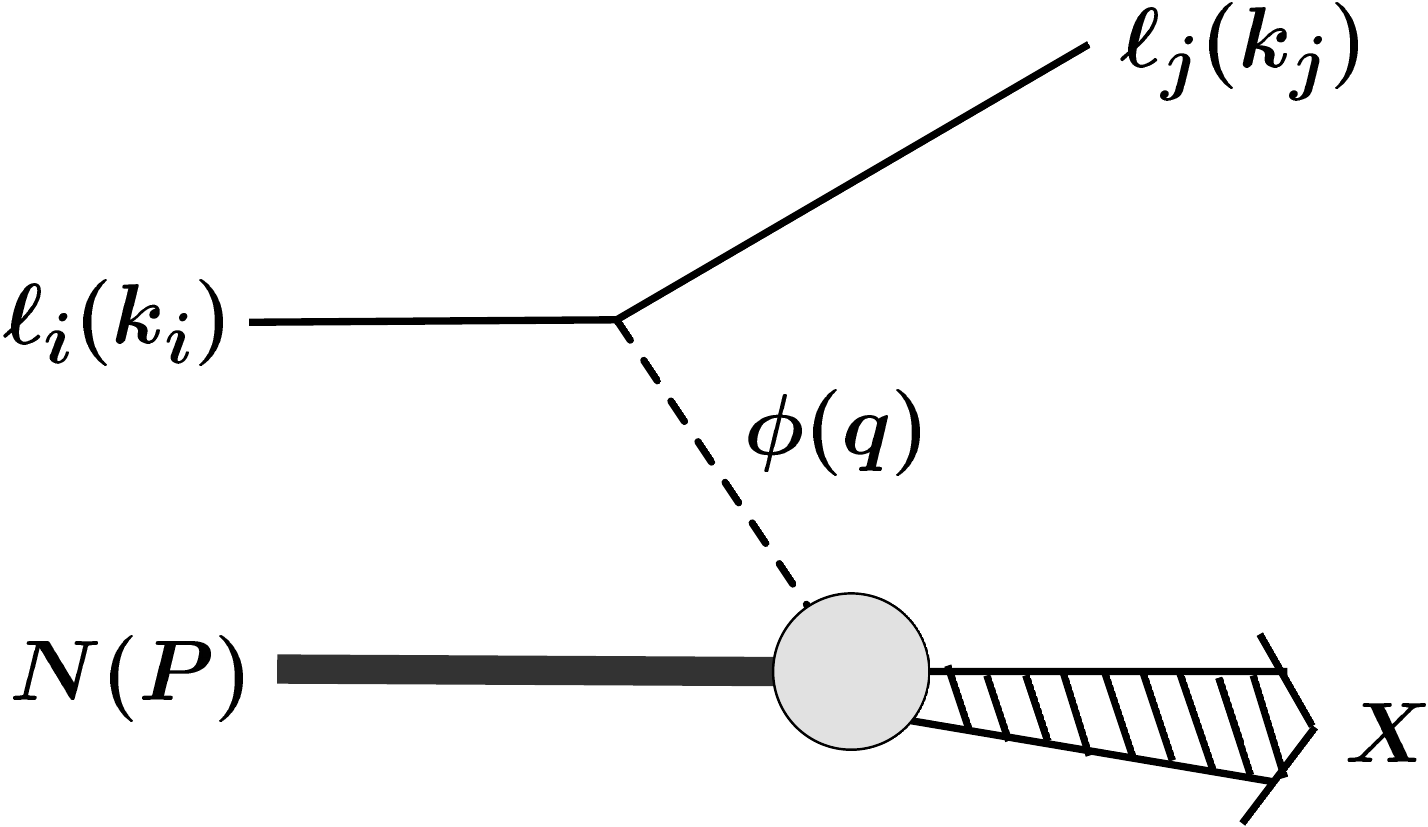}
\end{center}
\caption{ 
The CLFV lepton-nucleon scattering induced by $t$-channel exchange of 
the (pseudo-)scalar mediator. The initial and final lepton momenta are $k_i$ and  $k_j$, respectively, and the nucleon momentum is $P$. }
\label{Fig:clfv_dis_0}
\end{figure}

In Fig.~\ref{Fig:clfv_dis_0} schematic diagram for the process 
$\ell_i N\to\ell_j X_H$  is shown, where initial lepton $\ell_i$ with a 
momentum $k_i$ and a nucleon $N$ with a momentum $P$  are scattered by exchanging the CLFV mediator $\phi$  with a momentum $q=k_i-k_j$. 
The final states are lepton $\ell_j $ with momentum $k_j$ 
 and an arbitrary hadronic system $X_H$ which contains heavy quarks. 
The amplitude, where the CLFV mediator $\phi$ is exchanged in $t$-channel, is factorized into leptonic and hadronic amplitudes. 
 The  cross section $\sigma$ consists 
 of the leptonic and hadronic parts, which are respectively denoted by $L_\phi$ 
 and $F_\phi$, and is written in the following form:
\begin{align}
\frac{d\sigma}{dxdy}
\left(\ell_iN\to\ell_j X\right) 
=
\frac{y\left|\rho_{qq}^\phi \right|^2 L_{\phi}(Q^2) \, F_\phi(x,Q^2)}
{16\pi\left(Q^2+m_\phi^2\right)^2},
\end{align}
where $m_\phi$ is a mediator mass, and the dimensionless variables $x, y$ 
are defined by
\begin{eqnarray}
x =\frac{Q^2}{2P\cdot q},
\hspace{1cm}
y=\frac{2P\cdot q}{2P\cdot k_i},
\end{eqnarray}
with $Q^2=-q^2=xy(s-m_N^2)$. Here $s=(P+k_i)^2$ is the collision energy squared of the initial lepton-nucleon system, and $m_N$ is  the nucleon mass.

The leptonic part is given by
\begin{eqnarray}
L_{\phi} &=& \big( |\rho_{ij}^\phi|^2 +|\rho_{ji}^\phi|^2\big)
 \big(Q^2+m_i^2+m_j^2\big)
 +4\, {\rm Re} \left( \rho_{ij}^\phi \rho_{ji}^\phi\right) m_i m_j,
\end{eqnarray}
with $m_{i(j)}$ being the initial (final) lepton mass,
and the hadronic part is called structure function written 
in a convolution form as 
\begin{eqnarray}
F_\phi(x,Q^2) &=&
\sum_{k}
\int_0^1 \frac{d\xi}{\xi} \  C_{\phi, \, k}\left(\frac{x}{\xi}\right) f_{k/N}(\xi, \mu_f^2),
\end{eqnarray}
where $k\in \{g, q, \bar{q} \}$ is a parton which contributes to the process $\phi k \rightarrow X$.  
The $C_{\phi, k}$ is a coefficient function calculable in perturbative
QCD, while the parton distribution function (PDF) $f_{k/N}(x, \mu_f^2)$
is a nonperturbative object which describes a probability of parton 
$k$ having momentum fraction $\xi$ inside the nucleon $N$ at a factorization
scale $\mu_f^2$.
The $\mu_f$-dependence is governed by  renormalization group equation, so called  Dokshitzer-Gribov-Lipatov-Altarelli-Parisi (DGLAP) equation \cite{Gribov:1972ri, Altarelli:1977zs} :
\begin{eqnarray}
\mu_f^2\frac{\partial~}{\partial \mu_f^2} f_{k/N}(\xi, \mu_f^2) &=& 
\sum_{l} 
\frac{\alpha_s(\mu_f^2)}{2\pi}
\int_{\xi}^1 \frac{d\eta}{\eta} 
{\cal P}_{kl} \left(\frac{\xi}{\eta}\right)f_{l/N}(\xi,\mu_f^2) 
+{\cal O}(\alpha_s^2),
\end{eqnarray}
where $l$ runs over possible quark flavors and gluon, and 
${\cal P}_{kl}$ is a splitting function at one-loop level.
Conventionally the factorization scale is chosen as $\mu_f^2\sim Q^2$ 
to be the same order as the hard scale
in the process. For the heavy quark production, there are two hard scales $Q^2, m_q^2$ in the process, and we will adopt a 
more refined scale setting (see Eq.~\eqref{eq:mu_Q}).

\subsection{Cross sections and distribution}
In theory discussion it is sometimes convenient to use $x, Q^2$ as
independent variables instead of $x,y$. The conversion formula is given by 
\begin{eqnarray}
\frac{d^2\sigma}{dxdQ^2}(\ell_i N\to \ell_j X)
&=&
\left(\frac{1}{xs} \right)
\frac{d^2\sigma}{dxdy}(\ell_i N\to \ell_j X).
\end{eqnarray}
Integrated over $x$ one obtains the differential cross section $d\sigma/dQ^2$ as 
\begin{eqnarray}
\frac{d\sigma}{dQ^2}
&=&
\frac{|\rho_{qq}^\phi|^2 }{16\pi s^2}
\frac{Q^2 L_\phi(Q^2) M_\phi(s,Q^2)}{(Q^2+m_\phi^2)^2},
\label{eq:ds_dQ2}
\end{eqnarray}
where $M_\phi$ is the second inverse moment of the structure function defined by
\begin{eqnarray}
M_\phi(s, Q^2) 
&\equiv & 
\int_{x_{\rm min}(s, Q^2)}^{x_{\rm max}(Q^2)} x^{-2}\, F_\phi(x,Q^2) dx.
\label{eq:inverse_moment}
\end{eqnarray}
The integration region $[x_{\rm min}(s, Q^2), x_{\rm max}(Q^2)]$ 
depends on partonic processes,  and the $x_{\rm min}(s, Q^2)$ 
depends on $s$ (see \eqref{eq:xmin}), which introduces the collision energy dependence in the inverse moment.  A derivation of the physical region for the CLFV DIS is given in the Appendix.

As a direct observable in collider experiments,
we study the momentum distribution of the final lepton. 
To make our discussion concrete we give a cross 
section formula for $eN\to \tau X$ for fixed target 
experiment where the initial nucleon is at the rest.
The initial nucleon and electron momenta are 
parametrized by  $k_e=(E_e, 0, 0, E_e)$ and $P=(m_N, 0,0,0)$, 
respectively. 
Here we ignored the electron mass, and 
the nucleon mass is set to be $m_N=938\, {\rm MeV}$. 
The electron beam energy $E_e$ is related to the collision 
energy by $E_e=\sqrt{(s-m_N^2)/2m_N}$.
The final $\tau$-momentum at the nucleon rest frame is parametrized as $k_\tau = \left(E_\tau,  p_T, 0, p_Z\right)$
 where $p_T, p_Z$ are related to the dimensionless parameters $x, y$ as
\begin{eqnarray}
p_Z= (1-y) E_e - xy m_N -\frac{m_\tau^2}{2E_e},
\hspace{5mm}
p_T = \sqrt{(1-y)^2 E_e^2 -m_\tau^2 -p_Z^2},
\end{eqnarray}
with $E_\tau=(1-y)E_e$.
Then the $\tau$-momentum distribution at the nucleon rest frame is given by
\begin{eqnarray}
\frac{d^{\, 2}\sigma}{dp_T\, dp_Z}(\ell_i N\to \ell_j X)
&=&
\bigg[\frac{2p_T}{E_\tau y (s-m_N^2)} \bigg]
\frac{d^2\sigma}{dxdy}(\ell_i N\to \ell_j X),
\label{eq:mom-dist}
\end{eqnarray}
where a Jacobian factor is multiplied in the right-hand side of 
Eq.~\eqref{eq:mom-dist}  to convert the independent variables from $(x,y)$ to $(p_T, p_Z)$. 
\begin{figure}[htbp]
\centering
\includegraphics[width=0.45\linewidth]{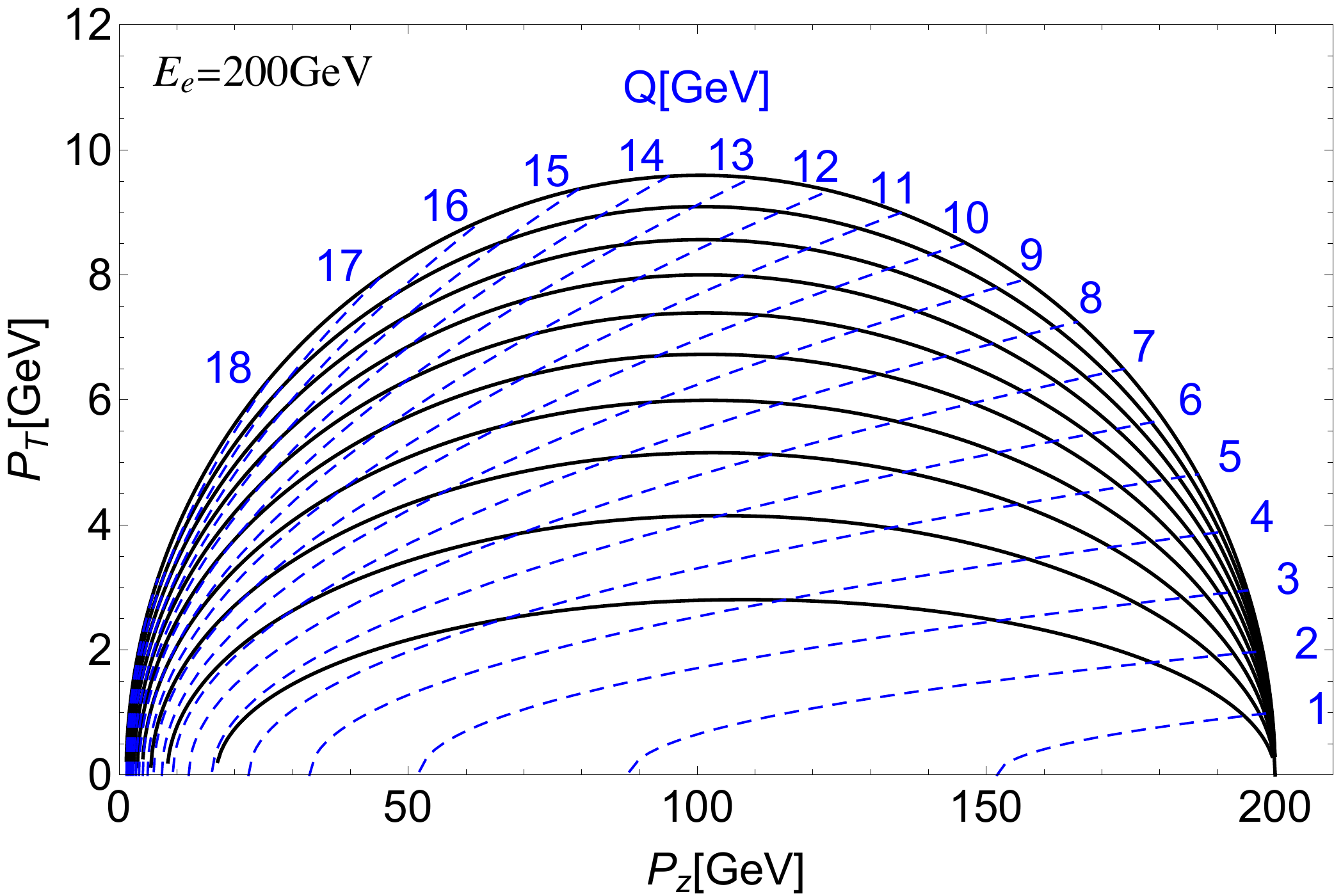}
\includegraphics[width=0.45\linewidth]{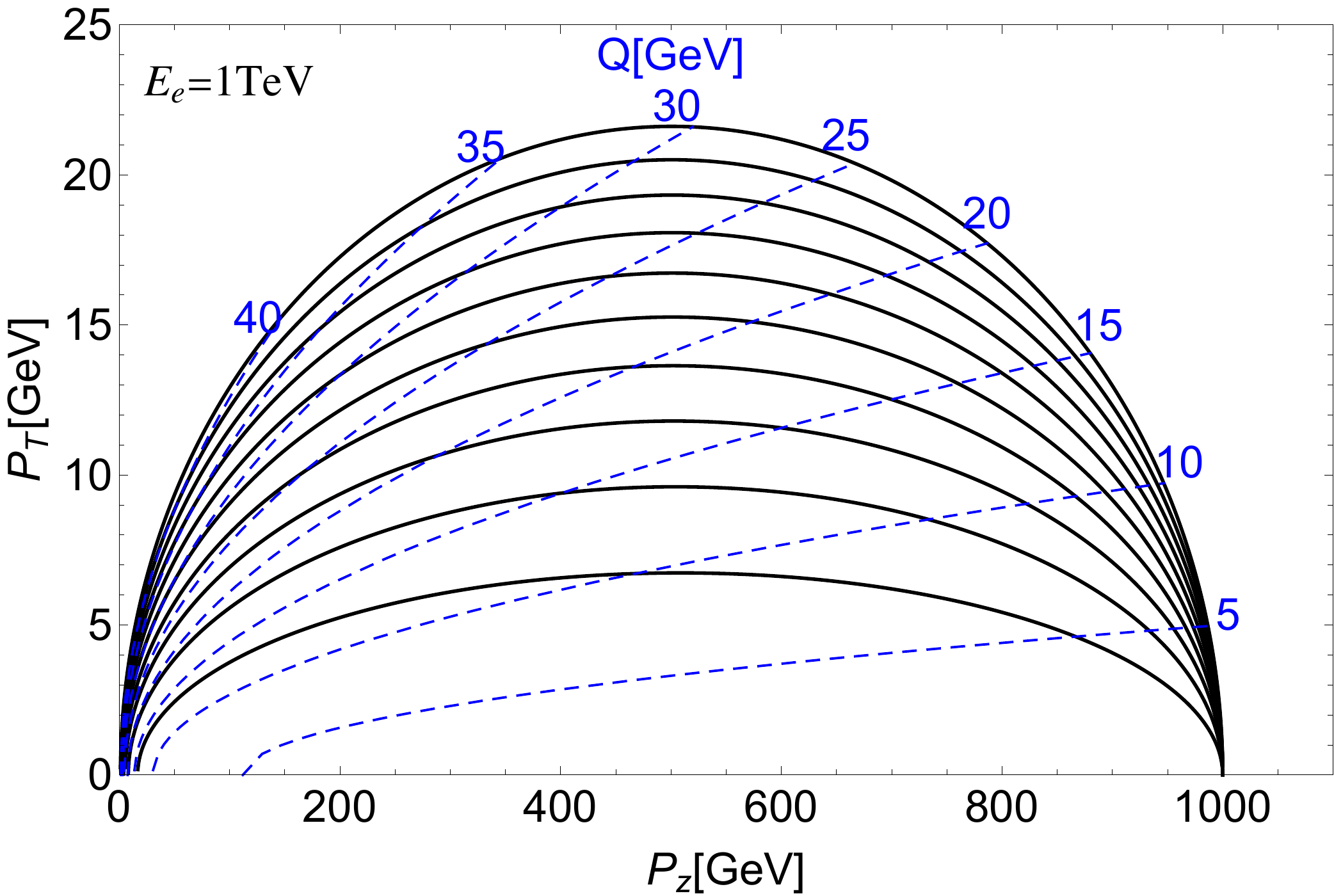}
\caption{
The physical region of the $\tau$-momentum 
for the process $e N\to \tau X$ in a fixed target experiment.
The electron beam energy is $E_e=200\, {\rm GeV}$(left panel)/$1\, {\rm TeV}$(right panel).
The contour lines of fixed $x$ (black lines) 
and fixed $Q$ (dashed blue lines) are plotted in the momentum space $(p_z, p_T)$.
}
\label{Fig:mom-space}
\end{figure}
In Fig.~\ref{Fig:mom-space} the physical region of the $\tau$-momentum 
$(p_Z,p_T)$ is plotted for the beam energies $E_e=200\, {\rm GeV}$ and $1\,  {\rm TeV}$ respectively in the left and right panels.
The black lines are the contours for $x=0.1, 0.2,\cdots, 1.0$ from smaller to larger arcs,  and the blue dashed lines are the contours for $Q$ values indicated in the plots. For CLFV signal searches in  the fixed target experiments it is of great importance to have a reliable theory prediction that covers all the physical regions of Fig.~\ref{Fig:mom-space}.  

\subsection{Heavy quark contribution to structure function}
\begin{figure}[h!]
\centering
\includegraphics[width=6cm]{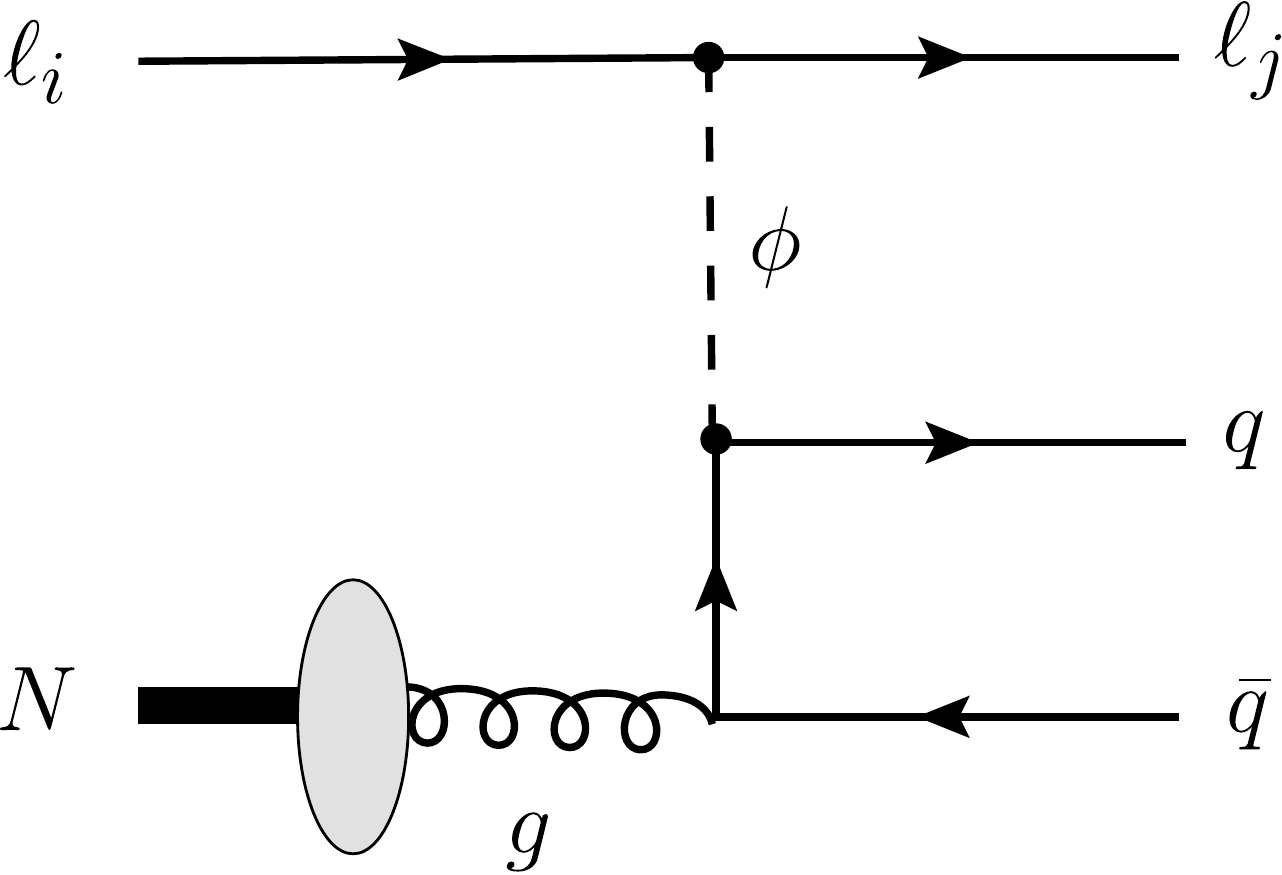}
\caption{A Feynman diagram for the heavy quark pair production
in the CLFV lepton-nucleon scattering.
}
\label{Fig:fd_diagram_02}
\end{figure}

In the following, we consider the heavy quark production 
via the CLFV scalar and pseudoscalar interactions. 
To be concrete we assume that  the heavy quark $q$ is bottom or 
charm quark, which is much heavier than the nucleon, and treat other lighter quarks as massless. 
Formulae derived here can be also applied to top quark, but the threshold of top quark pair production is too high, and we do not discuss its phenomenology in the present 
article. The heavy quark mass is so large that its 
intrinsic partonic content inside the nucleon is zero.
Yet the heavy quark can be produced in pair with its 
anti-quark via a gluon splitting $g\to q \bar{q}$ and 
subsequently $q$ or $\bar{q}$ is scattered by $\phi=S, P$ 
via the quark-mediator interaction of ${\cal L}_\phi$.  
In Fig.~\ref{Fig:fd_diagram_02} an example Feynman diagram
is shown for such a process.

The corresponding hadronic part $F_\phi$ which contributes
 to the $\ell_i N\to \ell_j q\bar{q}X$ is given by
\begin{eqnarray}
F_\phi^{\mM}(x, Q^2) &=&
\int_0^1 \frac{d\xi}{\xi} \ C_{\phi, g}^{\mM}\left(\frac{x}{\xi}, \frac{Q^2}{m_q^2}\right) f_{g/N}(\xi, \mu_f^2),
\label{eq:F_M}
\end{eqnarray}
where $C_{\phi, g}^{\mM}$ is a heavy quark contribution to the 
coefficient function. The superscript M indicates that 
the quantity is computed in massive scheme, 
where the heavy quark mass is retained in the computation:
\begin{eqnarray}
C_{S, g}^{\mM}\left(\frac{x}{\xi}, \frac{Q^2}{m_q^2}\right)
&=&
\frac{\alpha_s}{2\pi}
\frac{T_F}{(Q^2+w^2)^2 }
\Theta\left(w^2 -4m_q^2\right)
\bigg\{
 2  K w^2 (Q^2+4m_q^2)
 \nonumber\\
 &&\hspace{1cm}
 +\bigg[ Q^2(Q^2 +4m_q^2) +(w^2-4m_q^2)^2  \bigg]  \ln \frac{1+K}{1-K}
 \bigg\},
 \label{eq:CS_massive}
\\
C_{P, g}^{\mM}\left(\frac{x}{\xi},\frac{Q^2}{m_q^2}\right)
 &=&
\frac{\alpha_s}{2\pi}
\frac{T_F}{(Q^2+w^2)^2 }
\Theta\left(w^2 -4m_q^2\right)
\bigg\{
2 K w^2 Q^2 
\nonumber\\
&&
\hspace{1cm}
+ \bigg[ Q^2(Q^2 +4m_q^2) +w^4  \bigg]   \ln \frac{1+K}{1-K}  \bigg\},
 \label{eq:CP_massive}
\end{eqnarray}
with $T_F=1/2$ and  $\Theta(x)$ being the Heaviside step function. 
 The $K=\sqrt{1-4m_q^2/w^2}$ is the speed of the 
 heavy quark in the center-of-mass frame of the
 produced heavy quark pair with $w^2=(q+\xi P)^2=Q^2(\xi/x-1)$ being the 
invariant mass of $q \bar{q}$. 
Thus, the step function can be rewritten as
$\Theta(w^2-4m_q^2) =\Theta\left(\xi - \chi \right)$ 
with 
\begin{eqnarray}
\chi\equiv\left(1+\frac{4m_q^2}{Q^2}\right) x.
\label{eq:chi}
\end{eqnarray}
The appearance of $\chi$, not $x$, 
is an important mass effect due to the threshold of pair production of $q\bar{q}$. 
This leads $\chi$-rescaling prescription \cite{Tung:2001mv,Kretzer:2003it} based on an idea of so-called slow-rescaling 
\cite{Barnett:1976ak}, which will be discussed later.

\subsection{SACOT scheme}

\begin{figure}[h!]
\centering
(a)~~\includegraphics[width=4cm]{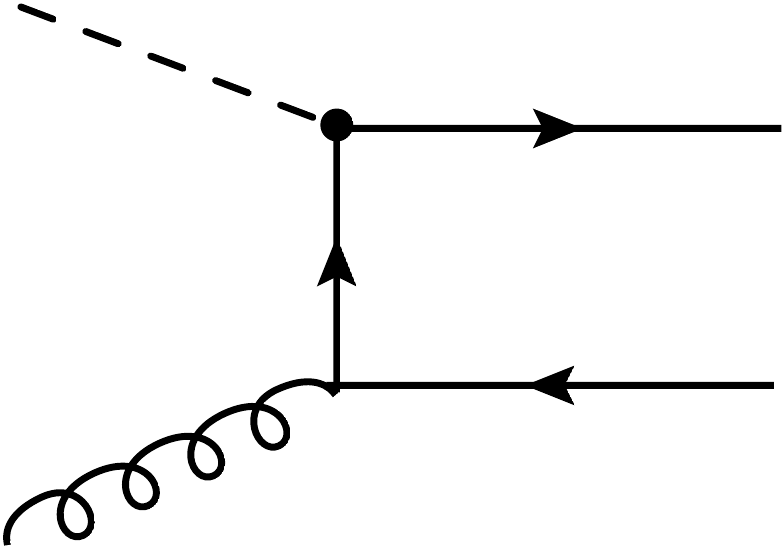}\hspace{2cm}
(b)~~\includegraphics[width=4cm]{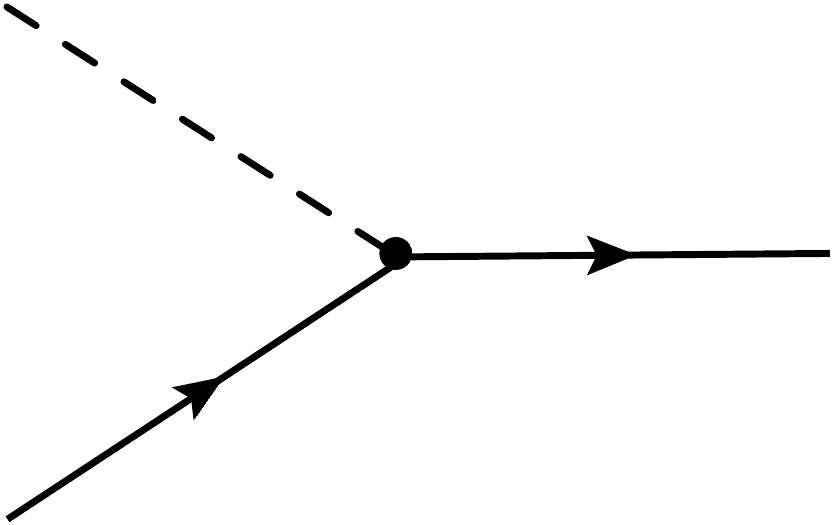}
\caption{Feynman diagrams for (a) the heavy
 quark pair production in scalar-gluon fusion, and (b) heavy quark excitation.}
\label{Fig:fd_diagrams_Ceff}
\end{figure}

In Eqs.\eqref{eq:CS_massive} and \eqref{eq:CP_massive}, the heavy quark 
contribution to the coefficient function is calculated retaining the heavy quark 
mass, which corresponds to the heavy quark pair production in 
scalar-gluon fusion (a) in  Fig.~\ref{Fig:fd_diagrams_Ceff}, 
and  its behavior around the threshold $w^2\sim 4m_q^2$ is correctly described order by order in perturbative expansion of 
$\alpha_s$.  However
when $Q^2 \gg m_q^2$ 
there appears a large logarithm $\ln(Q^2/m_q^2)$ in $C_{\phi, g}^{\mM}$, 
 and its appearance deteriorates the perturbative computation 
for $C_{\phi, g}^{\mM}$.  It can be seen by expanding $C_{\phi, g}^{\mM}$ in the limit $(m_q^2/Q^2) \to 0$:
\begin{eqnarray}
C_{\phi, g}^{\mM}
&\approx & 
\int_0^1  \frac{d\eta}{\eta}
\bigg[ C_{\phi, q}^{\mZM}\left(\frac{x}{\eta}\right) +C_{\phi, \bar{q}}^{\mZM}\left(\frac{x}{\eta}\right)\bigg]
\,
\Theta(\xi-\eta)
\bigg[
\left( \frac{\alpha_s}{2\pi} \right) 
{\cal P}_{qg} \left(\frac{\eta}{\xi} \right)  
\ln\left( \frac{Q^2}{m_q^2} \right) 
\bigg], 
\label{eq:C_g_high_energy}
\end{eqnarray}
where ${\cal P}_{qg} (z)  = T_F \big[ z^2 +(1-z)^2 \big]$ is a gluon-quark splitting function,  
and $C_{\phi, q}^{\mZM}(x), C_{\phi, \bar{q}}^{\mZM}(x)$ are 
single heavy quark contributions to the coefficient function
 in the massless limit, corresponding to the heavy quark 
 excitation diagram (b) in Fig.~\ref{Fig:fd_diagrams_Ceff}:
\begin{eqnarray}
C_{\phi, q}^{\mZM}\left(x\right) =C_{\phi, \bar{q}}^{\mZM}\left(x\right)
= \frac{1}{2} 
\delta\left(1-x\right).
\end{eqnarray}
Therefore the high-energy limit of the 
structure function can be written as
\begin{eqnarray}
F_{\rm \phi}^{\mMZ}(x,Q^2)
 &=&
 \int_0^1 \frac{d\eta}{\eta}\
2C_{\phi, q}^{\mZM}\left(\frac{x}{\eta}\right)
\frac{\alpha_s}{2\pi}\ln\left(\frac{\mu_f^2}{m_q^2}\right)
\int_\eta^1 \frac{d\xi}{\xi} {\cal P}_{qg}\left( \frac{\eta}{\xi}\right) 
 f_{g/N}(\xi, \mu_f^2)
\label{eq:F_M0x}
\end{eqnarray} 
where the superscript M0 denotes the leading contribution 
 in the massless limit of the massive scheme structure function.
Namely the $F_\phi^{\mMZ}$ contains the mass singularity of
 the massive structure function, and $(F_\phi^{\mM}-F_\phi^{\mMZ})$ 
 is finite in the massless limit. 
Here, the original collinear logarithm 
existing in Eq.~\eqref{eq:C_g_high_energy} is separated as 
$\alpha_s \ln(Q^2/m_q^2)
=\alpha_s \ln(Q^2/\mu_f^2) +\alpha_s \ln(\mu_f^2/m_q^2)$,
and the mass singularity $\alpha_s \ln(\mu_f^2/m_q^2)$ is absorbed 
in Eq.~\eqref{eq:F_M0x}.

The relation between the high-energy limit of the massive quark 
contribution and the $F_\phi^{\mMZ}$ structure function
is nothing but the factorization theorem \cite{Collins:1998rz} 
of collinear singularities
for the QCD structure function, which can be utilized to resum the 
large collinear logarithms to all orders in $\alpha_s$ 
to make the structure functions stable at high energy. 
The extracted collinear logarithm $\alpha_s \ln(\mu_f^2/m_q^2)$ 
has a form,  that can be resummed to all orders in $\alpha_s$ by means of 
the standard DGLAP evolution equation, leading  to 
zero-mass (ZM) scheme structure function $F_\phi^{\mZM}$:
\begin{eqnarray}
F_\phi^{\mZM}(x, Q^2) &=&
\int_0^1 \frac{d\eta}{\eta}\
2 C_{\phi, q}^{\mZM} \left(\frac{x}{\eta}\right) 
\big[ f_{q/N} (\eta, \mu_f^2)+ f_{\bar{q}/N}(\eta, \mu_f^2) \big]
\nonumber\\
&=&
\frac{1}{2}\big[f_{q/N} (x, \mu_f^2)+f_{\bar{q}/N}(x, \mu_f^2)\big].
\label{eq:F_ZMx}
\end{eqnarray}
The heavy quark PDFs $f_{q/N}$ and $f_{\bar{q}/N}$
introduced in Eq.~\eqref{eq:F_ZMx} are generated by the gluon splitting into $q\bar{q}$ 
through the DGLAP evolution equation, and thus reduces to Eq.~\eqref{eq:F_M0x}
at the leading order in $\alpha_s$ expansion:
\begin{eqnarray}
f_{q/N}(x, \mu_f^2) 
&=& 
\frac{\alpha_s}{2\pi }
\ln\left(\frac{\mu_f^2}{m_q^2}\right)
\int_{x}^1 \frac{d\xi }{\xi}\  {\cal P}_{qg} 
\left(\frac{x}{\xi}\right) f_{g/N}(\xi, \mu_f^2)
+{\cal O}(\alpha_s^2).
\end{eqnarray}

So far we have defined three types of structure function $F_\phi^{\rm M}, F_\phi^{\mMZ}$
and $F_\phi^{\mZM}$. The M-scheme structure function is reliable 
in low-$Q^2$ 
region but unreliable in the high-$Q^2$ region due to the mass singularity, while 
the ZM scheme structure function is 
reliable in the high-$Q^2$ region but unreliable in the low-$Q^2$ region. Therefore 
these two are complementary to each other. According to these observations, a scheme for the structure function was constructed, which includes heavy quark
mass effects near $q\bar{q}$ threshold region and also the large
logarithm resummation making the structure function stable even 
at high-$Q^2$ region. The result is a new structure function which consists of three terms as 
\begin{eqnarray} 
F_\phi^{\rm SACOT}\left(x, Q^2\right)
&=&
 F_\phi^{\rm M}\left(x, Q^2\right)
+F_\phi^{\mZM}\left(x,Q^2\right)
-F_\phi^{\rm sub}(x,Q^2).
\label{eq:F_SACOTx}
\end{eqnarray}
Here the second and third terms are computed by setting 
heavy quark masses to zero, and this is called Simplified ACOT (SACOT) scheme \cite{Kramer:2000hn,Aivazis:1993pi}.
 The first term $F_\phi^{\rm M}$ is the contribution of a heavy quark pair to the structure function where the heavy quarks $q, \bar{q}$ are massive and produced 
 in the scalar-gluon fusion.
The second term $F_\phi^{\mZM}$ is the contribution of heavy quark excitations, and plays a role to resum the large-collinear logarithm
$\ln(Q^2/m_q^2)$ to all orders in $\alpha_s$ by use of 
the heavy quark and anti-quark PDFs $\{f_{q/N}, f_{\bar{q}/N}\}$. 
By the construction of $F_\phi^{\mZM}$, there is a double counting
of large-logarithm between $F_\phi^{\rm M}$ and $F_\phi^{\mZM}$, 
and this double counting effect should be subtracted by the last term:
\begin{eqnarray}
F_\phi^{\rm sub}(x,Q^2) 
&=&
F_\phi^{\mMZ}(x,Q^2).
\label{eq:F_subx}
\end{eqnarray}
The physical picture of the SACOT scheme is as follows:  the 
first term $F_\phi^{\rm M}$ contains all the mass effects order 
by order in the expansion of powers of $\alpha_s$, 
and the second term $F_\phi^{\rm ZM}$
adds the large logarithmic corrections of $(\alpha_s \ln(Q^2/m_q^2))^n$ for $n=1,\cdots, \infty$ to improve the high-$Q^2$ behavior 
avoiding the double counting by the last term $F_\phi^{\rm sub}$.

\subsection{Improvements for threshold behavior}
 Although we are working on the  leading order formulation for the 
 heavy quark structure function, there are a number of important 
effects, which can be incorporated in the present computation,
 on the threshold behavior of the structure functions.
We take into account such improvements here.
 
 The first such effect is one by so-called $\chi$-rescaling prescription\cite{Tung:2001mv,Kretzer:2003it}.
 The $\chi$-rescaling prescription aims to incorporate threshold kinematics
 of the heavy quark production into  the massless structure function (ZM) 
 and the massless limit (M0) of  the massive structure function using 
 $\chi(x,Q^2)$ introduced in Eq.\eqref{eq:chi}
  instead of $x$ variable. This defines structure functions in \ZMC~
   and \MZC~ schemes:
 \begin{eqnarray}
F_\phi^{\mZMC}(x,Q^2)&=& \frac{1}{2}\big[f_{q/N} (\chi(x, Q^2), \mu_f^2)+f_{\bar{q}/N}(\chi(x,Q^2), \mu_f^2)\big],
\label{eq:F_ZMchi}
\\
F_{\rm \phi}^{\mMZC}(x,Q^2)
 &=&
\frac{\alpha_s}{2\pi}\ln\left(\frac{\mu_f^2}{m_q^2}\right)
\int_{\chi(x,Q^2)}^1 \frac{d\xi}{\xi} {\cal P}_{qg}\left( \frac{\chi(x,Q^2)}{\xi}\right) 
 f_{g/N}(\xi, \mu_f^2).
 \label{eq:F_M0chi}
\end{eqnarray}
 The subtraction term with $\chi$-rescaling is similarly defined by
 $F_\phi^{\mSubC}(x,Q^2)=F_\phi^{\mMZC}(x, Q^2)$. 
With these structure functions \AC~ scheme is also defined
as 
\begin{eqnarray}
F_{\phi}^{\mAC}(x, Q^2) 
&\equiv  & 
F_\phi^{\rm M}\left(x, Q^2\right)
+
\big[ F_{\phi}^{\mZMC}\left(x,Q^2\right) 
       -F_{\phi}^{\mSubC}(x,Q^2)\big].
\label{eq:F_SACOTchi}
\end{eqnarray}
 
The second improvement is a choice of the factorization 
scale $\mu_f^2$. In the traditional DIS analysis the scale choice 
$\mu_f^2=Q^2$ is commonly used assuming $Q^2\gg m_q^2$. 
However we are interested in not only high-$Q^2$ but also
 the threshold region of the heavy quark production, especially the contribution from the 
gluon fusion of Fig.~\ref{Fig:fd_diagram_02}, for which $Q^2$
 can be the same order with $m_q^2$ or 
even smaller than $m_q^2$. For such a case, 
the scale choice $\mu_f^2=Q^2$ is not suitable,
 and one needs to take a proper physical scale of the process.
To ensure that the factorization scale 
does not become too low, we take the scale as $\mu_f^2=\mu_Q^2$ 
with 
\begin{eqnarray}
\mu_{Q}^2 &=& Q^2\, 
\big[
c\, (1-z_m)^n\,  \Theta(1-z_m) +z_m
\big],
\label{eq:mu_Q}
\end{eqnarray}
where $z_m=m_q^2/Q^2$, $n=2$ and $c=0.5$ are chosen following Ref.\cite{Aivazis:1993pi}.

The \AC~ structure function interpolates between massive 
 $F_\phi^{\mM}$ and massless $F_\phi^{\mZMC}$ structure functions. 
 Ideally $F_\phi^{\mAC}$ is supposed to reduce to the massive one in the  low-$Q^2$ region, while in the high-$Q^2$ region 
to the massless one.  
This expectation holds parametrically at each order in expansion in powers of $\alpha_s$, but numerically it can happen that 
the $F_\phi^{\mAC}$ does not converge well to $F_\phi^{\mM}$ near the heavy quark threshold $Q^2\sim m_q^2$. 
If the cancellation between $F_\phi^{\mZMC}$ and $F_\phi^{\mSubC}$ in low-$Q^2$ region is 
not effective, it must be due to unsuppressed higher-order terms in powers of $\alpha_s$ resummed into 
$F_\phi^{\mZMC}$.  Namely the terms $F_\phi^{\mZMC}-F_\phi^{\mSubC}={\cal O}(\alpha_s^{k} L^k)~(k\geq 2)$
are too large in the region where the massless approximation is not 
trustable. 
Easy solution to avoid this trouble is to suppress the $F_\phi^{\mZMC}-F_\phi^{\mSubC}$ 
in low-$Q^2$ region by hand. Thus we define improved structure functions
for \ZMC ~ and \MZC ~ schemes:
\begin{eqnarray}
F_{\phi, \rm thr}^{\mZMC}\left(x,Q^2\right)
&=&F_\phi^{\mZMC}\left(x,Q^2\right)S_{\rm thr}(m_q^2/Q^2),
\label{eq:F_ZMchi_thr}
\\
F_{\phi, \rm thr}^{\mMZC}\left(x,Q^2\right)
&=&
F_{\phi,\rm thr}^{\rm sub\chi} \left(x,Q^2\right)
=
F_\phi^{\mMZC}(x,Q^2)S_{\rm thr}(m_q^2/Q^2),
\label{eq:F_M0chi_thr}
\end{eqnarray}
where $S_{\rm thr}(m_q^2/Q^2)$ is a function which suppress the 
structure functions in \ZMC ~ and \MZC ~ schemes forcing them 
to smoothly match with correct threshold behavior. 
The functional form of $S_{\rm thr}(z)$ 
is somewhat arbitrary but the only requirement is $S_{\rm thr}(z)\stackrel{z\to 0}{\to} 1$ for the large-logarithm resummation for high $Q^2$. For simplicity we choose
\begin{eqnarray}
S_{\rm thr}(z)=
\left(1-z\right)^2  ~
\Theta\left(1-z\right),
\label{eq:S_thr}
\end{eqnarray}
in the same form introduced in Ref.\,\cite{Forte:2010ta}. 
Taking all the improvements we define the \AC~(thr) structure function by 
\begin{eqnarray}
F_{\phi, {\rm thr}}^{\mAC}(x, Q^2) &\equiv  & 
F_\phi^{\mM}\left(x, Q^2\right)
+
\big[ F_{\phi,\rm thr}^{\mZMC}\left(x,Q^2\right) 
       -F_{\phi,\rm thr}^{\mSubC}(x,Q^2)\big].
\label{eq:SACOTchi_thr}
\end{eqnarray}
The combination $\big[ F_{\phi, \rm thr}^{\mZMC} -F_{\phi, \rm thr}^{\mSubC}\big]$ ensures that 
the \AC~(thr) structure function reduces to massive one near heavy quark threshold. 
For the \AC~(thr) scheme,
we always adopt the scale setting $\mu_f^2=\mu_Q^2$ 
and the threshold factor $S_{\rm thr}(m_q^2/Q^2)$.

\section{Numerical analysis: Structure functions }

We analyze the structure functions for the scalar interaction, i.e. $\phi=S$.
The pseudoscalar case ($\phi=P$) is much the same as the scalar case, 
and we refrain from showing the numerical results for the pseudoscalar case. 
In the present article we use \text{CT14} LO PDFs \cite{Dulat:2015mca}, and present the numerics
for the proton with $m_N=938\, {\rm MeV}$ as a target nucleon.
The scale choice $\mu_f^2=\mu_Q^2$ is adopted for all the structure functions. 
\subsection{Effect of $\chi$-rescaling on the structure functions}
\begin{figure}[htbp]
\centering
\includegraphics[width=15cm]{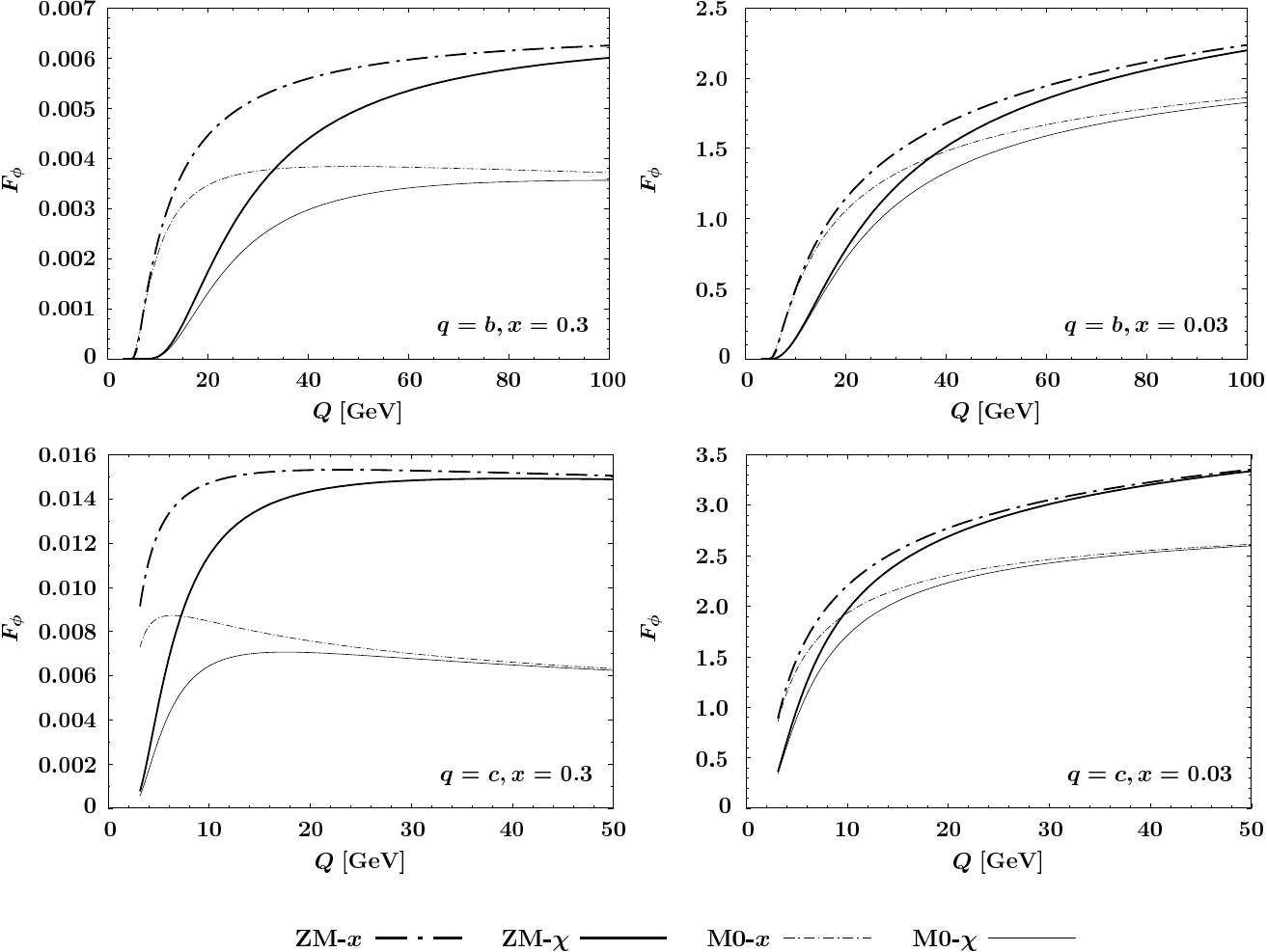}
\caption{
The structure functions $F_\phi$ computed 
in the \ZMC~, \ZMx~, \MZC~, and \MZx~ schemes
are plotted as functions of $Q$ for $x=0.3$ and $x=0.03$ 
in the left and right columns, respectively. The upper and lower rows show the contributions of the bottom ($q=b$) 
and charm ($q=c)$ quarks, respectively.}
\label{Fig:str_func_0}
\end{figure}

Let us first discuss the importance of $\chi$-rescaling for the 
structure functions in the ZM and M0 schemes in low-$Q^2$ region.
For the QCD structure functions effects of $\chi$-rescaling are
discussed in Refs. \cite{Tung:2001mv, Kretzer:2003it, Stavreva:2012bs}. 
To  make the difference explicit between the structure functions using 
$x$ and $\chi$ variables, we introduce the following notations: 
$F_\phi^{\mZMx}(=F_\phi^{\mZM}), F_\phi^{\mMZx}(=F_\phi^{\mMZ})$.
Hereafter we call them $x$-scheme structure functions when the  
distinction is necessary. 
 The ZM and M0 structure functions ($x$-scheme) in Eqs.\eqref{eq:F_M0x} and \eqref{eq:F_ZMx}
  do not contain any information on the heavy quark threshold. 
Thus there is no reason to trust $F_\phi^{\mZMx}, F_\phi^{\mMZx}$ near the heavy quark threshold, but it is still expected to have an improvement on 
 the threshold behavior near $Q^2=m_q^2$ by the $\chi$-rescaling.
In Fig.~\ref{Fig:str_func_0} the ZM and M0 structure 
functions are plotted as functions of $Q$. 
The structure functions 
for $x=0.3$ and $0.03$ are shown in the left and right columns, and those
for bottom and charm quarks are shown in the upper and lower rows.
The solid lines represent the \ZMC~ and \MZC~ structure
 functions, and the dot-dashed
 lines represent the  corresponding $x$-scheme structure functions.
One can see that the effect of $\chi$-rescaling
 is huge and plays an essential role for the threshold suppression
 near $Q\sim m_q$ ($m_b=4.75\, {\rm GeV}$ for bottom quark 
 and $m_c=1.3\, {\rm GeV}$ for charm quark).
The $x$-scheme structure functions are unrealistically large
 for $Q\sim m_q$, and it is remarkable that the $\chi$-rescaling improves the 
unphysical behavior of the massless structure functions nicely. 
 The effect of $\chi$-rescaling is decreasing in high-$Q^2$ 
 region, and the difference between the use of $x$ and
 $\chi$-rescaling is negligible at $Q=100\, {\rm GeV}$ 
 ($Q=50\, {\rm GeV}$) for bottom (charm) quark. 
 We conclude that the $\chi$-rescaling is effective only in the low-$Q^2$ region
 of $Q\leq 100\, {\rm GeV}~(Q\leq 50\, {\rm GeV})$ for 
 bottom (charm) quark.
 
\subsection{Massive vs. zero-mass schemes, and SACOT scheme}
\begin{figure}[htbp]
\centering
\includegraphics[width=13cm]{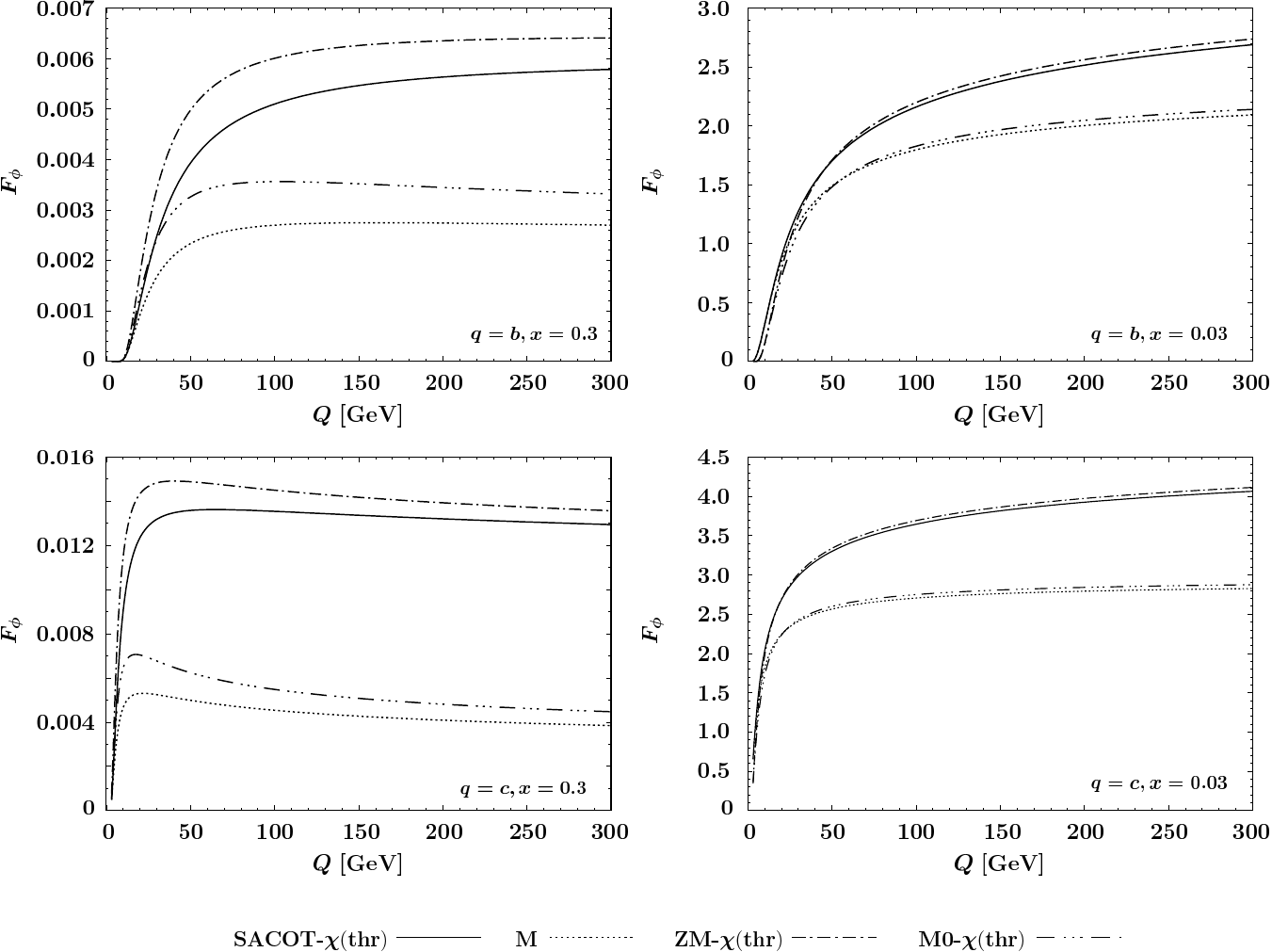}
\caption{
The structure functions $F_\phi$ in the 
ZM-$\chi$, M0-$\chi$, M, and SACOT-$\chi$(thr) schemes are plotted as functions 
of $Q$ for $x=0.3$ and $x=0.03$  in the left and right columns, respectively.
The upper and lower rows show the
contributions of bottom  ($q=b$) and charm  ($q=c$) quarks, respectively.}
\label{Fig:str_func_SACOT_HighQ}
\end{figure}

\begin{figure}[htbp]
\centering
\includegraphics[width=13cm]{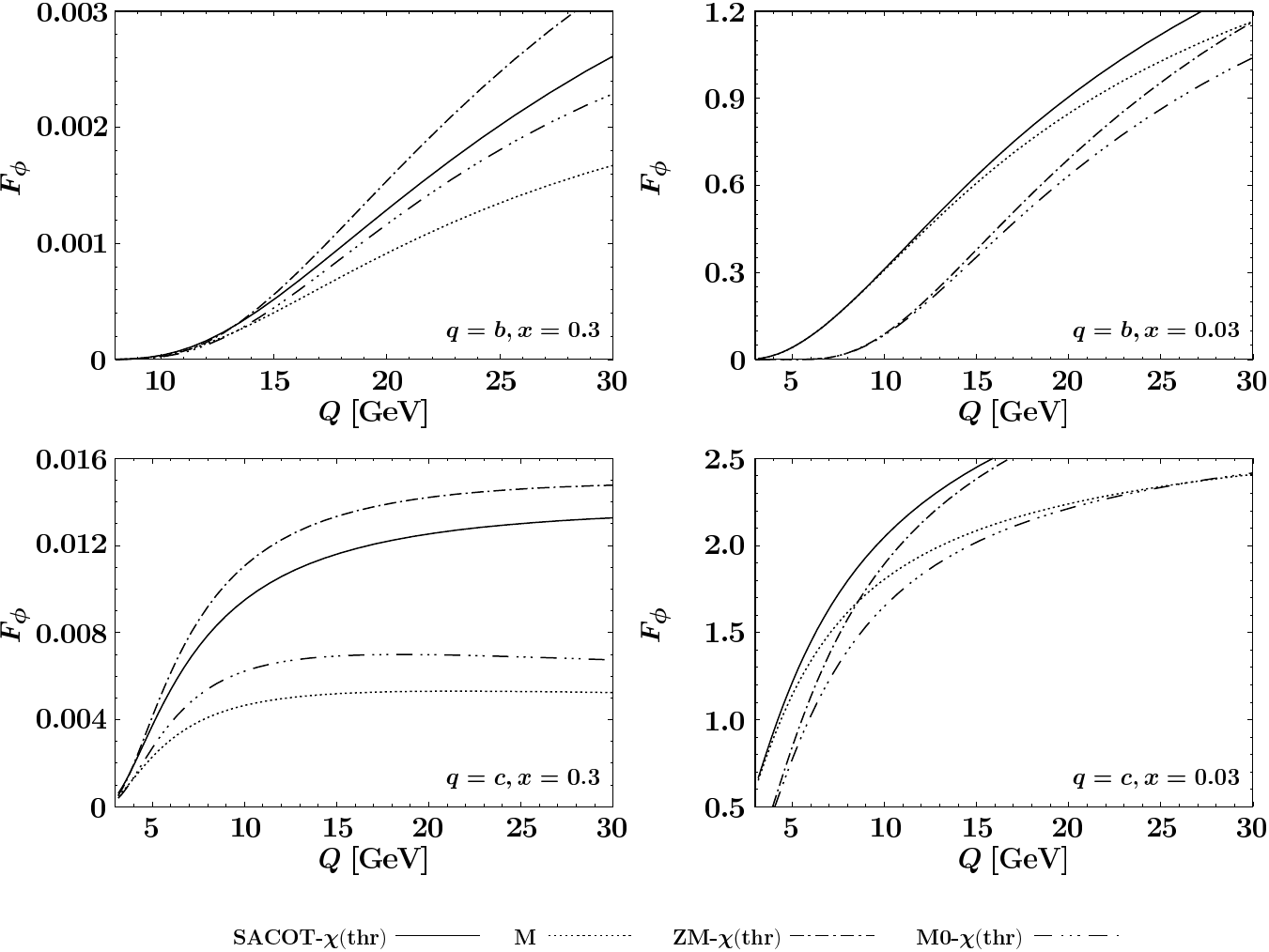}
\caption{
Same as Fig.~\ref{Fig:str_func_SACOT_HighQ} but for low-$Q^2$ region:
$x=0.3$ (left column) and $x=0.03$ (right column) for 
bottom ($q=b)$ and charm ($q=c$) quarks in the two rows.}
\label{Fig:str_func_SACOT_LowQ}
\end{figure}

Here we will compare the structure functions in M, \ZMC~(thr), and \AC~(thr) schemes (for the last two schemes $\chi$-rescaling and the threshold factor are adopted).
In Fig.~\ref{Fig:str_func_SACOT_HighQ} the heavy quark contribution to 
the structure functions are plotted as functions of $Q$ 
(up to $Q=300\, {\rm GeV}$) for $x=0.3$ and $x=0.03$.
For $x=0.3$ it is seen that  the curves for \AC~(thr) and \ZMC~(thr) are broadly
similar to each other for bottom and charm quarks, 
 and their values are much larger than the massive scheme result.
The difference between the curves of the \ZMC~(thr) and \AC~(thr) schemes
 is explained by the difference between the M and \MZC ~ schemes 
 because 
$ (F_{\phi,\rm thr}^{\mAC}-F_{\phi,\rm thr}^{\mZMC})=(F_\phi^{\rm M}-F_{\phi,\rm thr}^{\mMZC})$. 
Remember that $F_{\phi,\rm thr}^{\mMZC}$ ~ is the subtraction term of the \AC~(thr) scheme.
For $x=0.03$ this agreement between the \AC~(thr) and \ZMC~(thr) schemes
becomes tighter, because $(F_\phi^{\mM}-F_{\phi,\rm thr}^{\mSubC})_{x=0.03}\simeq 0$. 
It should be noted that the structure functions for $x=0.03$ are more than two orders of magnitude larger than those of $x=0.3$.
 
The same plots as Fig.~\ref{Fig:str_func_SACOT_HighQ},
 focusing on the range below $Q=30\, {\rm GeV}$,
 are shown in Fig.~\ref{Fig:str_func_SACOT_LowQ}.
It will be seen later that  the contributions of  
the structure functions to the CLFV DIS cross section 
for $E_e\leq 1\, {\rm TeV}$ are dominated by the $Q$ values
of this range. Therefore Fig.~\ref{Fig:str_func_SACOT_LowQ}
 is more relevant than Fig.~\ref{Fig:str_func_SACOT_HighQ} 
 for the fixed target experiments at present and in near 
 future. For bottom quark, 
 the M scheme structure function is closer to the \AC~(thr) scheme than 
 \ZMC~(thr) for $x=0.03$. For $x=0.3$ both the curves of M and \ZMC~(thr) 
 are away from the \AC~(thr) but their magnitudes are less than 
 percent level compared to those at $x=0.03$.  
 For charm quark at $x=0.03$, the M scheme structure function is closer to the \AC~(thr) for very low $Q$. 
 For $Q\geq 8.6\, {\rm GeV}$ the \ZMC~(thr) scheme  becomes closer to \AC~(thr)
especially for the large-$Q^2$ region.
 For charm quark at $x=0.3$, the magnitudes of the structure functions are less than a percent level compared to those at $x=0.03$. 

\subsection{Threshold factor}

In Figs.~\ref{Fig:str_func_SACOT_HighQ} and \ref{Fig:str_func_SACOT_LowQ}
the threshold factor $S_{\rm thr}$ had been taken into account 
for the structure functions $F_\phi^{\mAC}$, $F_\phi^{\mZMC}$, and $F_\phi^{\mMZC}$. 
The effect of the threshold factor is limited in low-$Q^2$ 
region: $S_{\rm thr}(m_b^2/Q^2)=0.60, 0.89, 0.95$ for 
$Q=10, 20, 30\, {\rm GeV}$ for bottom quark, and 
$S_{\rm thr}(m_c^2/Q^2)=0.87, 0.97, 0.99$ for $Q=5, 10, 15\, {\rm GeV}$ 
for charm quark. These values give 
about $40\%, 11\%, 5\%$  suppressions for the structure functions 
$F_{\phi, \rm thr}^{\mZMC}, F_{\phi, \rm thr}^{\mMZC}$ of 
bottom quark at $Q=10, 20, 30\, {\rm GeV}$ compared to those
without the threshold factor, and about $13\%, 3\%, 1\%$ suppressions
for the charm quark case at $Q=5, 10, 15\, {\rm GeV}$.
For the \AC~(thr) scheme 
the relative size of $(F_{\phi, \rm thr}^{\mMZC}-F_{\phi, \rm thr}^{\mZMC})=S_{\rm thr}(m_q^2/Q^2)(F_{\phi}^{\mMZC}-F_{\phi}^{\mZMC})$
 to $F_{\phi}^{\mM}$ determines the effect of $S_{\rm thr}$:
\begin{eqnarray}
\frac{F_{\phi, {\rm thr}, b}^{\mAC}}{F_{\phi, b}^{\mAC}}=
\bigg\{
\begin{array}{ll}
0.94, ~0.97, ~0.98 &(x=0.3)
\\
0.99, ~0.99, ~1.0 & (x=0.03)
\end{array}~~~
 \mbox{for}~Q=10, 20,30\, {\rm GeV}, 
\end{eqnarray}
for bottom quark, and 
\begin{eqnarray}
\frac{F_{\phi, {\rm thr}, c}^{\mAC}}{F_{\phi, c}^{\mAC}}=
\bigg\{
\begin{array}{ll}
0.95, ~0.98, ~0.99 &(x=0.3)
\\
0.99, ~1.0, ~1.0 & (x=0.03)
\end{array}~~~
 \mbox{for}~Q=5, 10, 15\, {\rm GeV}, 
\end{eqnarray}
for charm quark.  It is understood that a major part of the threshold suppression is 
already taken care by the $\chi$-rescaling, and the effect of $S_{\rm thr}$
became minor for the structure function in the \AC~(thr) scheme.

\section{Numerical Analysis: Cross sections}
In this section, we investigate how effective the 
SACOT-$\chi$ scheme and the others are for the description of CLFV process associated with the heavy quark pair productions. 
As a continuation of the previous section only the scalar case ($\phi=S$) will be studied. 
As is explained in the previous section, the structure functions are functions of $x$ and $Q^2$, and each scheme of the structure functions has validity regions for a specific $Q^2$ range.
However our concerns are the total cross sections $\sigma$ and differential distributions $d\sigma/d\bm{p}_\tau$ of the final $\tau$-momentum. There arises a question of which scheme is the most relevant, and which scheme is the most effective for the description of the CLFV DIS in the full kinematical range of the cross section, which will be answered in this section.
  
For definiteness we take electron and tau lepton 
as initial and final leptons, respectively, namely 
$i=e$ and $j=\tau$.
In the numerical  analysis we ignore the electron mass, and take the following mass values:
\begin{eqnarray}
m_b= 4.75\, {\rm GeV}, ~ m_c=1.3\, {\rm GeV},~m_\tau=1.78 \, {\rm GeV}.
\end{eqnarray}
The CLFV couplings $\rho_{ij}^\phi$ and quark-mediator coupling $\rho_{qq}^\phi$ are a priori not known and we set their values 
as
\begin{eqnarray}
|\rho^\phi_{qq}|^2=|\rho_{ij}^\phi|^2 +|\rho_{ji}^\phi|^2=1.
\label{eq:cc_set_to_1}
\end{eqnarray}
These coupling constants determine the normalization 
of the CLFV cross section, and therefore our numerical 
results need to be multiplied by mode-dependent prefactors 
to match them with experimental values to be measured. 
For the choice of the factorization scale $\mu_f^2$ we adopt  
$\mu_f^2=\mu_Q^2$
defined in Eq.~\eqref{eq:mu_Q}. In our numerical analysis, we have applied a kinematical cut of  $Q\geq 1.3\, {\rm GeV}$ 
and $W \equiv \sqrt{(P+q)^2}\geq 1.4\, {\rm GeV}$ to ensure that the processes we are considering are in 
perturbative and deep-inelastic r\'{e}gime, though it turned out that the effect of the cut is tiny and numerically 
negligible for the CLFV DIS associated with the heavy quark pair productions.

\subsection{Zero-mass schemes}

The ZM schemes are the most frequently used schemes
for DIS involving heavy quarks as well as the light quarks.
This is so even for the CLFV DIS associated with bottom and
 charm quark productions because of 
their computational simplicities.
However,  the use of massless approximation cannot 
be justified for low $Q^2$, 
and a reliable computational scheme should 
be the massive scheme there. Nevertheless, it is worthwhile to 
know limitations of  the ZM schemes for the CLFV cross sections.
Here we investigate the ZM-$x$ and ZM-$\chi$ schemes to clarify 
their applicability for the cross section in relatively low collision 
energies. As example cases, we simulate the cross section 
for $E_e=200\, {\rm GeV}$ and $E_e=1\, {\rm TeV}$. 
Here we do not include the threshold factor $S_{\rm thr}$ 
for the  ZM schemes because it cuts away the low-$Q^2$ 
region and the difference between \ZMx ~ and \ZMC ~ are naturally 
suppressed.
\begin{figure}[htbp]
\centering
\includegraphics[width=0.35\linewidth]
{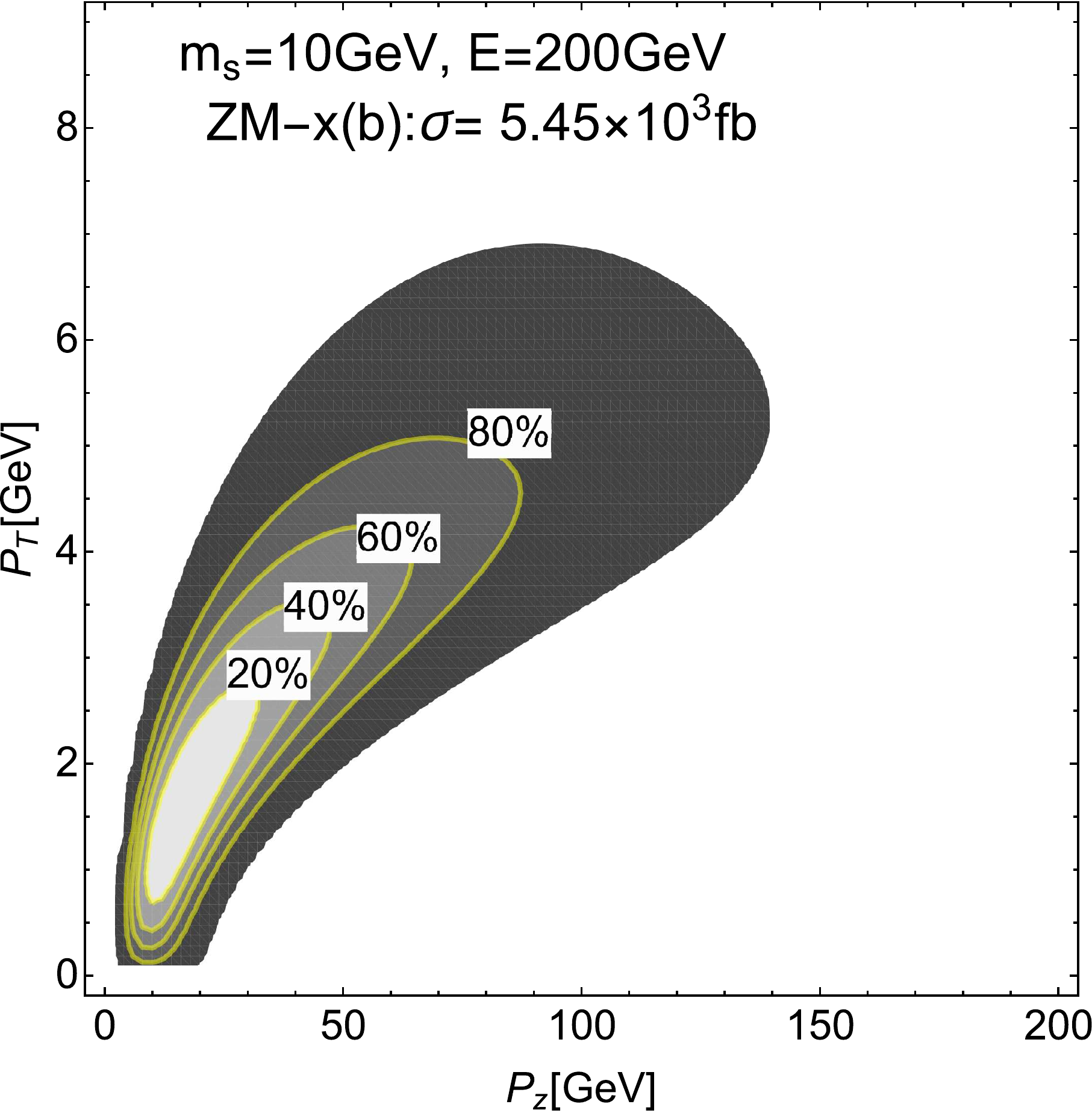}\hspace{1cm}
\includegraphics[width=0.35\linewidth]
{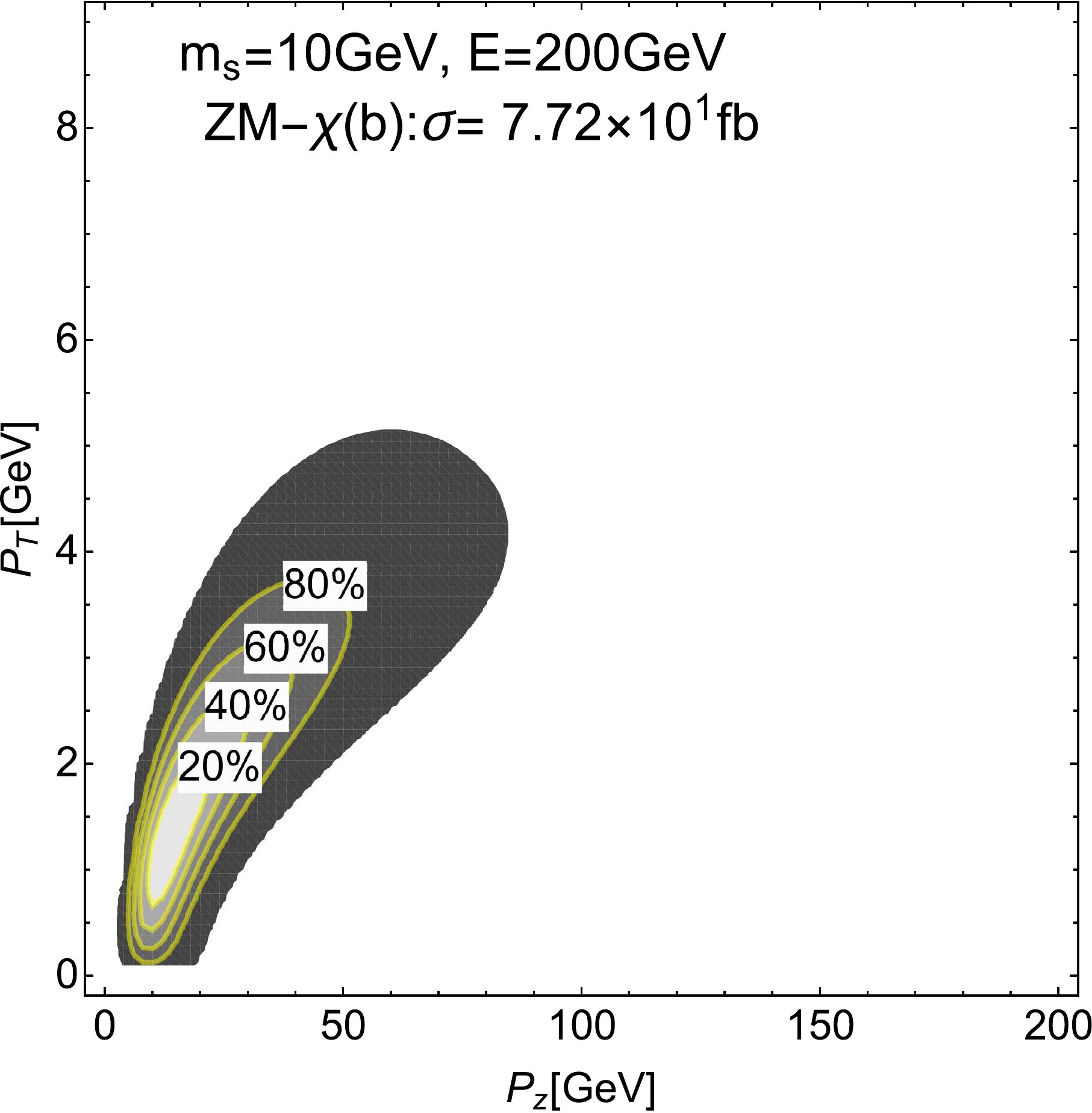}\\
\vspace{5mm}
\includegraphics[width=0.35\linewidth]
{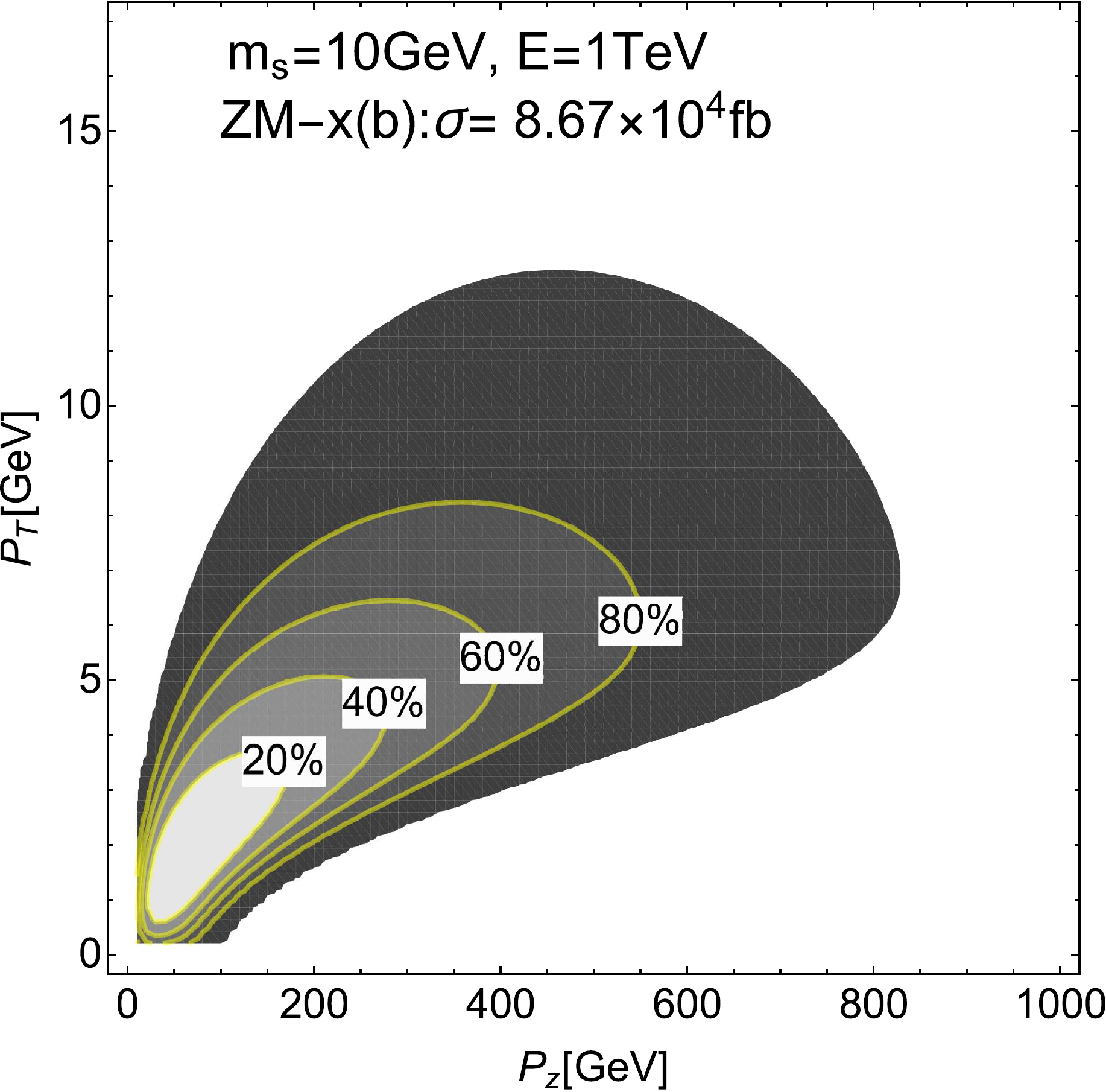}\hspace{1cm}
\includegraphics[width=0.35\linewidth]
{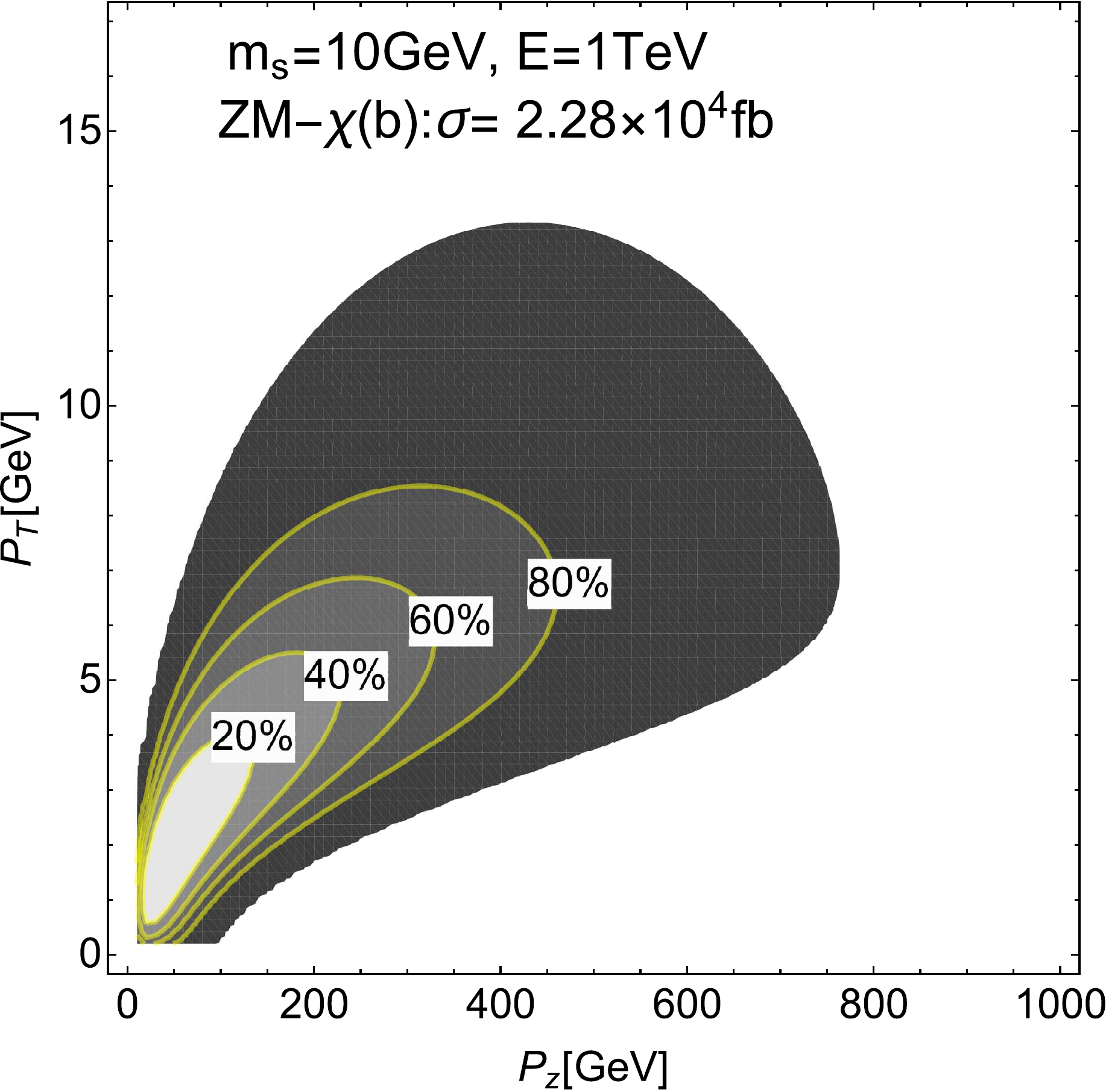}
\caption{
The $\tau$-momentum distributions Eq.~\eqref{eq:mom-dist} for 
the bottom quark production in the \ZMx ~(left column) and  
\ZMC~ (right column) schemes.
The distributions for the electron beam energies 
of $E_e=200\, {\rm GeV}$ (upper row) and $E_e=1\, {\rm TeV}$ 
(lower row) with scalar mass $m_S=10\, {\rm GeV}$ are shown.}
\label{Fig:b_dist_rescaling}
\end{figure}

In Fig.~\ref{Fig:b_dist_rescaling} we show the
$\tau$-momentum distributions for the CLFV DIS associated with bottom quark production for $E_e=200\, {\rm GeV}$ and $E_e=1\, {\rm TeV}$
 in the two rows.
The scalar mass is set to $m_S=10\, {\rm GeV}$. 
The left  and right columns show the distributions 
in the \ZMx~ and \ZMC~ schemes, respectively.
The contour lines are labeled by percentages ($20\%, 40\%, 60\%, 80 \%$)
of the cross section of the enclosed region normalized to their total cross section. 
The colored region contains $99\%$ of total events of the process.
The value of the scalar mass $m_S$, the beam energy $E_e$,  and the total cross section $\sigma$ are shown inside each panel.
For $E_e=200\, {\rm GeV}$ the effect of 
$\chi$-rescaling is huge suppression for the overall normalization $\sigma$, 
and the physical regions of $\chi$-scheme distributions are shrunk into a 
smaller region than those of $x$-schemes. 
For $E_e=1\, {\rm TeV}$ the effect 
of $\chi$-rescaling is still large for the overall normalization 
but weak compared to the case of $E_e=200\, {\rm GeV}$.
The ratio of the total cross section in $x$-scheme to 
that in $\chi$-scheme is $\sigma^{\mZMx}_b/\sigma^{\mZMC}_b\sim 70~(3.8)$
for $E_e=200\, {\rm GeV} ~(1\, {\rm TeV})$. The large enhancements
 of the total cross sections in $x$-schemes hold even in the case
  of heavy scalar mass. For instance,  taking $m_S=10^5\, {\rm GeV}$, 
   the cross section ratio is 
   $\sigma_b^{\mZMx}/\sigma_b^{\mZMC}\sim 73 ~(3.0)$ 
 for $E_e=200\, {\rm GeV}~(1\, {\rm TeV})$
for the bottom quark production.

\begin{figure}[htbp]
\centering
\includegraphics[width=0.35\linewidth]
{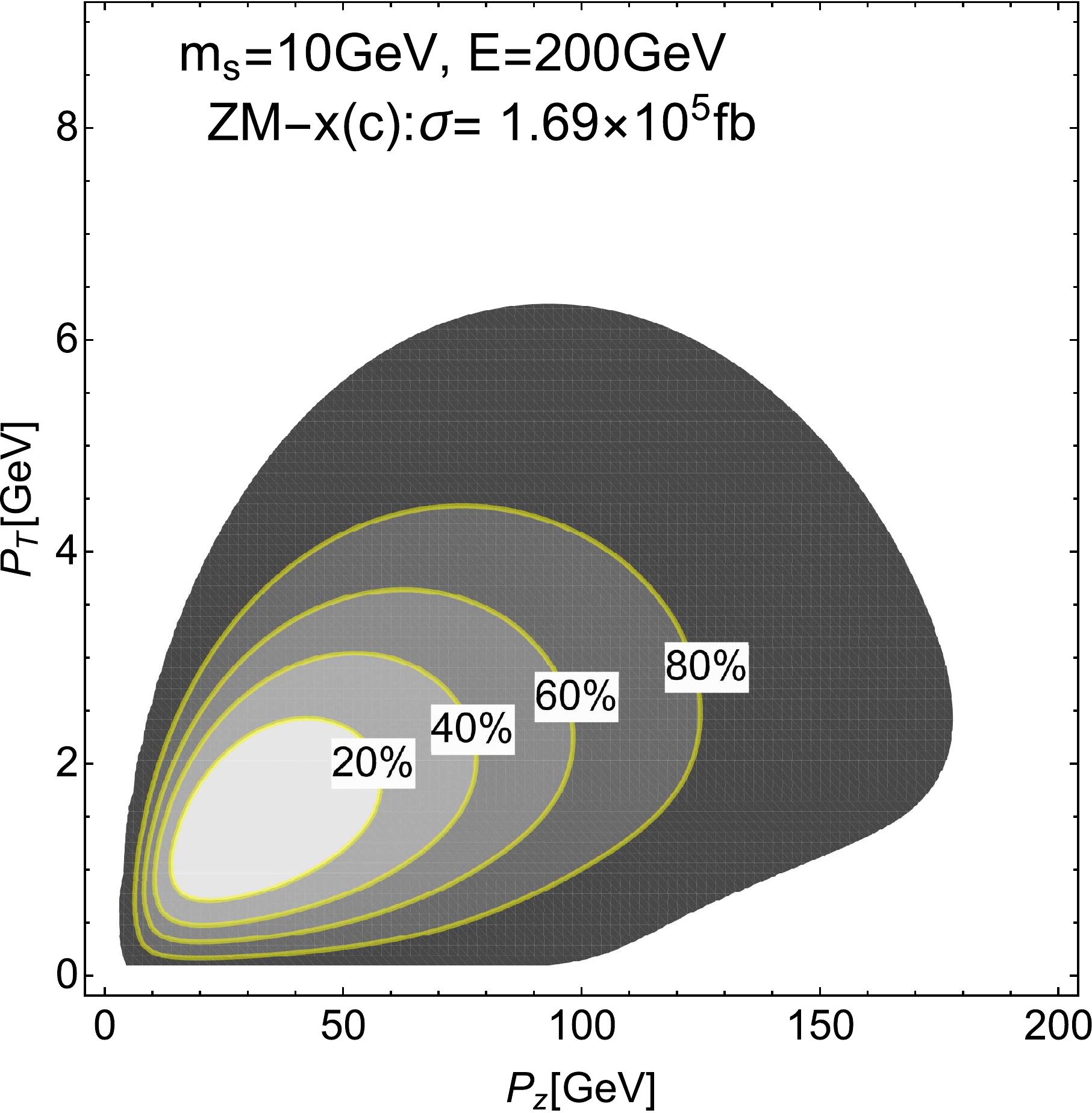}\hspace{1cm}
\includegraphics[width=0.35\linewidth]
{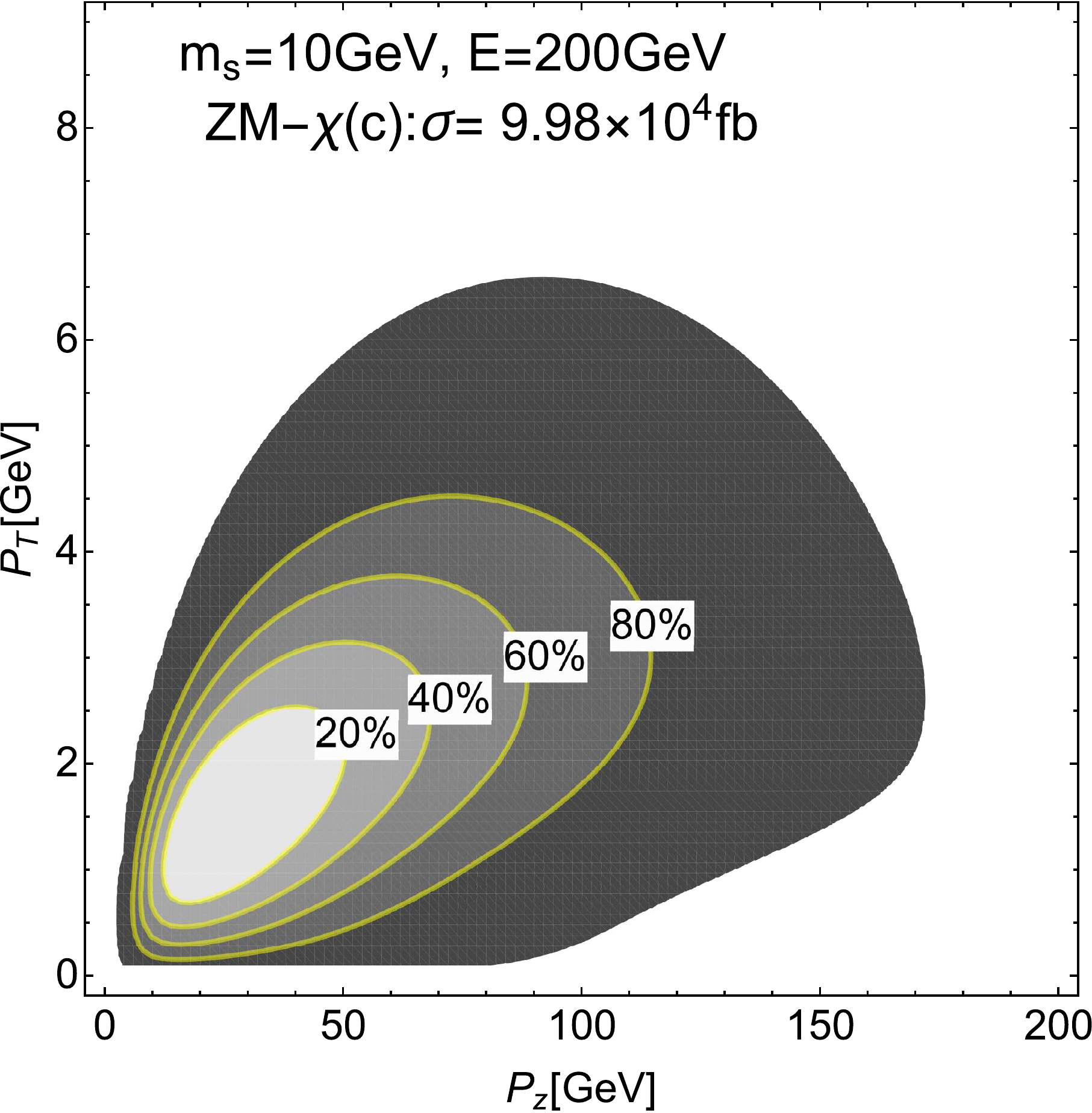}\\
\vspace{5mm}
\includegraphics[width=0.35\linewidth]
{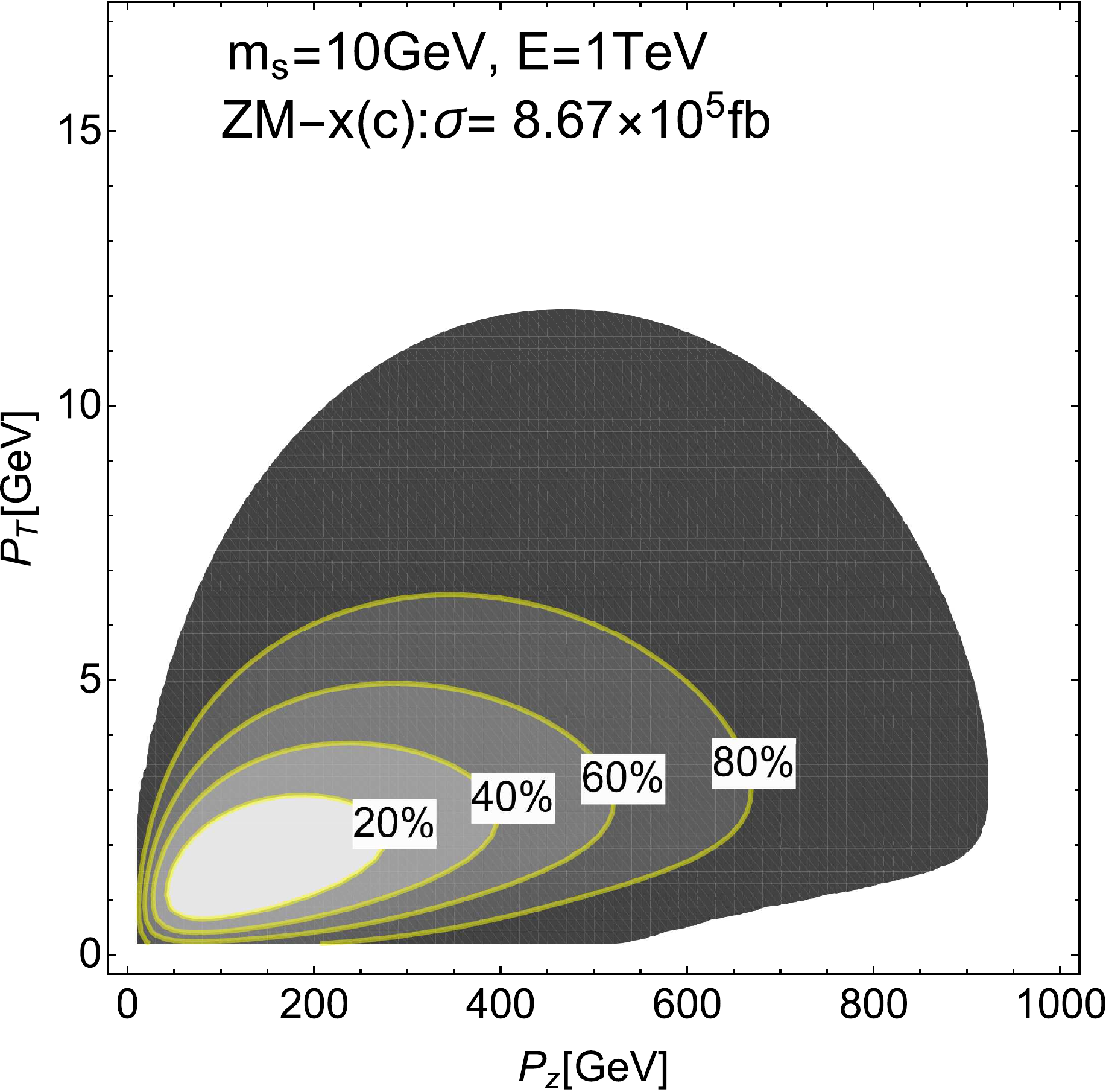}\hspace{1cm}
\includegraphics[width=0.35\linewidth]
{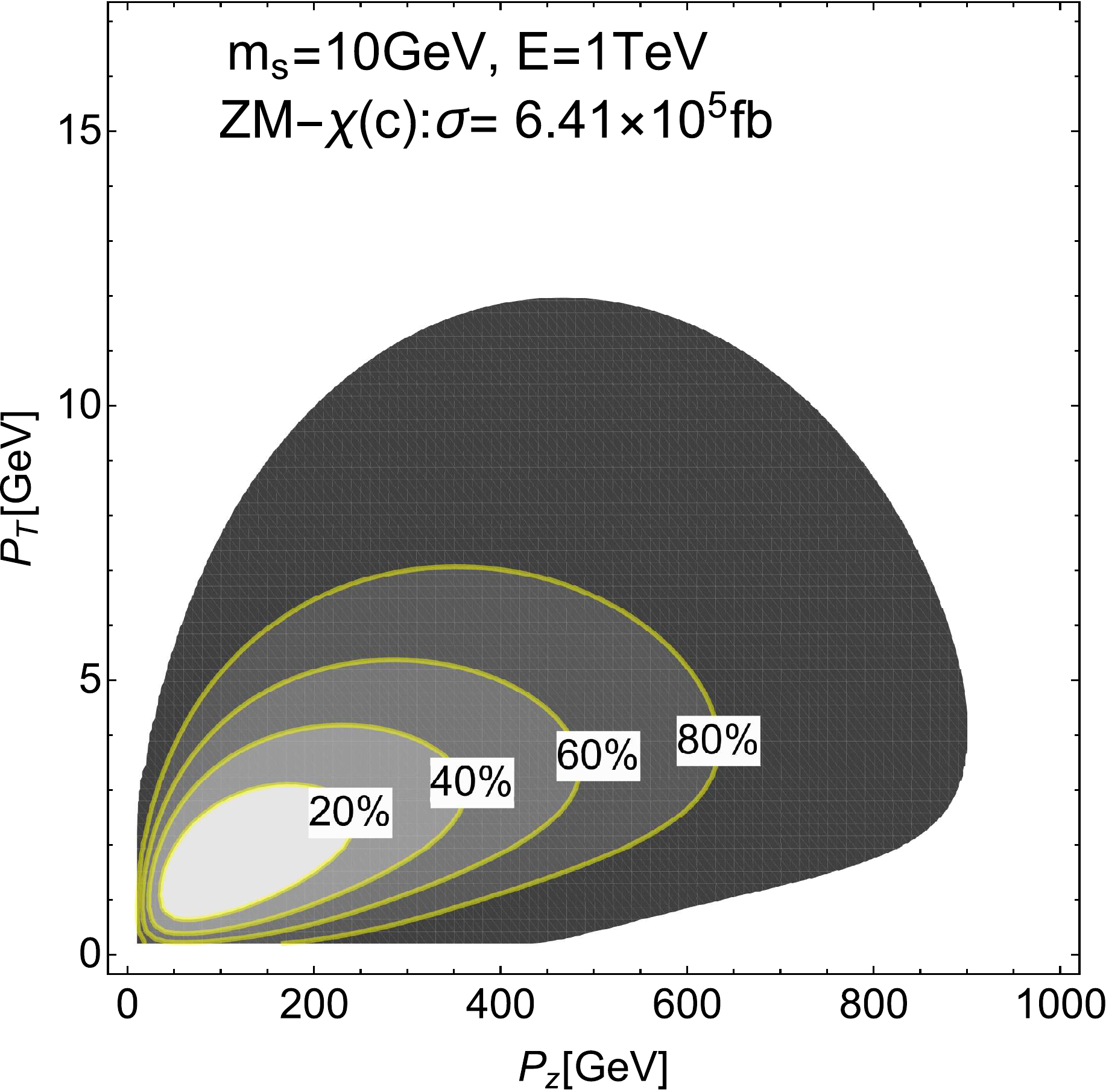}
\caption{
Same as Fig.~\ref{Fig:b_dist_rescaling}, but heavy quark is charm quark ($q=c$): ZM-$x$ (left column) 
and ZM-$\chi$ (right column) for $E_e=200\, {\rm GeV}$ 
and $E_e=1\, {\rm GeV}$ in the two rows.}
\label{Fig:c_dist_rescaling}
\end{figure}

In Fig.~\ref{Fig:c_dist_rescaling} we show the $\tau$-momentum
distributions for $m_S=10\, {\rm GeV}$ in the case of charm quark production. Comparing the $x$ and 
$\chi$ schemes, the shapes of the contour lines look quite similar to each other, but the effect of
 $\chi$-rescaling is still sizable for the normalization of total cross sections. 
The ratio is  $\sigma_c^{\mZMx}/\sigma_c^{\mZMC}\sim 1.7~(1.4)$
for $E_e=200\, {\rm GeV}~(1\, {\rm TeV})$. This value is not changed much 
even for the heavy scalar case. For instance, taking $m_S=10^{5}\, {\rm GeV}$, 
the ratio is $\sigma_c^{\mZMx}/\sigma_c^{\mZMC} \sim 1.6~(1.2)$ for $E_e=200\, {\rm GeV}~(1\, {\rm TeV})$.

\begin{figure}[htbp]
\centering
\includegraphics[width=8cm]
{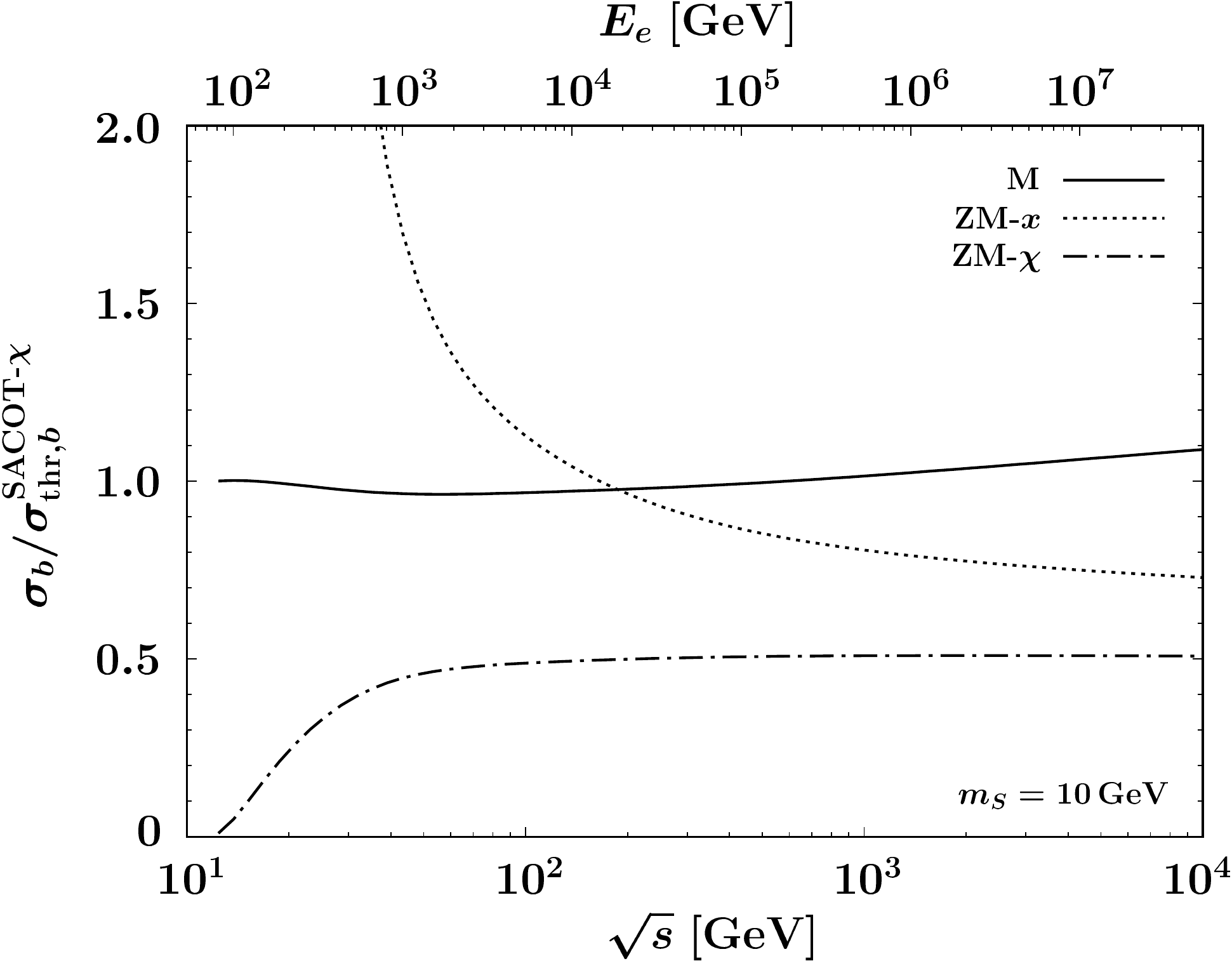}
\includegraphics[width=8cm]
{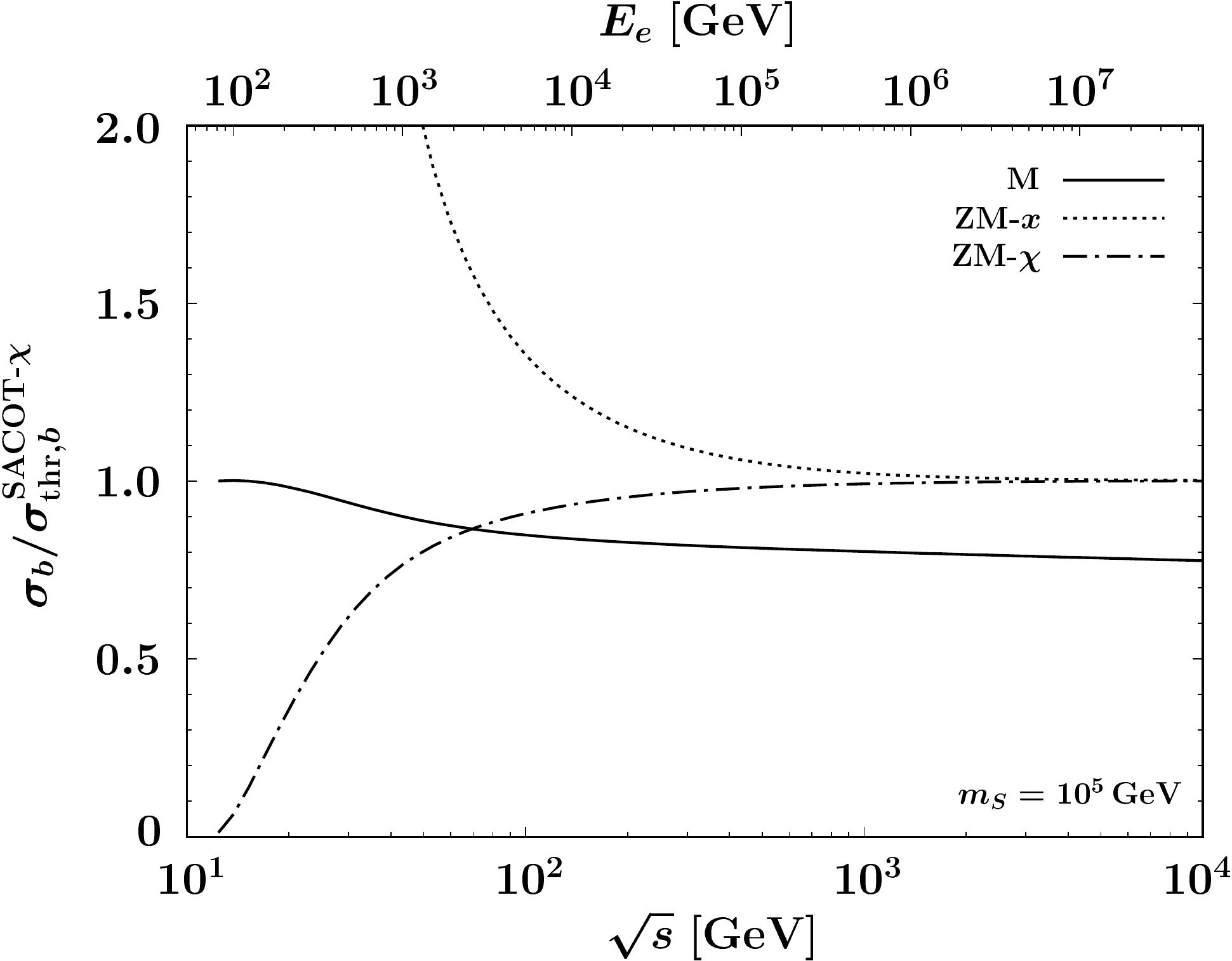}
\\
\vspace{8mm}
\includegraphics[width=8cm]
{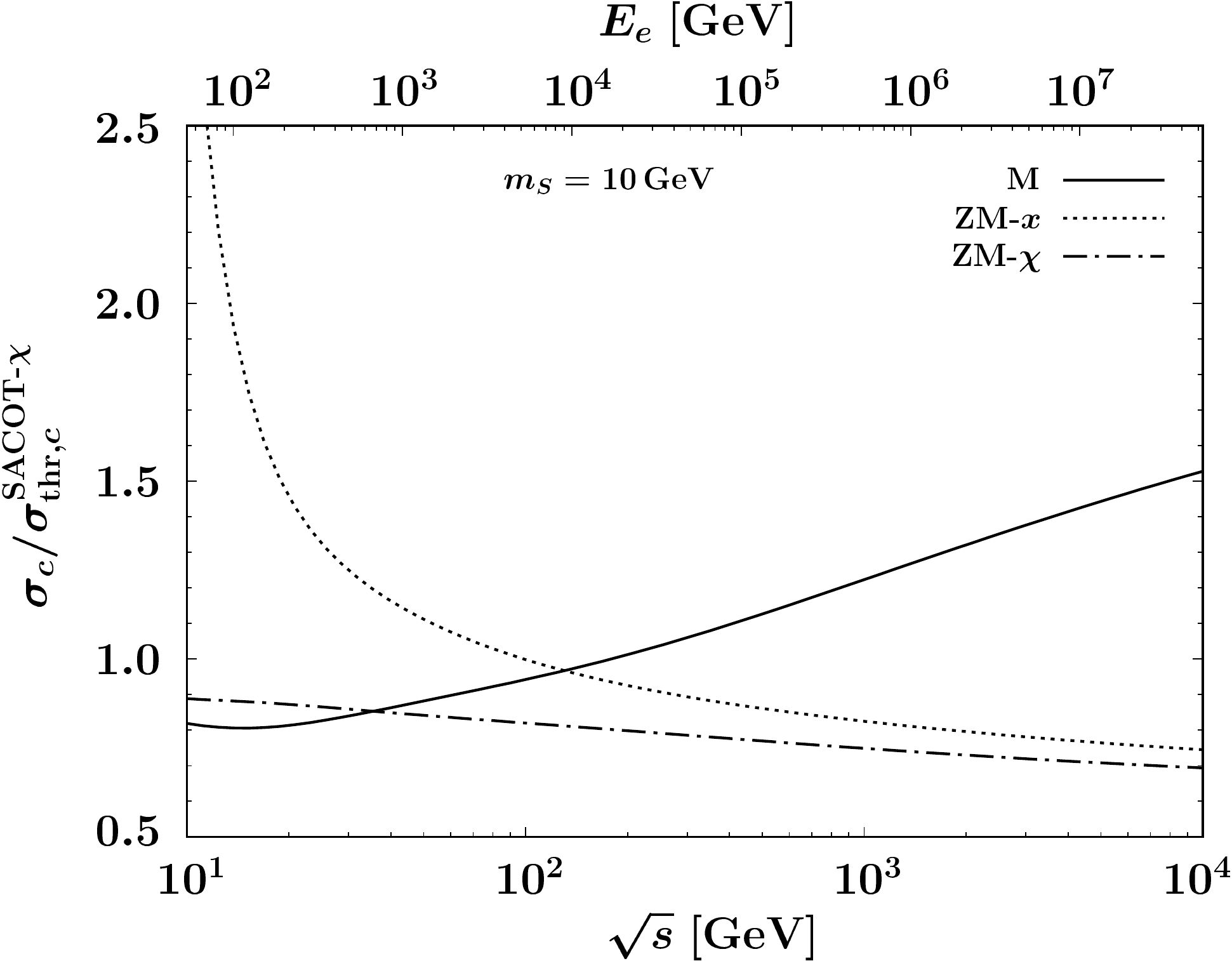}
\includegraphics[width=8cm]
{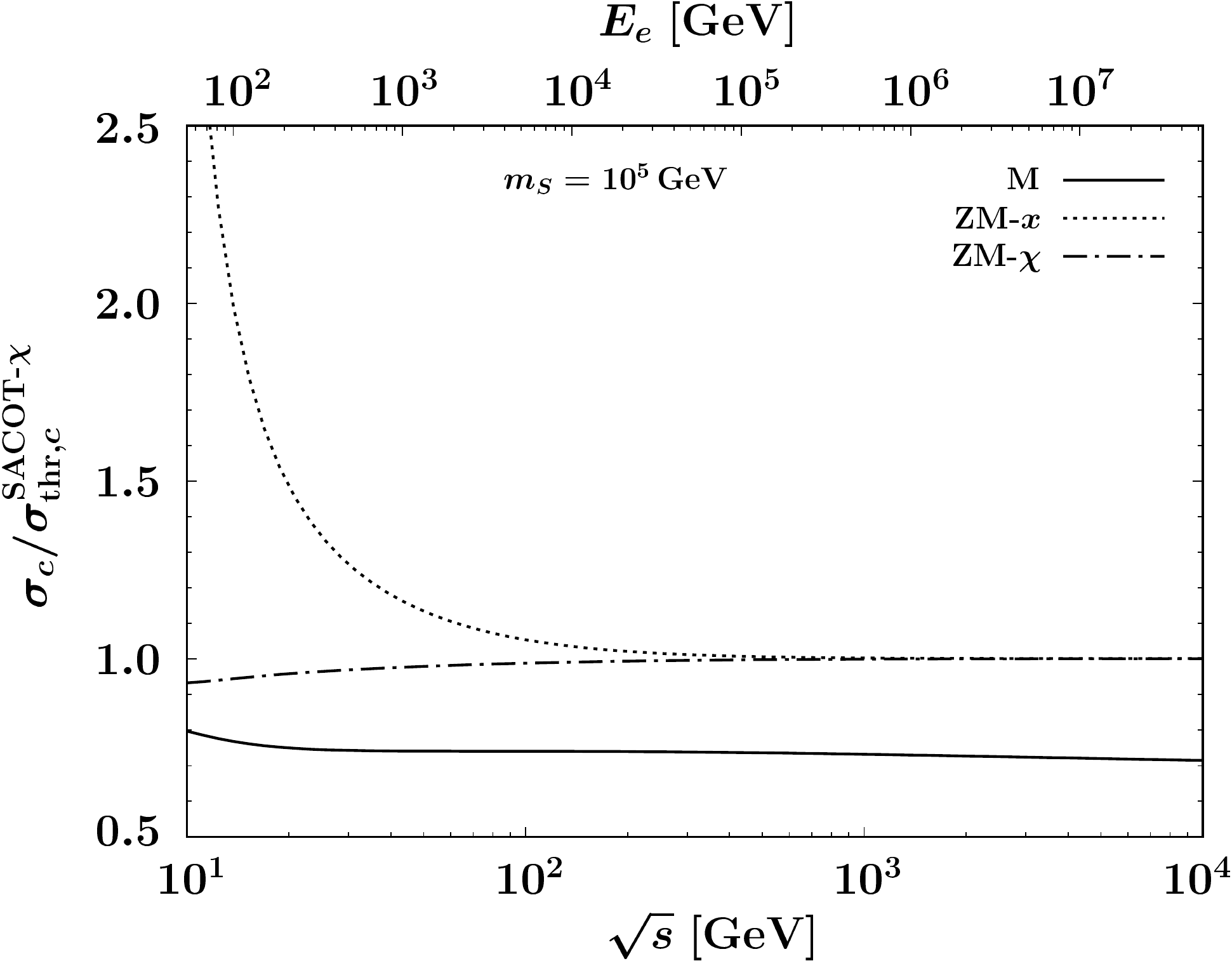}
\caption{
The cross sections  for the CLFV DIS in the M, \ZMx~, and \ZMC~ schemes
normalized to the \AC~(thr).
The scale of the upper horizontal axis in each panel shows 
the beam energy in the fixed target experiments. 
The results for bottom ($q=b$) and charm ($q=c$) quarks are shown in the upper and lower rows, 
and the results for the scalar masses $m_S=10\, {\rm GeV}$  and $m_S=10^5\, {\rm GeV}$ are shown in the left and right columns.}
\label{Fig:XS-scheme}
\end{figure}

In Fig.~\ref{Fig:XS-scheme} the total cross sections associated with 
bottom and charm quark productions are plotted respectively in 
the upper and lower rows, where each cross section is normalized to 
the value of \AC~(thr).   
The left and right columns are two cases of scalar mass, $m_S=10\, {\rm GeV}$ and $10^5\, {\rm GeV}$, respectively.
Numerical values of the cross sections are listed in Table \ref{tab:ZM_M}.
In the plots we observe the following:
\begin{itemize}
\item
For the bottom quark production the curves of 
M scheme are close to the one of 
\AC~(thr), i.e. $\sigma_{b}^{\rm M}/\sigma_{{\rm thr}, b}^{\mAC}\sim 1$.
The cross section in the \ZMC~ scheme for the small scalar mass 
$m_S=10\, {\rm GeV}$ is 
quite off from that of \AC~(thr) irrespective
 of the collision energy: 
 $\sigma_b^{\mZMC}/\sigma_{{\rm thr}, b}^{\mAC}  \lesssim 0.5.$
 For the case of the large scalar mass 
 $m_S=10^5\, {\rm GeV}$,  the \ZMC~ and \ZMx~ 
 curves are gradually approaching to the \AC~ , and around $\sqrt{s}\sim 10^3\, {\rm GeV}$  they meet at one point. Notably, the curve of the \ZMx~ scheme  for the bottom production is far off from that of \AC~(thr) in low collision energy.

 \item
Even for the charm quark production, inadequacy of 
the \ZMx~ scheme in low collision energy is the same 
as the case of bottom quark, but the behaviors of the 
cross sections in the M and \ZMC ~ schemes are quite different.
Specifics of the charm cross sections are as follows.
For the small scalar mass $m_S=10\, {\rm GeV}$ the M scheme curve grows with $\sqrt{s}$ and overshoots the \AC~(thr) at $\sqrt{s}\sim 10^2\,{\rm GeV}$, 
while for the large scalar mass $m_S=10^5\, {\rm GeV}$ it is almost constant and
    small by a sizable amount compared to the \AC~(thr).  
For the large scalar mass,
the value of the \ZMC~ is close to the one of \AC~(thr) for arbitrary collision energy, and it is expected that charm quark can be treated as massless provided that the $\chi$-rescaling 
is adopted and the scalar mass is large enough.
It will be shown in Fig.~\ref{Fig:ms4XS_vs_ms} that the 
scalar mass of $m_S\sim 50\, {\rm GeV}$ is large enough 
to validate the treatment of massless charm with $\chi$-rescaling adopted for it.
\end{itemize}

\begin{table}[htbp]
\small
\centering
\caption{The CLFV total cross sections associated with 
bottom and charm quark productions in ZM-$x$, ZM-$\chi$, and M schemes.
The coupling constants are set to one by Eq.~\eqref{eq:cc_set_to_1}.}
\vspace{5mm}
\begin{tabular}{|c|c|c|c|c|}
\hline
$m_S~[\rm {GeV}]$
& $E_e~{\rm [GeV]}$ 
& $\sigma_b^{\mZMx} ~{\rm [fb]}$  
& $\sigma_b^{\mZMC} ~{\rm [fb]}$ 
& $\sigma_b^{\rm M} ~{\rm [fb]}$  \\
\hline
& 
$10^2$ &  
$6.49\times 10^2$  & $1.02\times 10^{-2}$ & $2.08\times 10^{-1}$ 
\\
& 
$10^3$ &  
$8.67\times 10^4$  & $2.28\times 10^4$ & $4.93\times 10^4$ 
\\
$10$ & 
$10^4$ &  
$7.84\times 10^5$  & $3.72\times 10^5$ & $7.32\times 10^5$ 
\\
& 
$10^5$ &  
$3.24\times 10^6$  & $1.90\times 10^6$ & $3.72\times 10^6$
\\
& 
$10^6$ &  
$9.12\times 10^6$  & $5.87\times 10^6$ & $1.18\times 10^7$ 
\\
& 
$10^7$ &  
$2.11\times 10^7$  & $1.43\times 10^7$ & $2.99\times 10^7$ 
\\
\hline
& 
$10^2$ &  
$1.51\times 10^{-13}$  & $2.02\times 10^{-18}$ & $3.17\times 10^{-17}$ 
\\
& 
$10^3$ &  
$5.64\times 10^{-11}$  & $1.89\times 10^{-11}$ & $2.23\times 10^{-11}$ 
\\
$10^5$ & 
$10^4$ &  
$1.88\times 10^{-9}$  & $1.41\times 10^{-9}$ & $1.27\times 10^{-9}$ 
\\
& 
$10^5$ &  
$3.35\times 10^{-9}$  & $3.10\times 10^{-8}$ & $2.57\times 10^{-8}$ 
\\
& 
$10^6$ &  
$4.66\times 10^{-7}$  & $4.57\times 10^{-7}$ & $3.66\times 10^{-7}$ 
\\
& 
$10^6$ &  
$5.76\times 10^{-6}$  & $5.73\times 10^{-6}$ & $4.50\times 10^{-6}$ 
\\
\hline\hline
$m_S~[\rm {GeV}]$
& $E_e~{\rm [GeV]}$ 
& $\sigma_c^{\mZMx} ~{\rm [fb]}$  
& $\sigma_c^{\mZMC} ~{\rm [fb]}$ 
& $\sigma_c^{\rm M} ~{\rm [fb]}$    \\
\hline
& 
$10^2$ &  
$6.57\times 10^4$ & $2.99\times 10^4$ & $2.74\times 10^4$ 
\\
& 
$10^3$ &  
$8.67\times 10^5$ & $6.41\times 10^5$ & $6.59\times 10^5$ 
\\
$10$ & 
$10^4$ &  
$4.10\times 10^6$ & $3.46\times 10^6$ & $4.15\times 10^6$ 
\\
& 
$10^5$ &  
$1.15\times 10^7$ & $1.02\times 10^7$ & $1.46\times 10^7$ 
\\
& 
$10^6$ &  
$2.45\times 10^7$ & $2.24\times 10^7$ & $3.83\times 10^7$ 
\\
& 
$10^7$ &  
$4.68\times 10^7$ & $4.33\times 10^7$ & $8.68\times 10^7$ 
\\
\hline
& 
$10^2$ &  
$1.00\times 10^{-11}$ & $4.75\times 10^{-12}$ & $3.86\times 10^{-12}$ 
\\
& 
$10^3$ &  
$3.03\times 10^{-10}$ & $2.55\times 10^{-10}$ & $1.93\times 10^{-10}$ 
\\
$10^5$ & 
$10^4$ &  
$4.92\times 10^{-9}$ & $4.70\times 10^{-9}$ & $3.51\times 10^{-9}$ 
\\
& 
$10^5$ &  
$6.37\times 10^{-8}$ & $6.31\times 10^{-8}$ & $4.66\times 10^{-8}$ 
\\
& 
$10^6$ &  
$7.52\times 10^{-7}$ & $7.50\times 10^{-7}$ & $5.48\times 10^{-7}$ 
\\
& 
$10^6$ &  
$8.45\times 10^{-6}$ & $8.45\times 10^{-6}$ & $6.09\times 10^{-6}$ 
\\
\hline
\end{tabular}
\label{tab:ZM_M}
\end{table}

\subsection{\AC~ and its components}

In this subsection we study the cross section and the $\tau$-momentum 
distribution in the \AC~(thr) scheme and its components, for which the 
$\chi$-rescaling and the threshold factor $S_{\rm thr}$ are adopted.
Here and hereafter the word ``components'' denotes the three 
contributions, M, \ZMC~(thr), and \MZC~(thr), which constitute 
the \AC~(thr) scheme.

\vspace{5mm}
\begin{figure}[htbp]
\centering
\includegraphics[width=0.3\linewidth]
{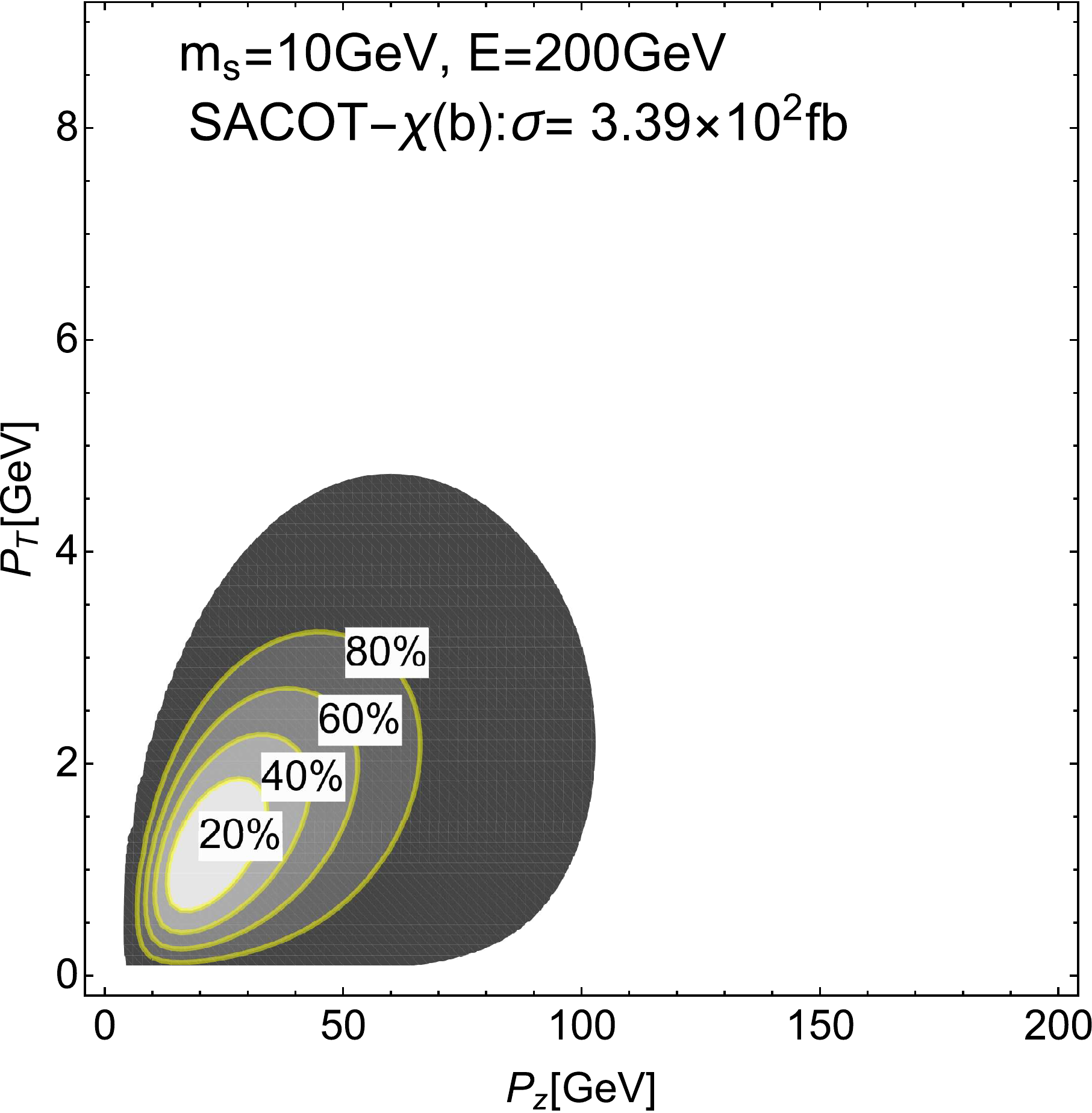}\hspace{1cm}
\includegraphics[width=0.3\linewidth]
{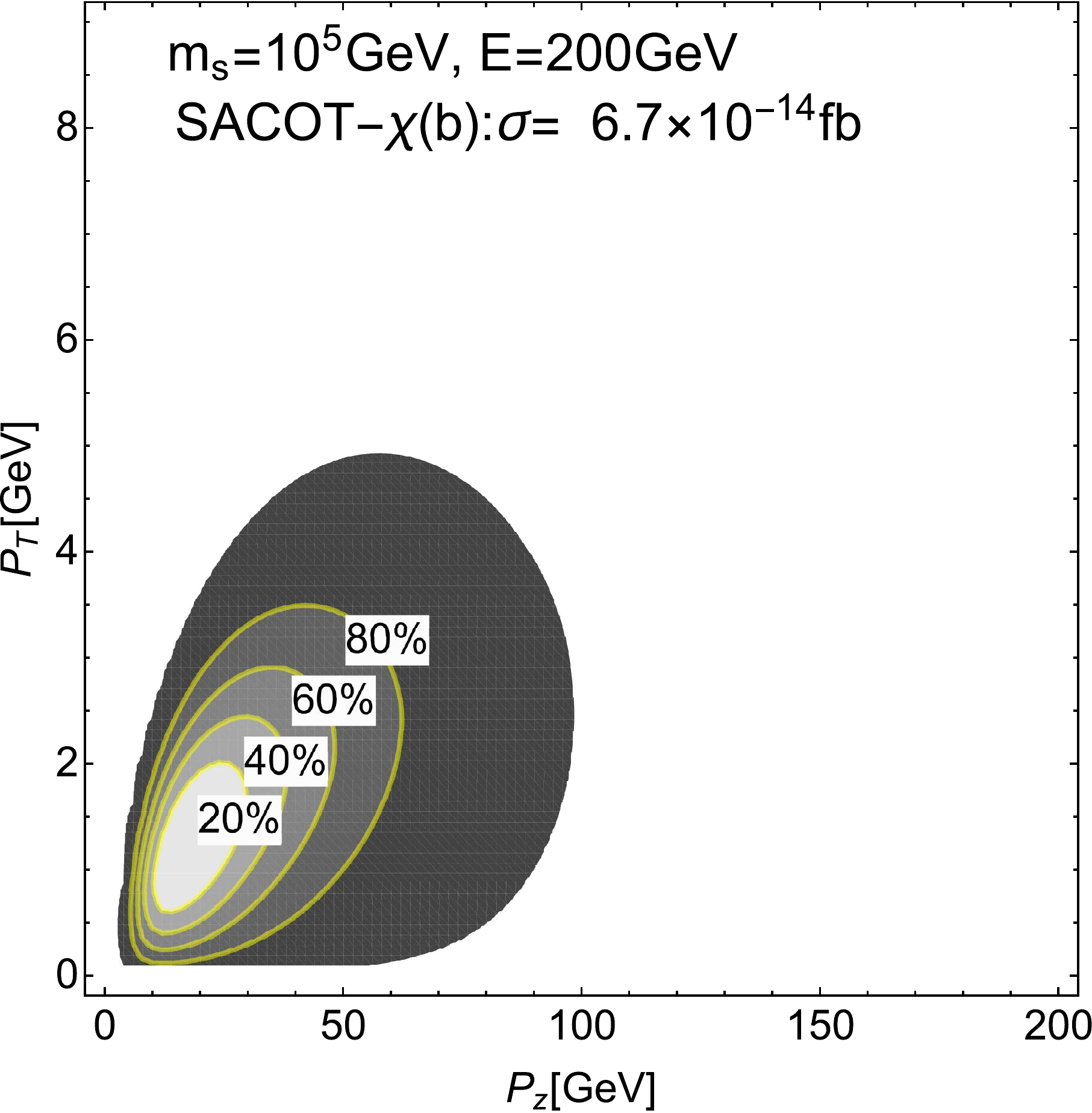}\\
\vspace{2mm}
\includegraphics[width=0.3\linewidth]
{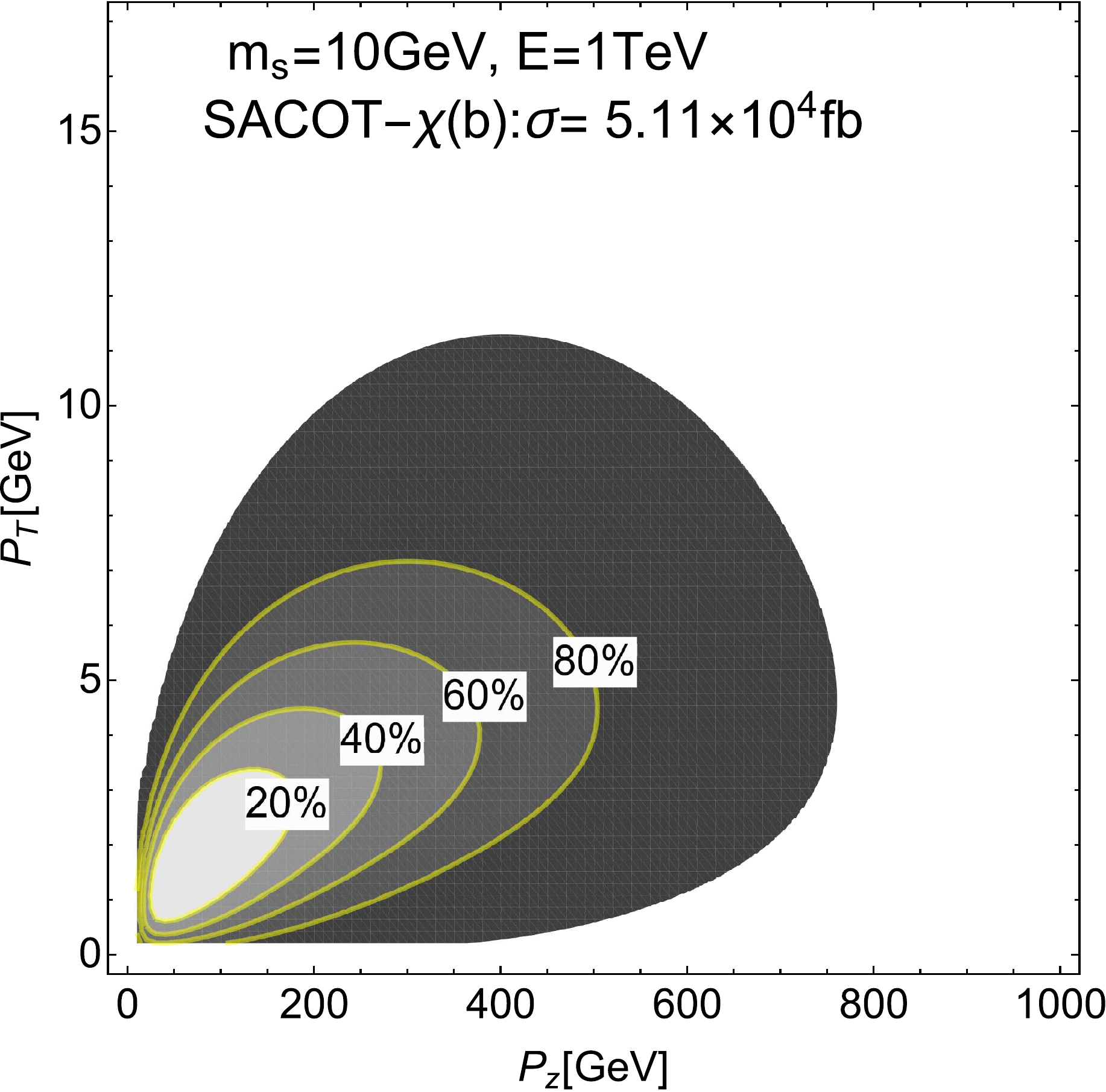}\hspace{1cm}
\includegraphics[width=0.3\linewidth]
{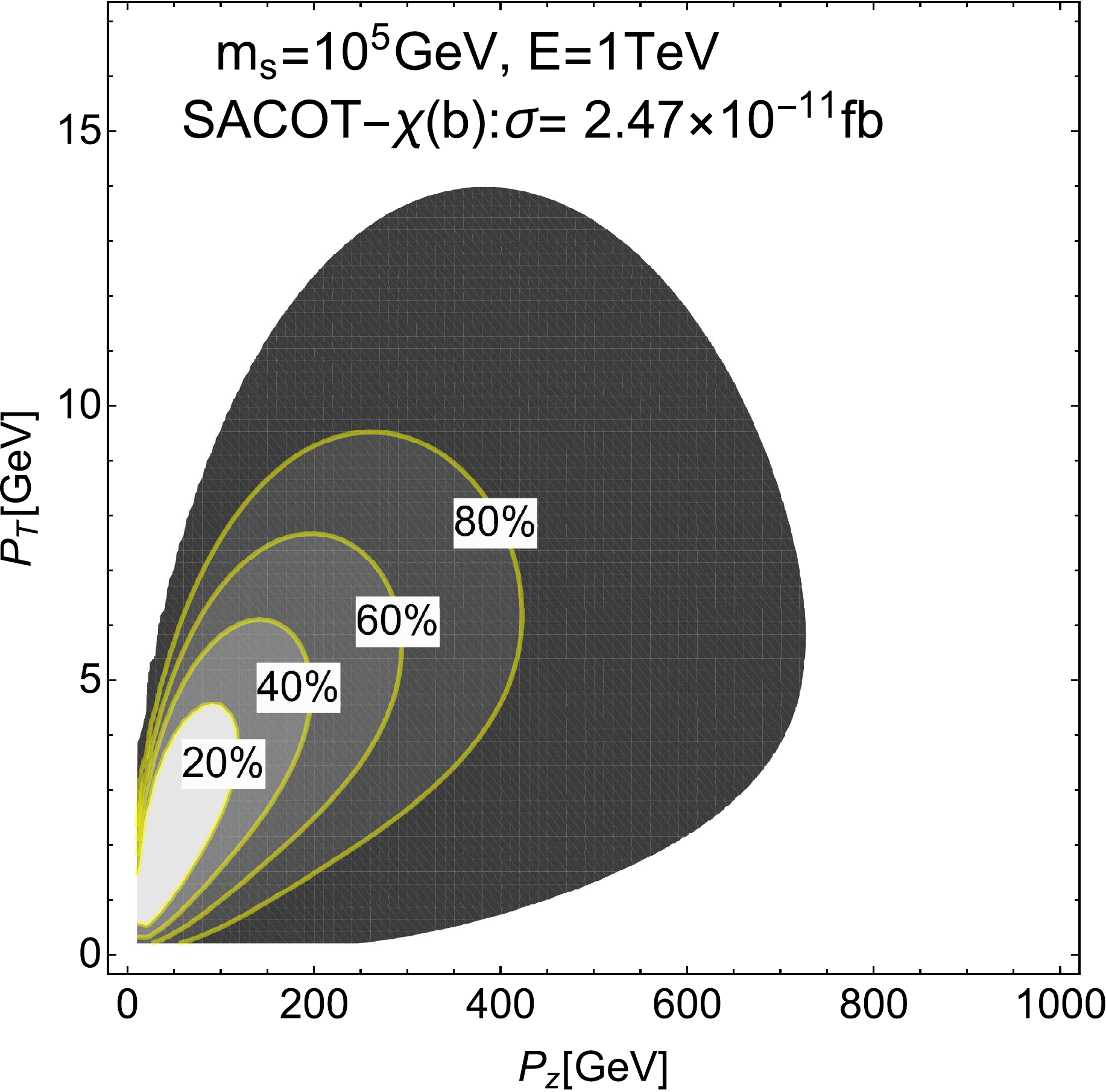}
\caption{
The $\tau$-momentum distribution in the CLFV DIS associated 
with bottom quark production in the \AC~ (thr) scheme.
The threshold factor $S_{\rm thr}$ is included.
The results for the scalar masses $m_S=10\, {\rm GeV}$
 and $m_S=10^5\, {\rm GeV}$ are shown in 
 the left and right columns, and the results for 
 the initial electron beam energies $E_e=200\, {\rm GeV}$  and 
$E_e=1\, {\rm TeV}$ are shown in the two rows.}
\label{Fig:dist_bACOT}
\end{figure}
\begin{figure}[htbp]
\centering
\includegraphics[width=0.3\linewidth]
{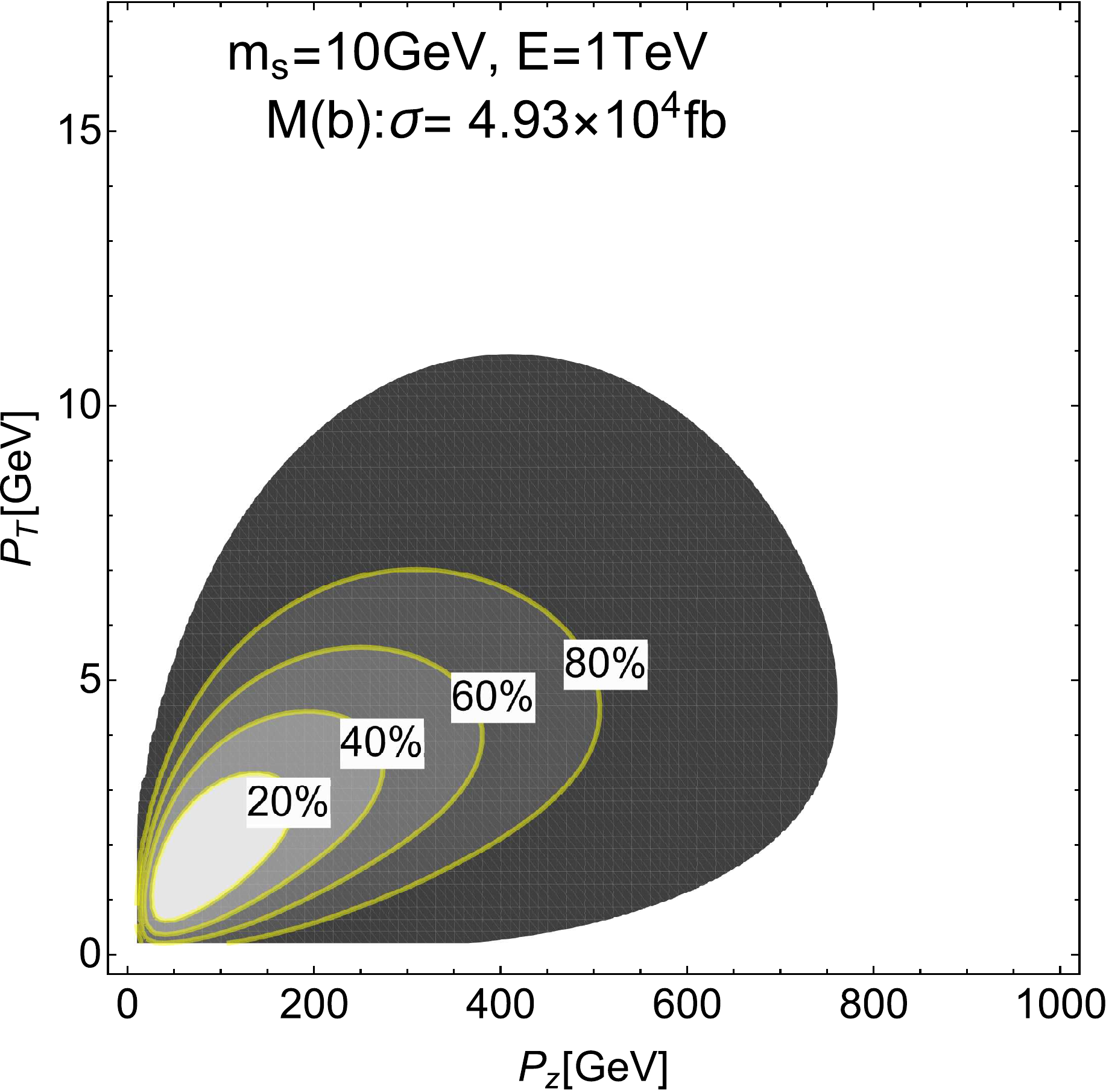}
\includegraphics[width=0.3\linewidth]
{cplt_m10GeV_E1TeV_bZMc.pdf}
\includegraphics[width=0.3\linewidth]
{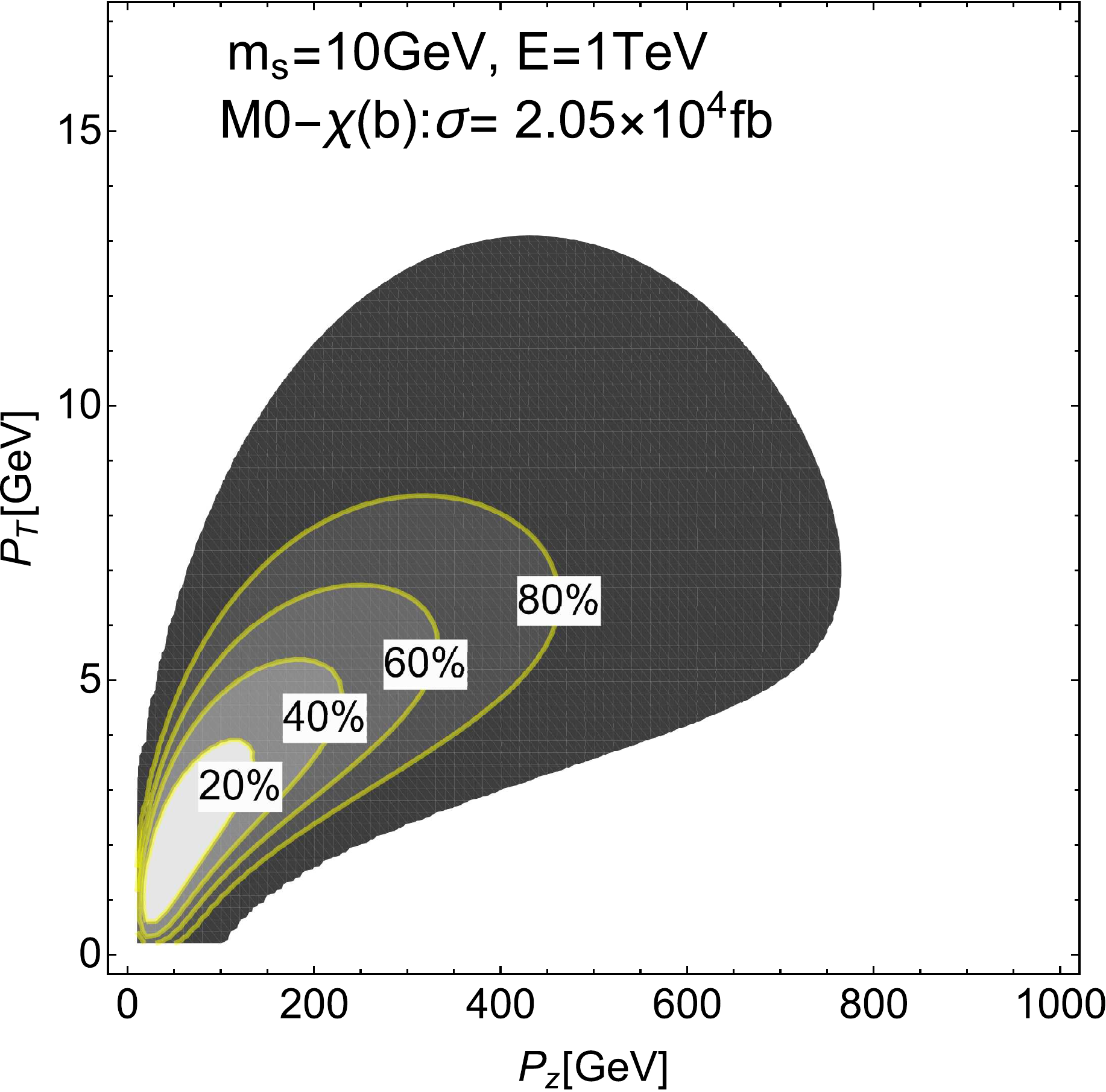}\\
\vspace{2mm}
\includegraphics[width=0.3\linewidth]
{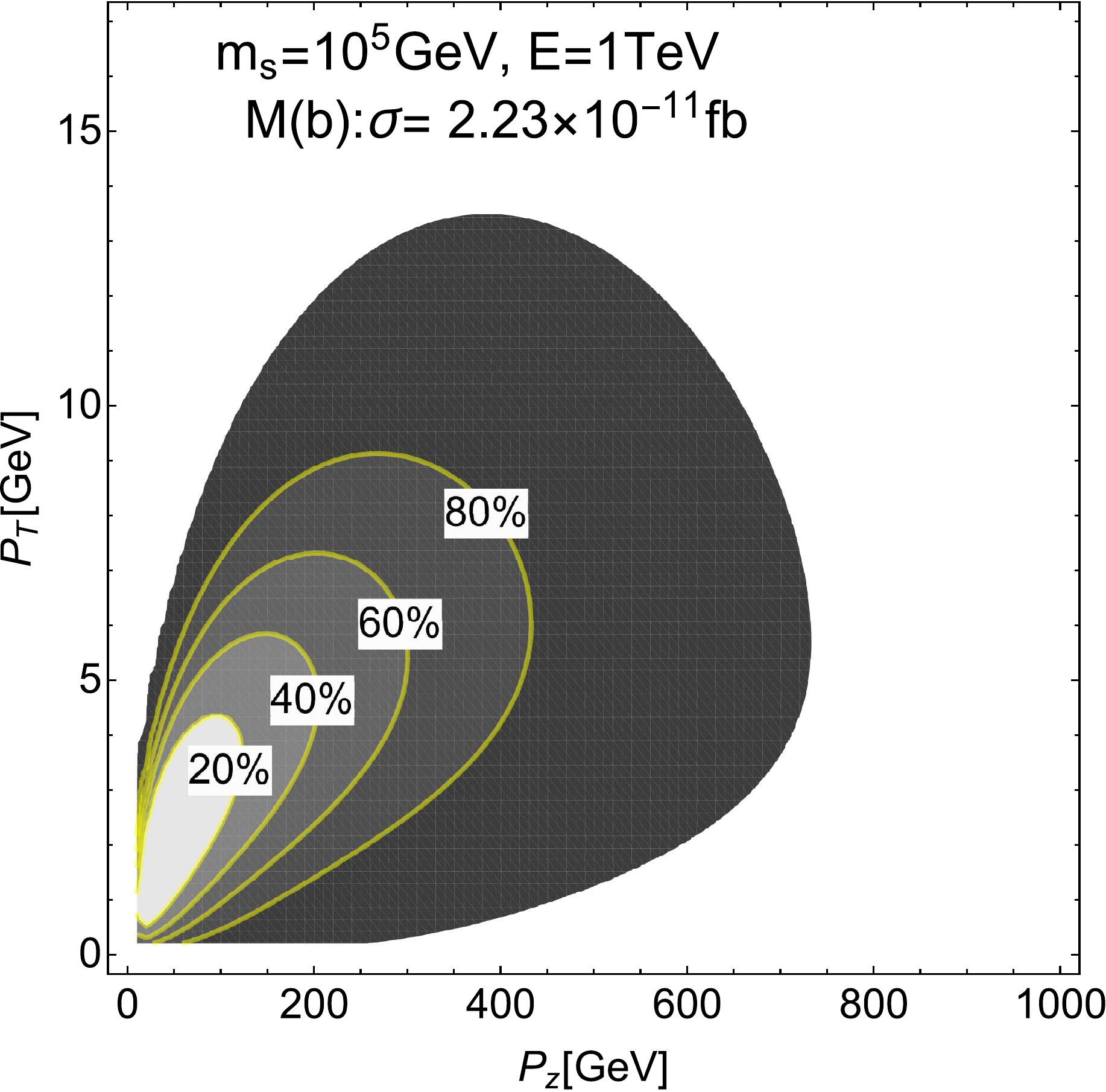}
\includegraphics[width=0.3\linewidth]
{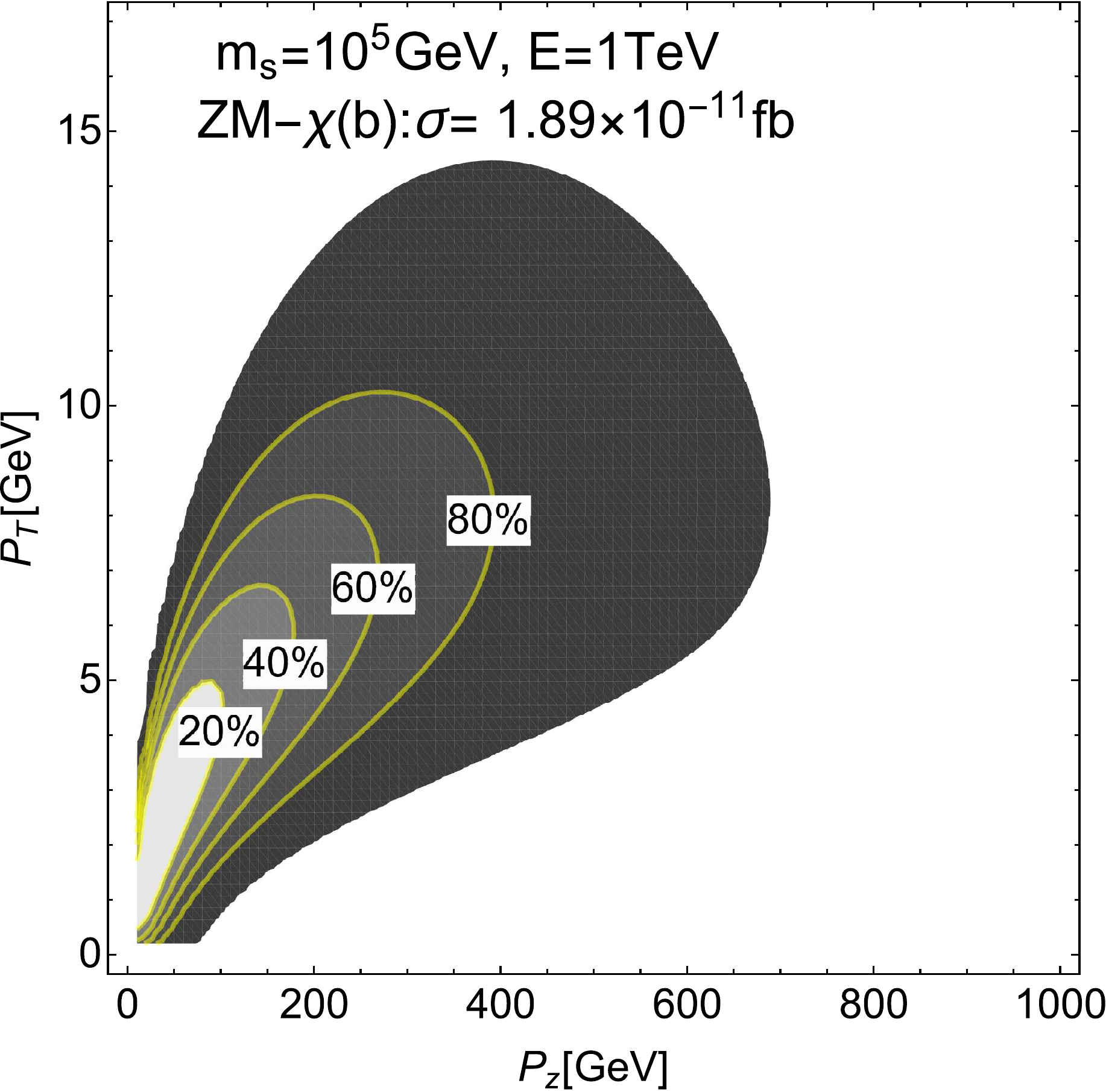}
\includegraphics[width=0.3\linewidth]
{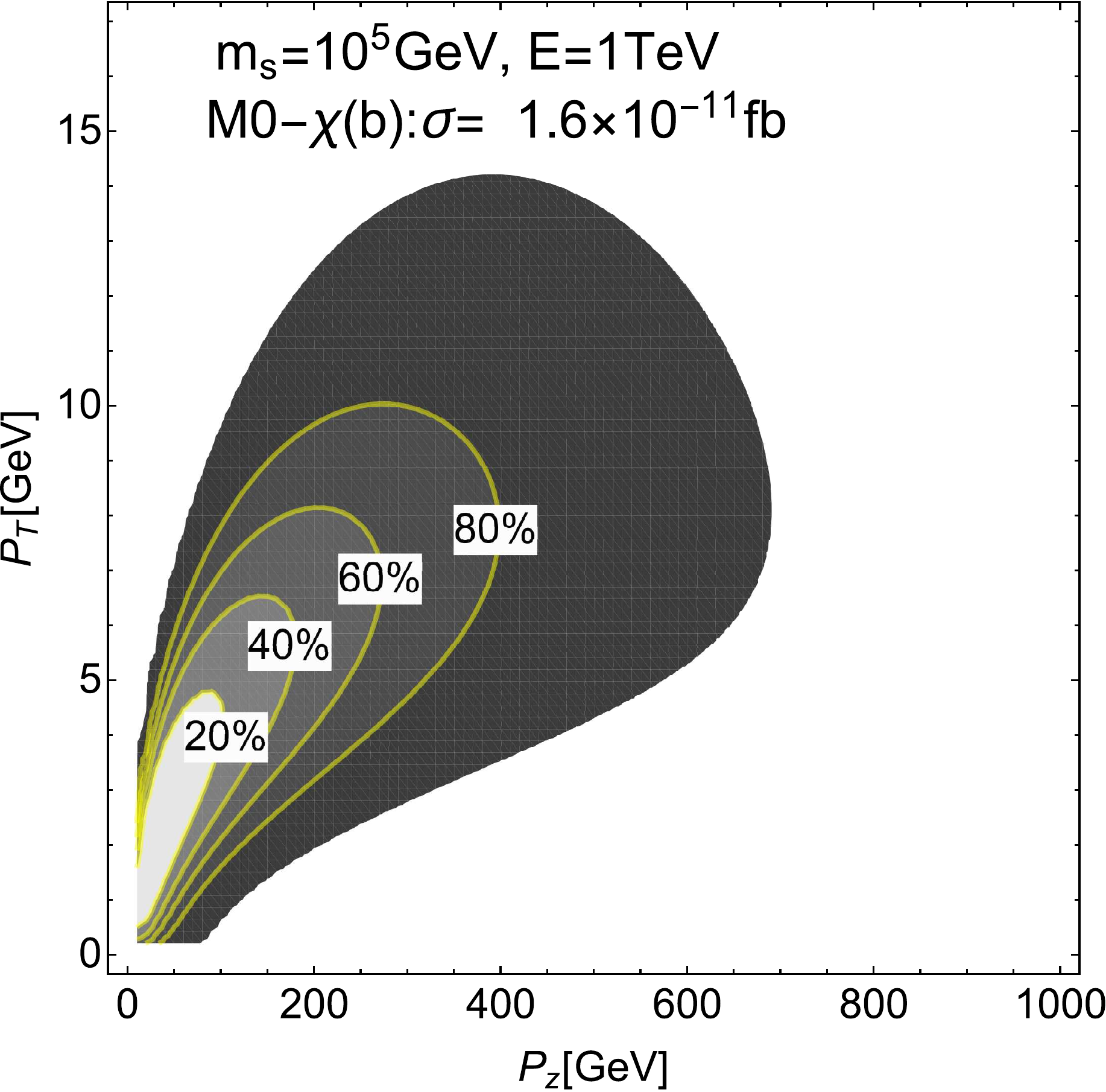}
\caption{
The $\tau$-momentum distribution 
for the components of \AC~ scheme for 
the bottom quark production at $E_e=1\, {\rm TeV}$. 
The results for  the M, \ZMC~, and \MZC~ without the threshold factor
are shown respectively in the first, second, and third columns. 
The scalar masses are 
$m_S=10\, {\rm GeV}$ and $m_S=10^5\, {\rm GeV}$ 
in the upper and lower rows.}
\label{Fig:dist_bACOT_components}
\end{figure}

In Fig.~\ref{Fig:dist_bACOT} we show the $\tau$-momentum 
distribution in the CLFV DIS associated with the bottom quark 
production in the \AC~(thr) scheme, where 
a combination of energies $E_e=200\, {\rm GeV},~1\, {\rm TeV}$
and the scalar masses $m_S=10\, {\rm GeV},~10^{5}\, {\rm GeV}$ 
is taken for each plot. 
One can see that the scalar mass does 
not affect much the shape of distribution  for $E_e=200\, {\rm GeV}$,
but the overall normalization. For $E_e=1\, {\rm TeV}$ there are 
 sizable differences in the shape of the distribution between $m_S=10\, {\rm GeV}$ and $10^5\, {\rm GeV}$. Decomposing the cross section of \AC~(thr) scheme into 
the components one obtains the contributions:
\begin{equation}
\small
\begin{split}
\mbox{(i)}~~~~&
\sigma_{b}^{\mAC} =
\big[3.37+0.32(0.77)-0.30(0.75)\big]\times 10^{\, 2}\, {\rm fb}~~~~=3.39(3.38)\times 10^{\, 2}\, {\rm fb},
\\
\mbox{(ii)}~~~&
\sigma_{b}^{\mAC} =
\big[4.93+1.53(2.28)-1.35(2.05)\big]\times 10^{\, 4}\, {\rm fb}~~~~=5.11(5.15)\times 10^{\, 4} \, {\rm fb},
\\
\mbox{(iii)}~~&
\sigma_{b}^{\mAC} =
\big[6.61+1.02(2.24)-0.93(2.12)\big]\times 10^{-14}\, {\rm fb}~=6.70(6.73)\times 10^{-14} \, {\rm fb},
\\
\mbox{(iv)}~~&
\sigma_{b}^{\mAC} =
\big[2.23+1.47(1.89)-1.22(1.60)\big]\times 10^{-11}\, {\rm fb}~=2.47(2.52)\times 10^{-11}\, {\rm fb},
\end{split}
\label{eq:SACOT_vs_components_b}
\end{equation}
where the first/second/third number in the square parenthesis represents
the M/\ZMC~/\MZC~ component with (without) the threshold factor $S_{\rm thr}$ for
cases
 (i) $E_e=200\, {\rm GeV},~ m_S=10\, {\rm GeV}$, 
 (ii) $E_e=1\, {\rm TeV},~ m_S=10\, {\rm GeV}$, 
 (iii) $E_e=200\, {\rm GeV},~ m_S=10^5\, {\rm GeV}$, and
 (iv) $E_e=1\, {\rm TeV},~ m_S=10^5\, {\rm GeV}$. 
 It is observed that the values of the \ZMC~ and \MZC~ components 
  receive sizable threshold suppressions by the $S_{\rm thr}$, 
  but the \AC~ cross section is approximately the same irrespective of inclusion 
of the threshold factor. It is because  
the contributions of the \ZMC~(thr) and \MZC~(thr) are the same size
and cancel each other in the combination
 of $\big[F_{\phi,\rm thr}^{\mZMC}-F_{\phi, \rm thr}^{\mSubC}\big]$ 
 in the \AC~(thr) scheme.
  
In Fig.~\ref{Fig:dist_bACOT_components} we show the 
$\tau$-momentum distribution of the components for 
the bottom quark production at $E_e=1\, {\rm TeV}$, 
but without the threshold factor $S_{\rm thr}$.
The upper and lower rows are the results for $m_S=10\, {\rm GeV}$ 
and $m_S=10^5\, {\rm GeV}$ respectively,  
and the first column is the result of M scheme, and the second and third columns are the results of the \ZMC~ and \MZC~ schemes respectively.
It turns out that the largest contribution 
is coming from the massive scheme cross section.
These observations lead that the massive scheme cross section is 
effective and nearly equal to the \AC~(thr) in the range of 
collision energy up to $E_e=1{\rm TeV}$ ($\sqrt{s}\simeq 45\, {\rm GeV}$).
Effectiveness of massive scheme cross section in a wider range 
 of collision energy will be discussed later (see Fig.\ref{Fig:acot_bc}).

\begin{figure}[htbp]
\centering
\includegraphics[width=0.3\linewidth]
{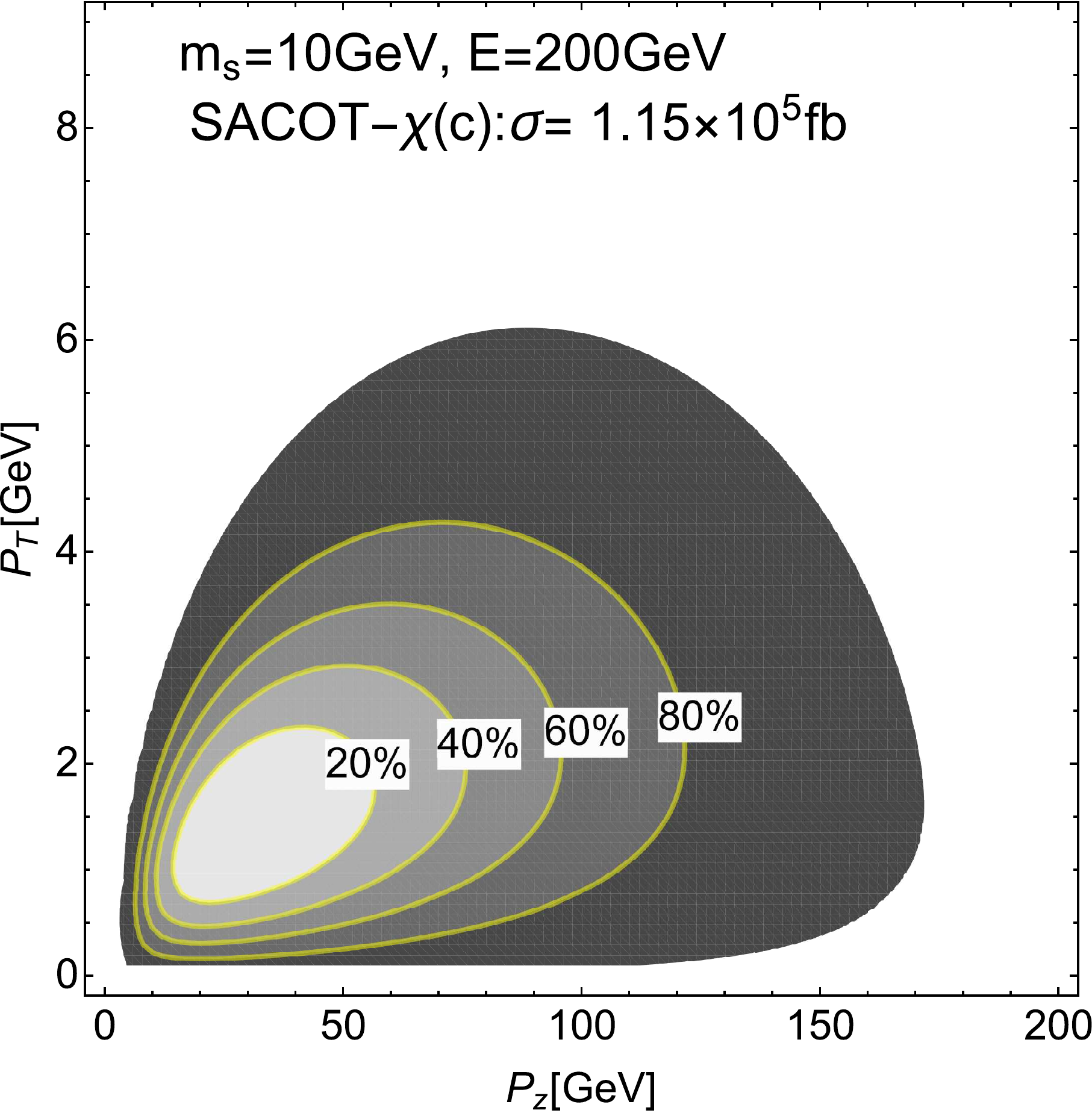}\hspace{1cm}
\includegraphics[width=0.3\linewidth]
{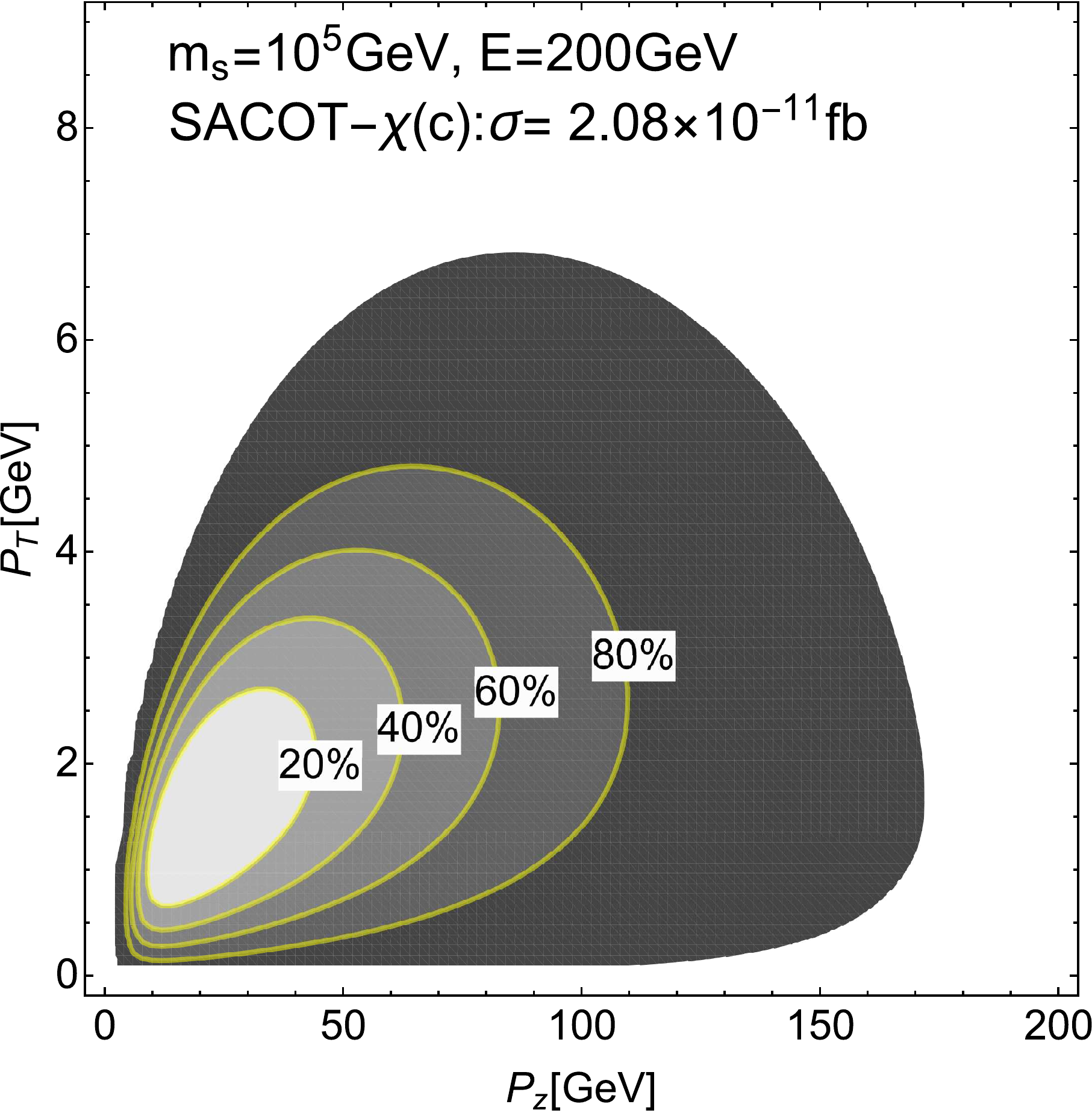}\\
\vspace{2mm}
\includegraphics[width=0.3\linewidth]
{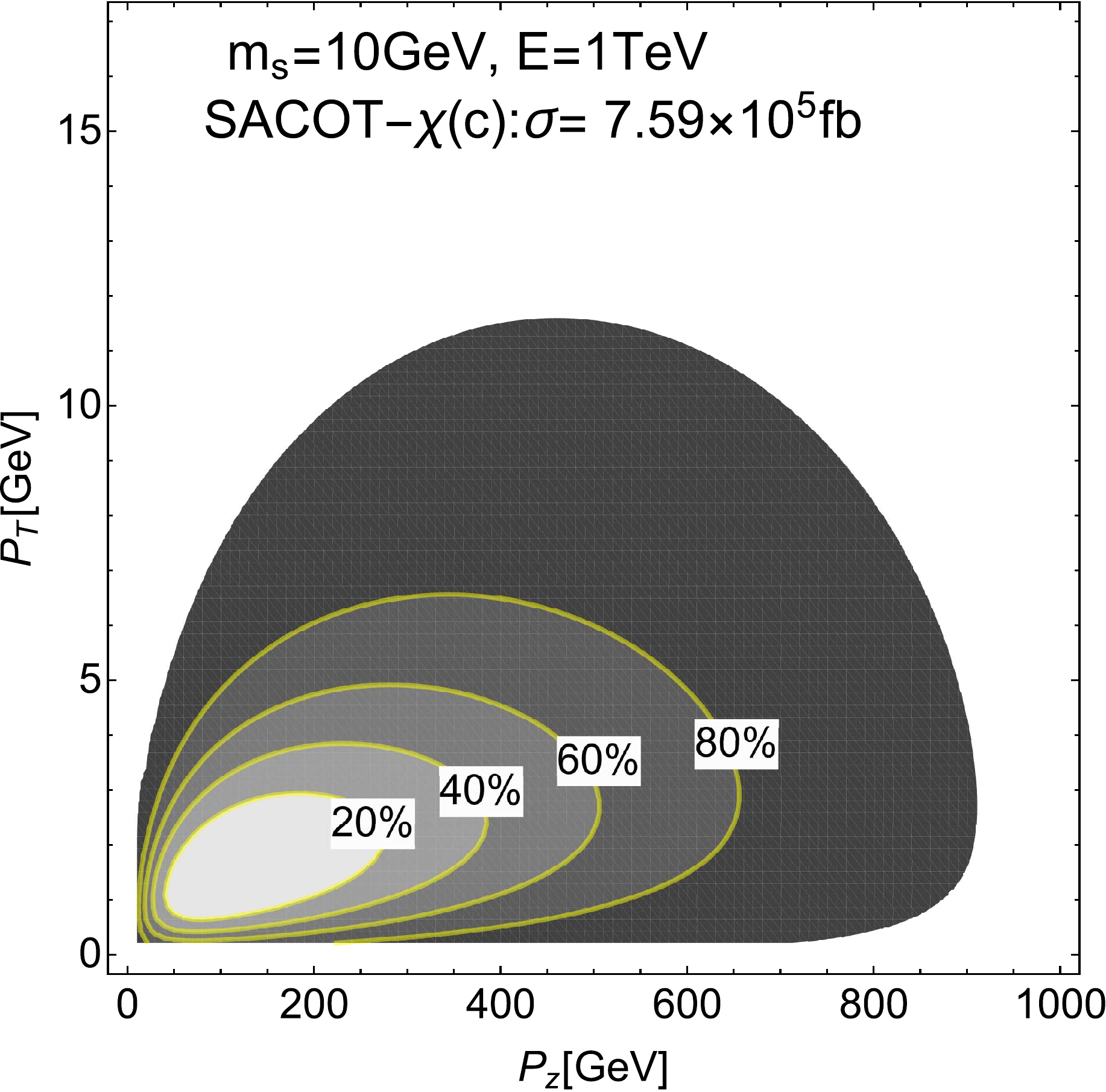}\hspace{1cm}
\includegraphics[width=0.3\linewidth]
{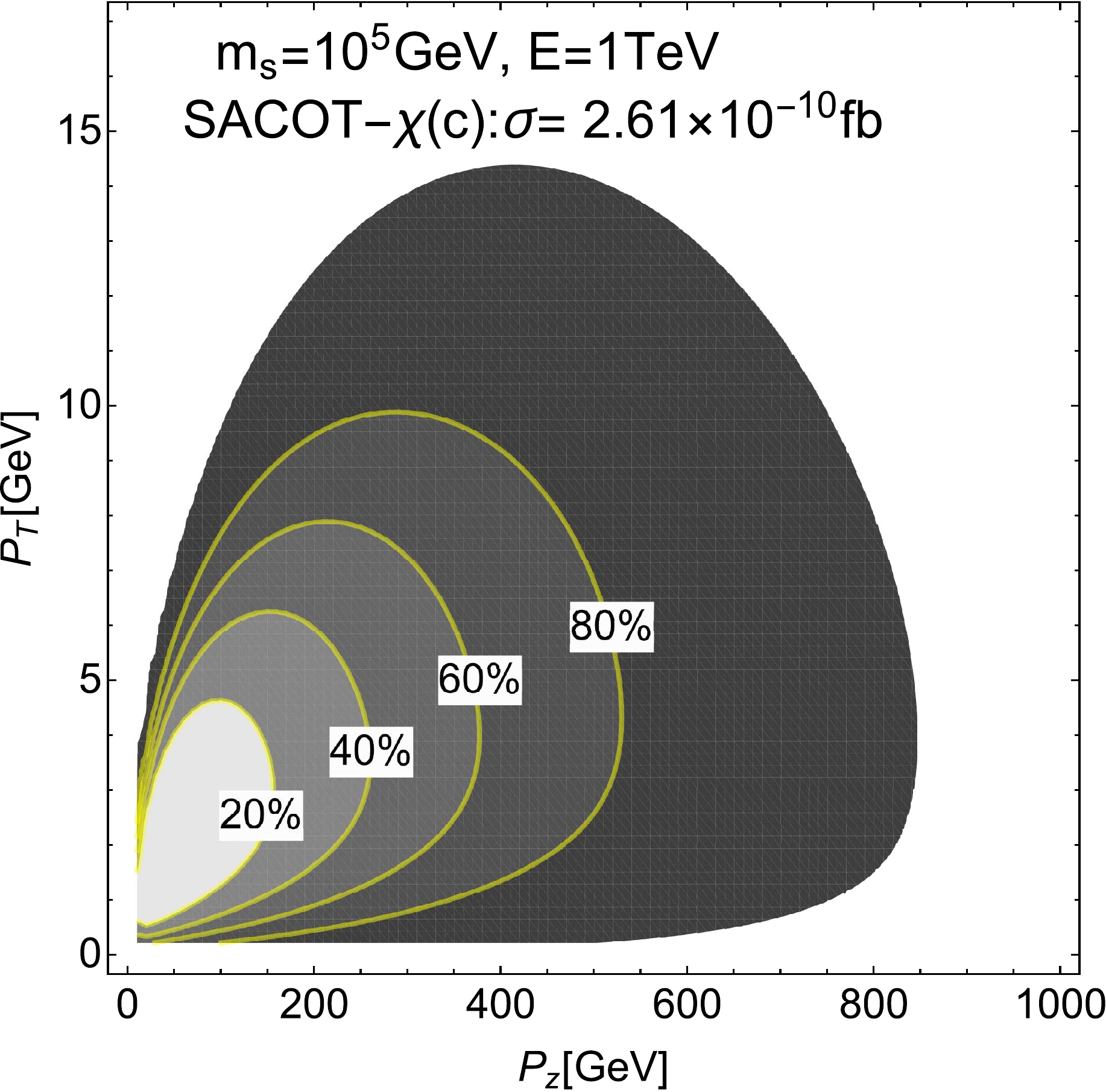}
\caption{
Same as Fig.~\ref{Fig:dist_bACOT}, but for the charm quark productions:
$m_S=10\, {\rm GeV}$ (left column) and $m_S=10^5\, {\rm GeV}$ 
(right column) for $E_e=200\, {\rm GeV}$ and $E_e=1\, {\rm TeV}$
in the two rows.
}
\label{Fig:dist_cACOT}
\end{figure}
\begin{figure}[htbp]
\centering
\includegraphics[width=0.3\linewidth]
{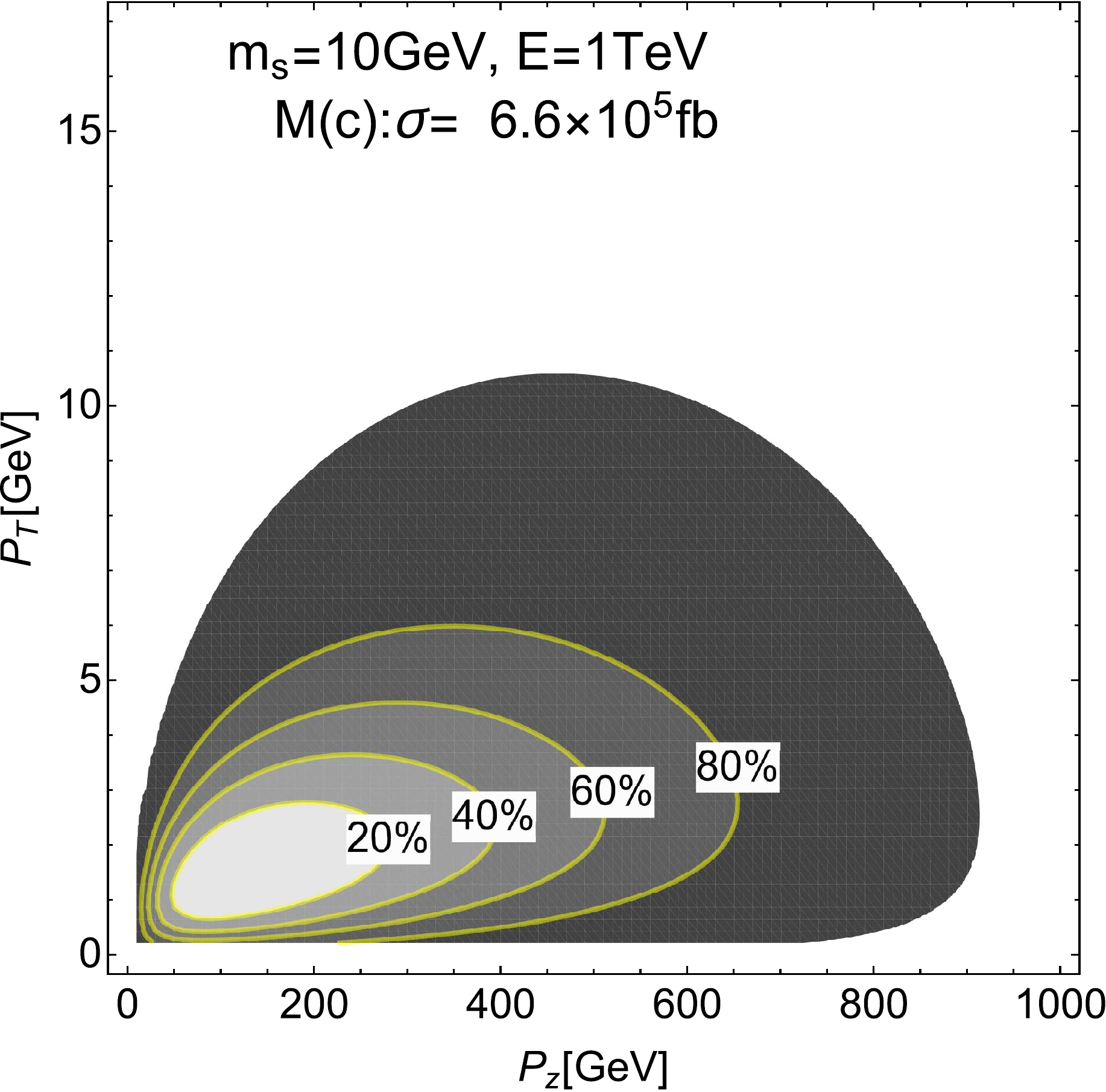}
\includegraphics[width=0.3\linewidth]
{cplt_m10GeV_E1TeV_cZMc.pdf}
\includegraphics[width=0.3\linewidth]
{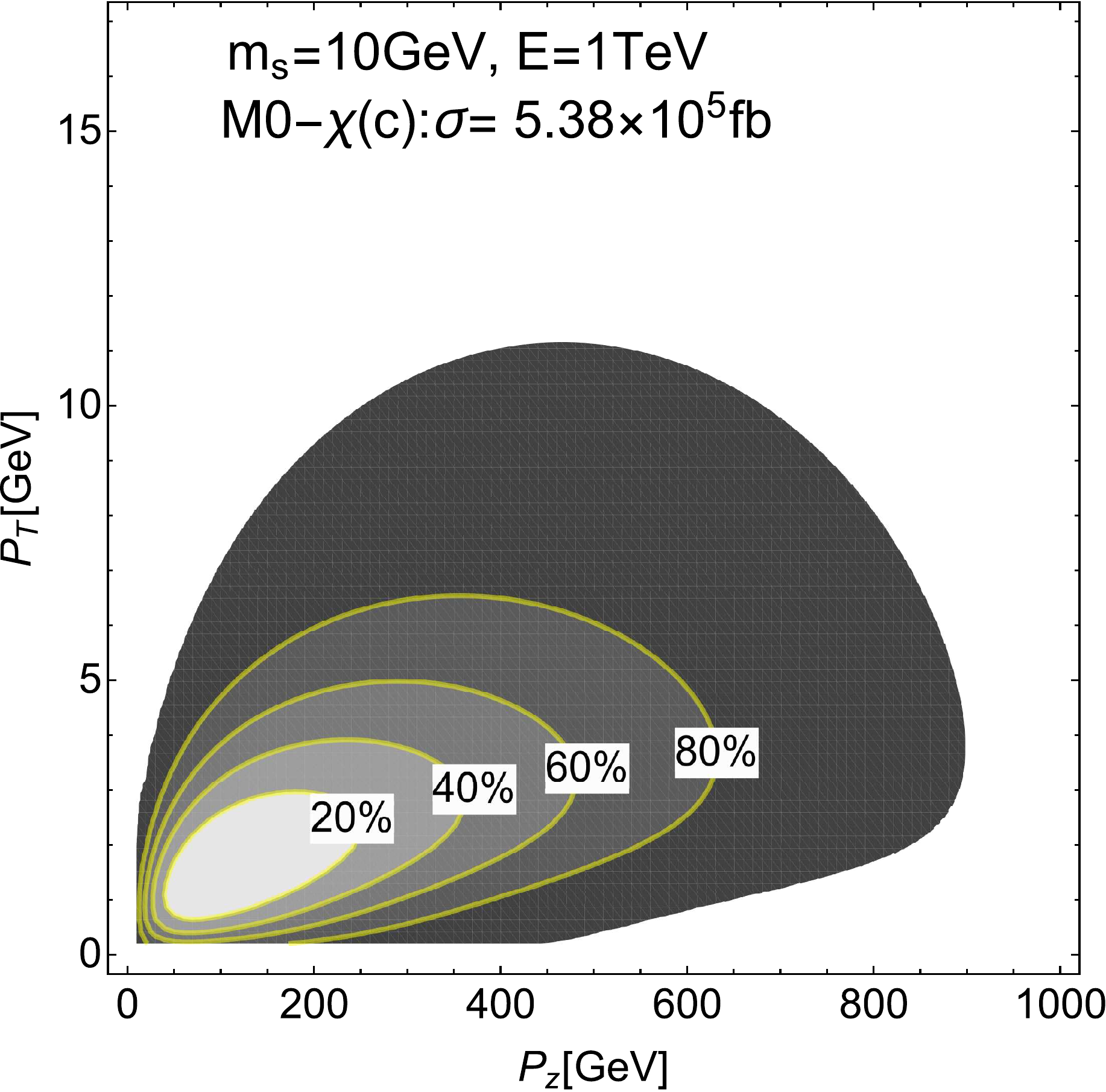}
\\
\vspace{2mm}
\includegraphics[width=0.3\linewidth]
{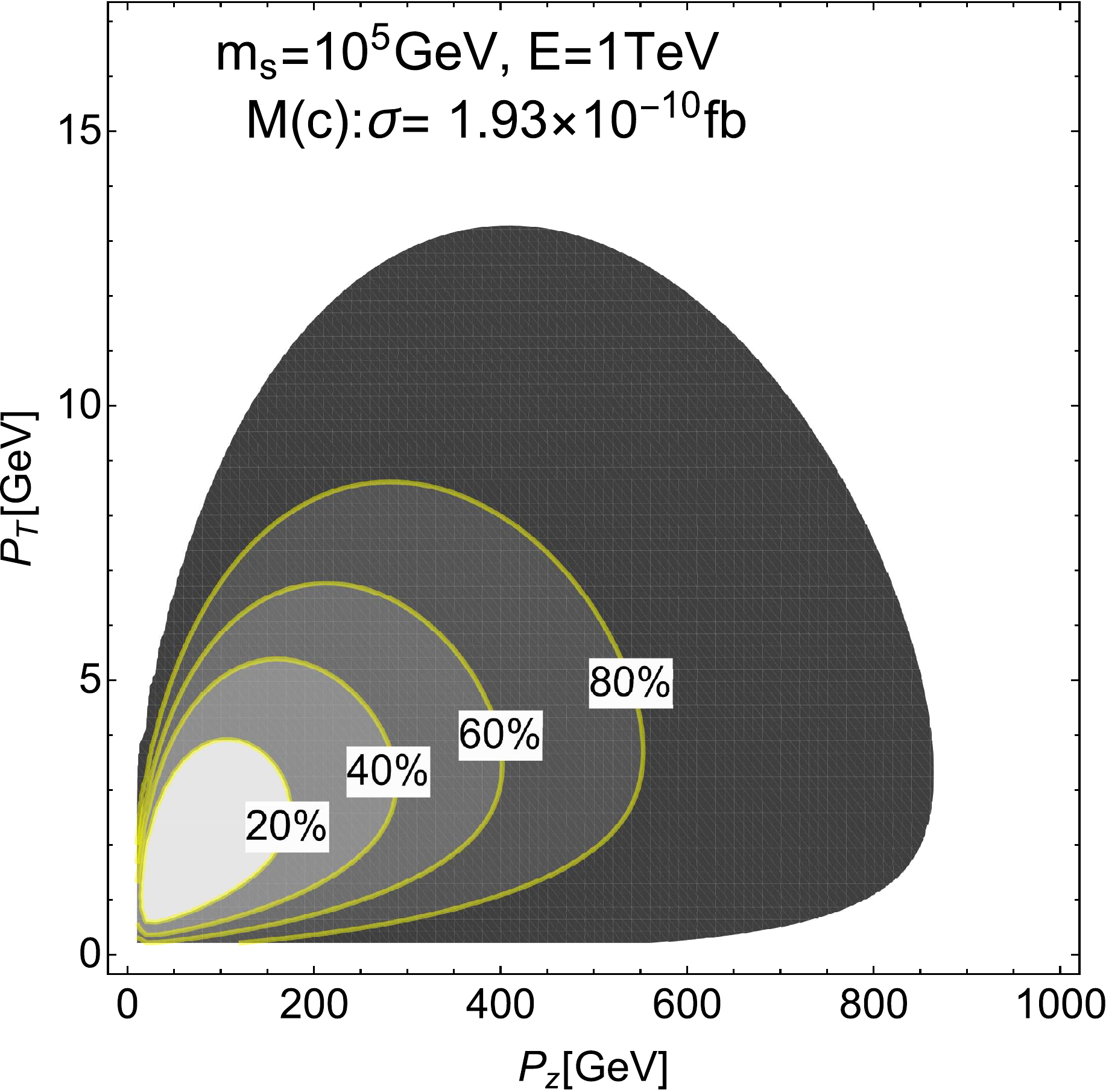}
\includegraphics[width=0.3\linewidth]
{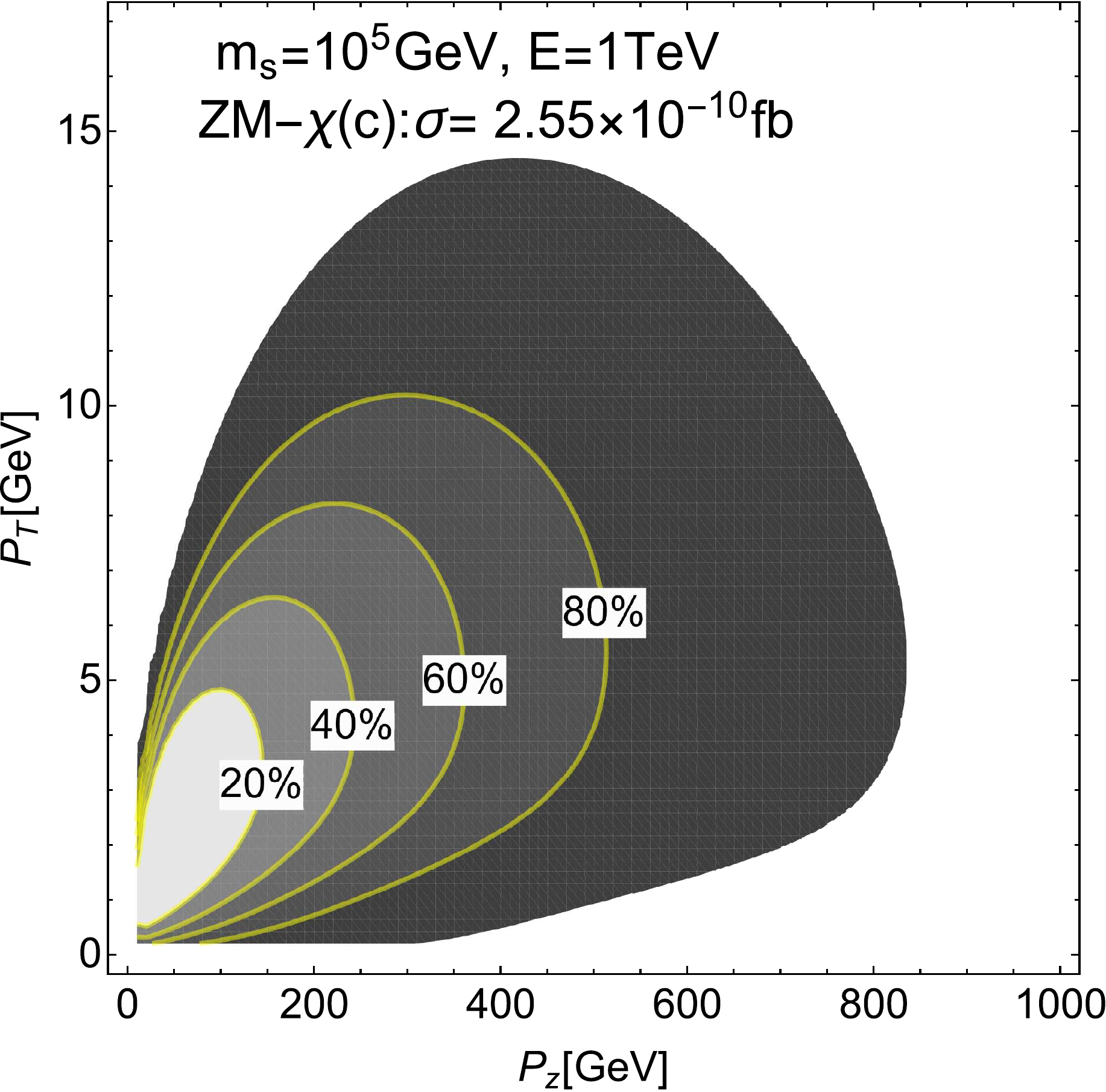}
\includegraphics[width=0.3\linewidth]
{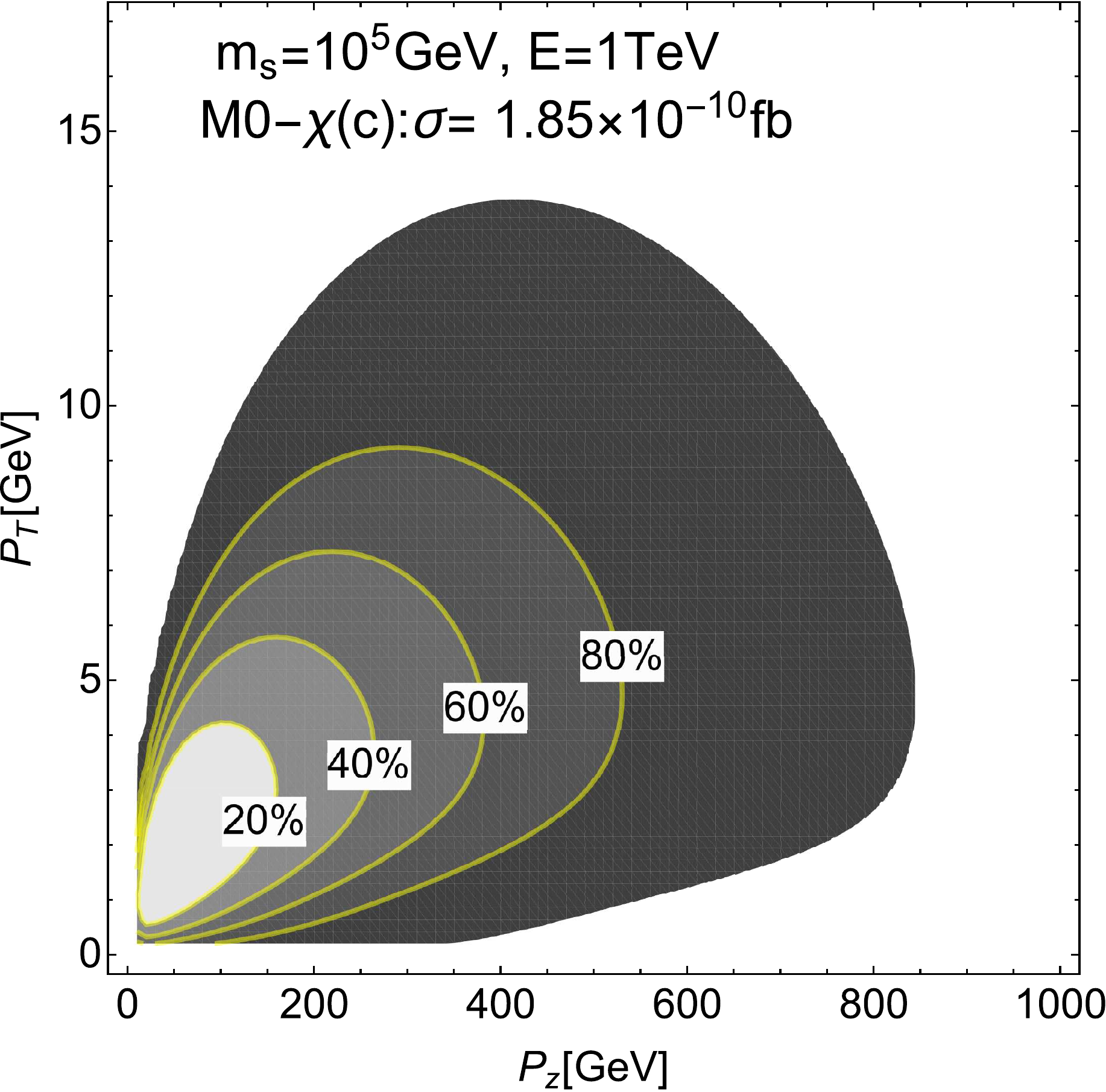}
\caption{
Same as Fig.~\ref{Fig:dist_bACOT_components}, but 
for the charm quark productions.
The results for the M, \ZMC~, and \MZC~ are shown in the three columns, 
and the results for $m_S=10\, {\rm GeV}$ and $m_S=10^5\, {\rm GeV}$ 
are shown in the two rows.}
\label{Fig:dist_cACOT_components}
\end{figure}

In Fig.~\ref{Fig:dist_cACOT} the $\tau$-momentum distribution in 
CLFV DIS associated with charm quark production is shown
for $E_e=200\, {\rm GeV}$ and $E_e=1\, {\rm TeV}$.  
The distribution is more concentrated in the high-$p_Z$ region 
compared to one of the bottom quark production
because the charm quark ($m_c=1.3\, {\rm GeV}$) is more relativistic
 than the bottom quark ($m_b=4.75\, {\rm GeV}$).
The size of scalar mass affects the normalization and 
also the shape of the distribution.
The cross sections in terms of components are given by:
\begin{equation}
\small
\begin{split}
\mbox{(i)}~~
 \sigma_{{\rm thr}, c}^{\mAC} &=
\big[9.30+8.64(9.98)-6.49(7.60)\big]\times 10^{\, 4}\, {\rm fb}~~~~~=1.15(1.17)\times 10^{\, 5}\, {\rm fb},
\\
\mbox{(ii)}~~
\sigma_{{\rm thr}, c}^{\mAC} &=
\big[6.60+5.85(6.41)-4.86(5.38)\big]\times 10^{\, 5}\, {\rm fb}~~~~~=7.59(7.63)\times 10^{\, 5}\, {\rm fb},
\\
\mbox{(iii)}~~
\sigma_{{\rm thr}, c}^{\mAC} &=
\big[1.56+1.78(1.99)-1.26(1.43)\big]\times 10^{-11}\, {\rm fb}~~=2.08(2.12)\times 10^{-11}\, {\rm fb},
\\
\mbox{(iv)}~~
\sigma_{{\rm thr}, c}^{\mAC} &=
\big[1.93+2.43(2.55)-1.75(1.85)\big]\times 10^{-10}\, {\rm fb}~~=2.61(2.62)\times 10^{-10}\, {\rm fb},
\end{split}
\label{eq:SACOT_vs_components_c}
\end{equation}
 where the first/second/third number in the square parenthesis represents
the M/\ZMC~/\MZC~ contribution
 with (without) the threshold factor 
for (i) $E_e=200\, {\rm GeV},~ m_S=10\, {\rm GeV}$, 
 (ii) $E_e=1\, {\rm TeV}, ~m_S=10\, {\rm GeV}$, 
 (iii) $E_e=200\, {\rm GeV},~ m_S=10^5\, {\rm GeV}$, 
 and (iv) $E_e=1\, {\rm TeV},~m_S=10^5\, {\rm GeV}$.
 In the components of \AC~(thr) cross section,
the rates of M and \ZMC~(thr) cross sections are the 
same size. For the charm quark production, dominance of only one component does not
hold for $E_e=200\, {\rm GeV}$ and $E_e=1\, {\rm TeV}$.
 
It should be remembered that we have fixed the  CLFV couplings 
by Eq.~\eqref{eq:cc_set_to_1},  which control the overall normalization of the cross section. Thus it is of great importance to find a sensitivity 
 to the scalar mass in the shape of the distribution, which can be utilized
for a detailed study to discriminate the structure of the CLFV
 interactions, namely for the separation of effects of coupling constants and  of scalar mass. 

\subsection{\AC~: $Q$-distribution}

Here we rewrite the cross section as
\begin{eqnarray}
\frac{d\sigma}{dQ} 
&=&  
\frac{1}{m_\phi^4}\,  W_\phi(m_\phi^2, Q^2) \widetilde{M}_\phi(s, Q^2),
\\
\widetilde{M}_\phi(s,Q^2)
&=& \frac{(Q^2)^{\frac{3}{2}}(Q^2+m_\tau^2)}{8\pi s^2} M_\phi(s, Q^2),
\\
W_\phi(m_\phi^2, Q^2) &=& \left(\frac{m_\phi^2}{Q^2+m_\phi^2}\right)^2,
\end{eqnarray}
where we have introduced a modified inverse moment 
$\widetilde{M}$ and a weighting factor $W_\phi$. 
The mediator mass can be set to $m_\phi=m_S$.
Note that the product of $W_\phi$ and $\widetilde{M}_\phi$ 
gives $d\sigma/dQ$ not $d\sigma/dQ^2$ (up to $m_\phi^4$).
The inverse moment $\widetilde{M}_\phi(s, Q^2)$ is more adequate
 than the structure function to see which region of $Q$ 
 contributes to the cross section.
 
\begin{figure}[htbp]
\centering
\includegraphics[width=16cm]{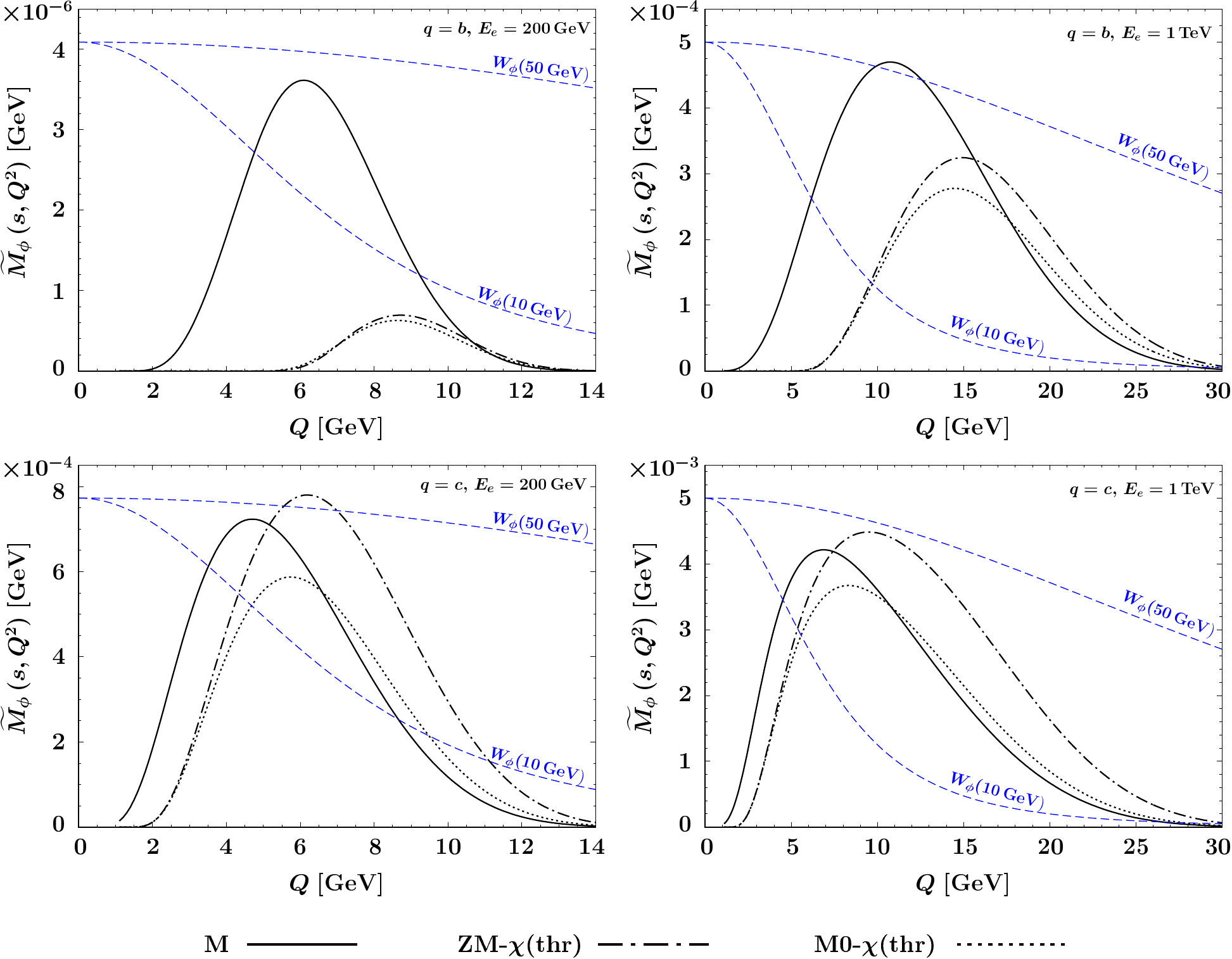}
\caption{
The inverse moment $\widetilde{M}_\phi(s, Q^2)$ 
for each component of the \AC~(thr), and the weighting factor
$W_\phi(m_S^2, Q^2)$ for $m_S=10\, {\rm GeV}$ and 
$m_S=50\, {\rm GeV}$. For visualization the weighting factor 
normalized to appropriate values are plotted.
The results for bottom quark ($q=b$) and charm quark ($q=c$) are 
shown respectively in the upper and lower rows, 
and the beam energies are $E_e={\rm 200\,  GeV}$ and 
$E_e={\rm 1\, TeV}$ in the left and right columns.}
\label{Fig:IM} 
\end{figure}

In Fig.~\ref{Fig:IM}, 
the inverse moments $\widetilde{M}_\phi(s,Q^2)$ are plotted as functions of $Q$ for the bottom and charm quark productions in upper and lower
rows, respectively. The results for $E_e=200\, {\rm GeV}$ and $E_e=1\, {\rm TeV}$ are shown in the left and right columns, respectively.
In the plots one can see the following features of the inverse moment:
\begin{itemize}
\item The support of the M scheme inverse moment starts at low $Q$:  
the curve of the $\widetilde{M}_\phi$ in the M scheme
 rises from 0 at $Q\sim 2\, {\rm GeV}$ ($Q\sim 1\, {\rm GeV}$) 
for the bottom (charm) case. On the other hand, 
the curve of the $\widetilde{M}_\phi$ in the \ZMC~(thr) scheme
rises from 0 at $Q\sim 6\, {\rm GeV}$ 
($Q\sim 2\, {\rm GeV}$) for the bottom (charm) case.
Thus the cross section of the M scheme is superior in magnitude to that 
of the \ZMC~(thr) in the very low-$Q^2$ region.

\item The larger the beam energy $E_e$ 
the higher the maximum $Q$ for the support 
of the inverse moment in all the schemes. 
The relative size of \ZMC~(thr) to that of the M scheme 
tends to grow with the beam energy.
\end{itemize}
These features due to the dynamics of the QCD
 (properties of the structure functions) 
together with the $Q$-dependence of the weighting function $W_\phi(m_\phi^2, Q^2)$
determine the relative importance of the M scheme contribution
to that of \ZMC~(thr) as a function of $m_\phi^2$ and $s$.

\begin{figure}[htbp]
\centering\
\includegraphics[width=15cm]{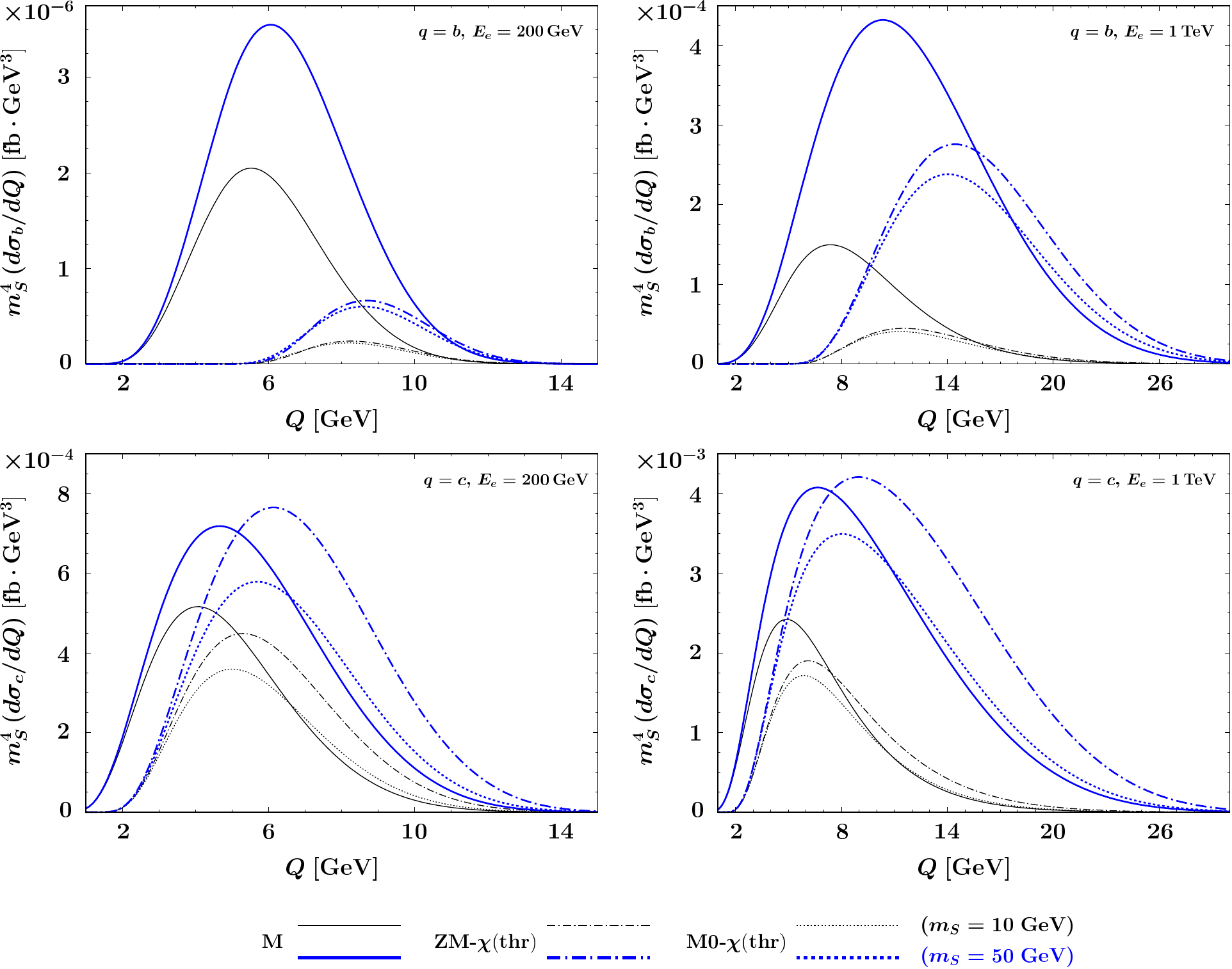}
\caption{
The CLFV DIS cross sections $m_S^4\, (d\sigma/dQ)$ 
are plotted as functions of $Q$.
The results for bottom quark ($q=b$) and charm quark ($q=c$)
are shown respectively in the upper and lower rows, 
and the beam energies are 
$E_e=200\, {\rm GeV}$ and $E_e=1\, {\rm TeV}$ in the 
left and right columns.
The scalar masses
are $m_S=10\, {\rm GeV}$(black lines)
 and $m_S=50\, {\rm GeV}$(blue lines).
}
\label{Fig:m4_ds_dQ}
\end{figure}

In Fig.~\ref{Fig:m4_ds_dQ}, the cross section $m_S^4\,  d\sigma/dQ$ is plotted as a function of $Q$ for CLFV DIS associated with the heavy quark 
production. A prefactor $m_S^4$ is multiplied to scale out the leading 
$m_S$-dependence in the large-$m_S$ limit.
The results for the bottom ($q=b$) and charm ($q=c$) quark productions are presented in the upper and lower rows, and the results for 
the beam energies of $E_e=200\, {\rm GeV}$ and $E_e=1\, {\rm TeV}$ are 
presented in the left and right columns.
Two representative cases $m_S=10\, {\rm GeV}$ (black curves)
and $m_S=50\, {\rm GeV}$ (blue curves) for the scalar mass are simulated. 
Note that $m_S=50\, {\rm GeV}$ is sufficiently large 
compared to typical momentum transfer $Q$ for $E_e=200\, {\rm GeV}$ and $E_e=1\, {\rm TeV}$ (see Figs.~\ref{Fig:IM}, \ref{Fig:ms4XS_vs_ms}),
thus it corresponds approximately to a case of the large-$m_S$ limit.
In the plots we observe the following:
\begin{itemize}

\item For the bottom quark production, 
the supports of the cross sections of M and \ZMC~(thr) schemes 
start respectively at $Q\sim 2\, {\rm GeV}$ and $Q\sim 6\, {\rm GeV}$. 
The contribution from the region ${\cal R}_Q=\{Q|~2\,{\rm GeV} \leq Q\leq 6\, {\rm GeV}\}$
to the M scheme cross section is given by $\sigma^{\mM}({\cal R}_R)=\int_{{\cal R}_Q}(d\sigma^{\mM}/dQ) dQ$ and it is large, while the contribution from the same region to \ZMC~(thr)
is $\sigma^{\mZMC}({\cal R}_Q)=0$. 
This yields significantly large contribution exclusively 
to the M scheme cross section, 
leading the dominance of the M component in the \AC~(thr) cross section
for $E_e=200\, {\rm GeV}$ and $E_e=1\, {\rm TeV}$. 
The difference $(\sigma_{\phi, {\rm thr}}^{\mZMC}-\sigma_{\phi, {\rm thr}}^{\mMZC})$ is approximately 0 except the case of
 $m_S=50\, {\rm GeV}$ for $E_e=1\, {\rm TeV}$.
The effect of the $Q^2$ evolution of the bottom PDF 
to the \AC~(thr) cross section is negligible for 
$m_S=10\, {\rm GeV}$ irrespective of the beam energy, 
and it is moderate for $m_S=10^5\, {\rm GeV}$ with $E_e=1\, {\rm TeV}$.

\item For the charm quark production,  all the components are approximately equal 
in magnitude, and the difference $(\sigma_{\phi, {\rm thr}}^{\mZMC}-\sigma_{\phi, {\rm thr}}^{\mMZC})$ is large, thus $(\sigma_{\phi, {\rm thr}}^{\mAC} - \sigma_{\phi}^{\mM})$ is also large. This means
that the effect of $Q^2$ evolution for the charm PDF is large, 
therefore the deviation of M component  from the \AC~(thr)
 becomes sizable. Breakdown of the M scheme dominance
  occurs depending on the value of scalar mass and also
   on beam energy $E_e$. Actually the cross section in \ZMC~(thr)
becomes dominant for large scalar mass $m_S=50\, {\rm GeV}$
(see also the case of $m_S=10^5\, {\rm GeV}$ in Eq.~\eqref{eq:SACOT_vs_components_c}).

\item The \MZC~(thr) curves are close to 
the \ZMC~(thr) curves in the  low-$Q^2$ region, 
while  in the higher $Q$ they approach to the 
M scheme curves.  This is expected by its construction;
 the subtraction term (=\MZC~(thr)) should agree with 
 M scheme contribution for $Q^2\gg m_q^2$ because the leading mass
  singularities are the same in both schemes, and the subtraction term 
  should agree with \ZMC~(thr) for $Q^2 \sim m_q^2$ because 
  the heavy quark PDFs are born in the region $\mu_f^2\sim m_q^2$ and 
 $f_{q/N}(x,\mu_f^2)$ are not evolved much. 
Indeed it is seen in all the plots that curves of \MZC~(thr) interpolate
the two schemes from low-$Q^2$ to high-$Q^2$ region between 
\ZMC~ and M schemes. 
\end{itemize}

There is one important difference between the CLFV DIS mediated by the massive 
scalar $\phi$ and standard neutral current DIS $e N\to e X$ where the 
massless photon is exchanged.
The total cross section for the CLFV DIS contains the contribution from 
all the $Q$ region, and the relative importance of the low-$Q^2$ region 
versus large-$Q^2$ region is controlled not only by the collision energy $\sqrt{s}$ but also
by the scalar mass $m_S$. Here the scalar mass $m_S$ plays a role of 
cut-off for the momentum transfer $Q$, and the contribution from the  region of $Q\gtrsim m_S$ to the CLFV cross section is suppressed. This is contrasted 
with the normal DIS where the $Q$ region which contributes to the cross section is controlled 
solely by the collision energy $\sqrt{s}$.

\subsection{\AC~: Total cross section}

\begin{figure}[htbp]
\centering
\includegraphics[width=8cm]{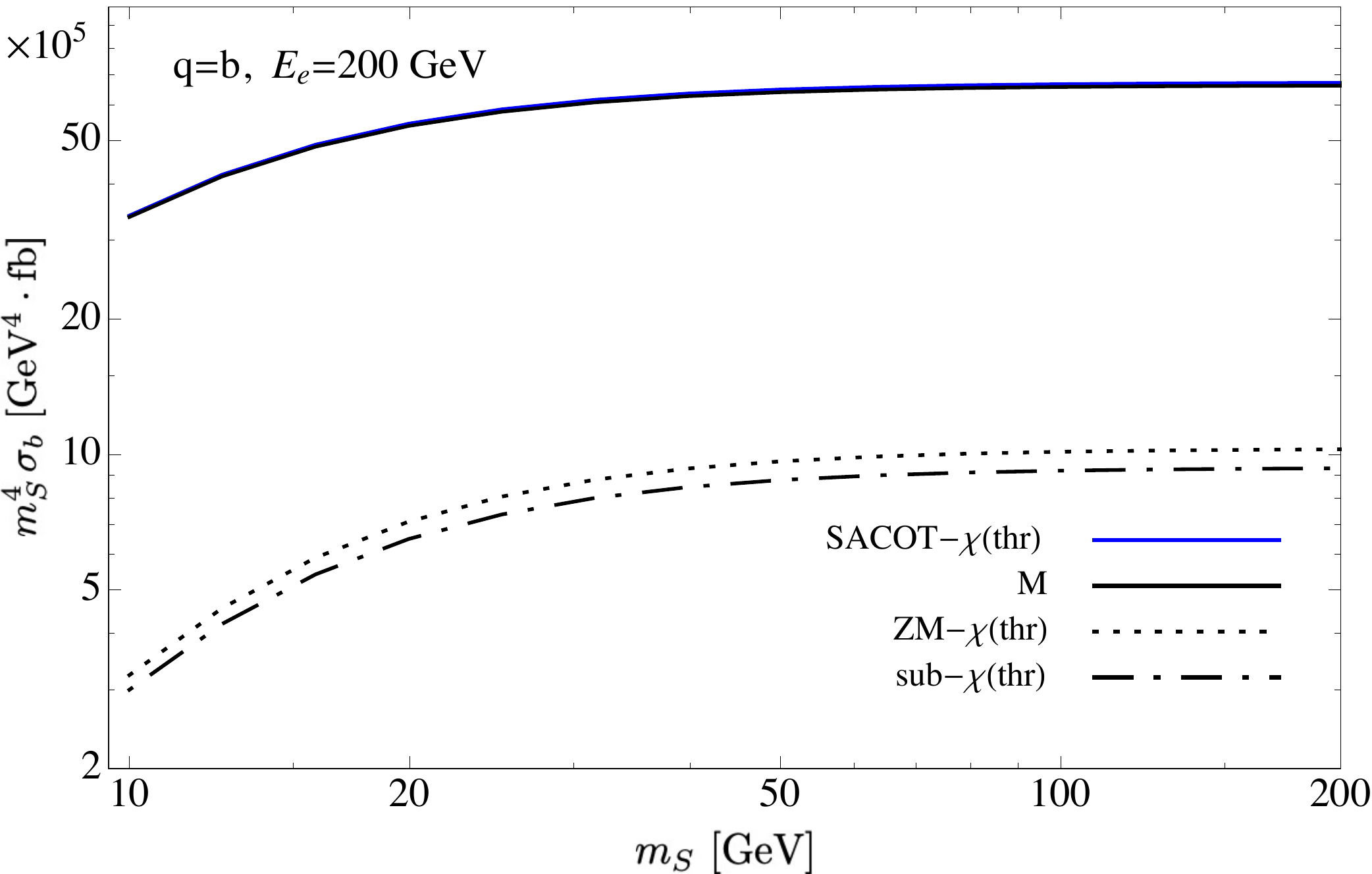}
\includegraphics[width=8cm]{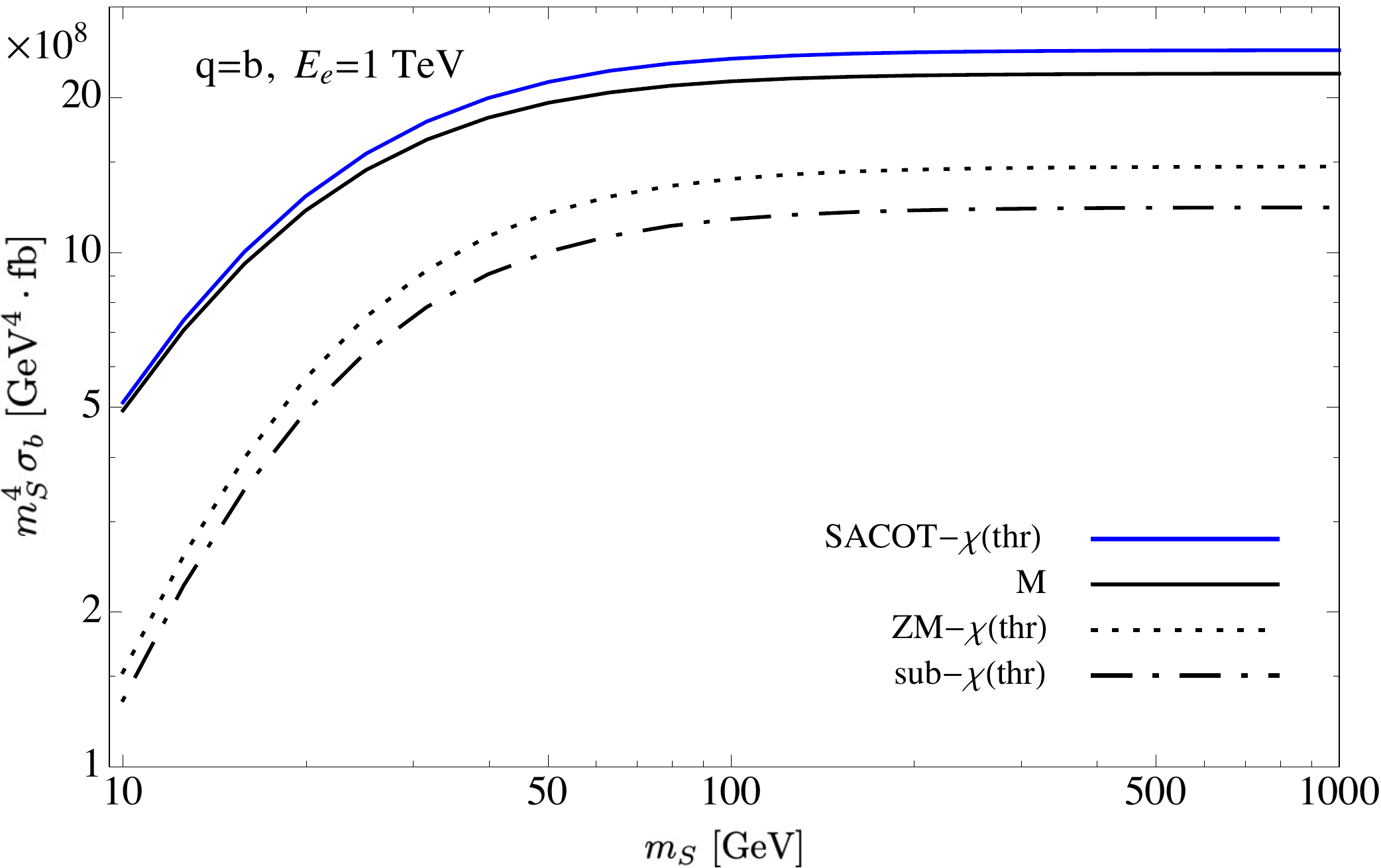}
\\
\vspace{5mm}
\includegraphics[width=8cm]{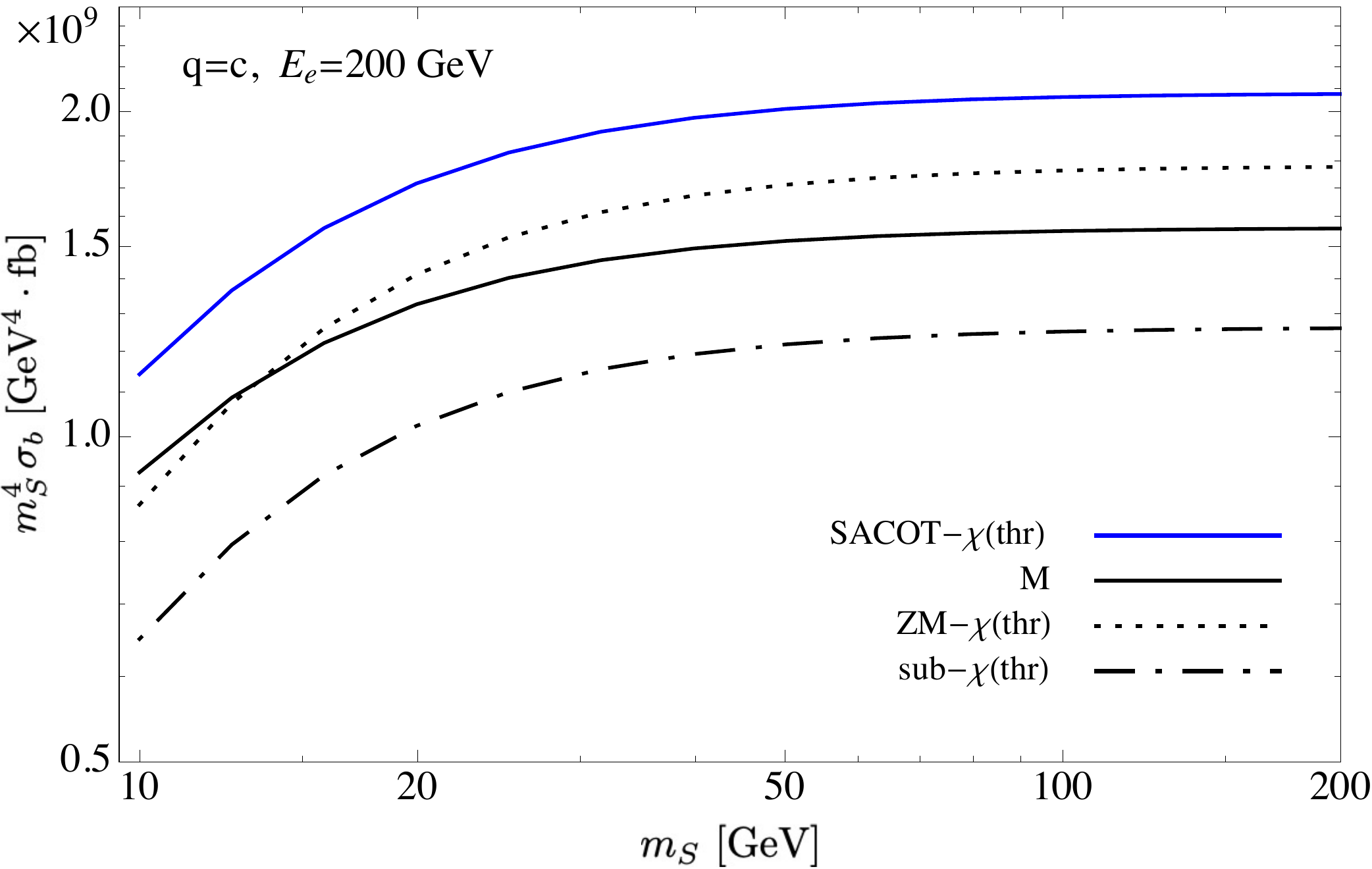}
\includegraphics[width=8cm]{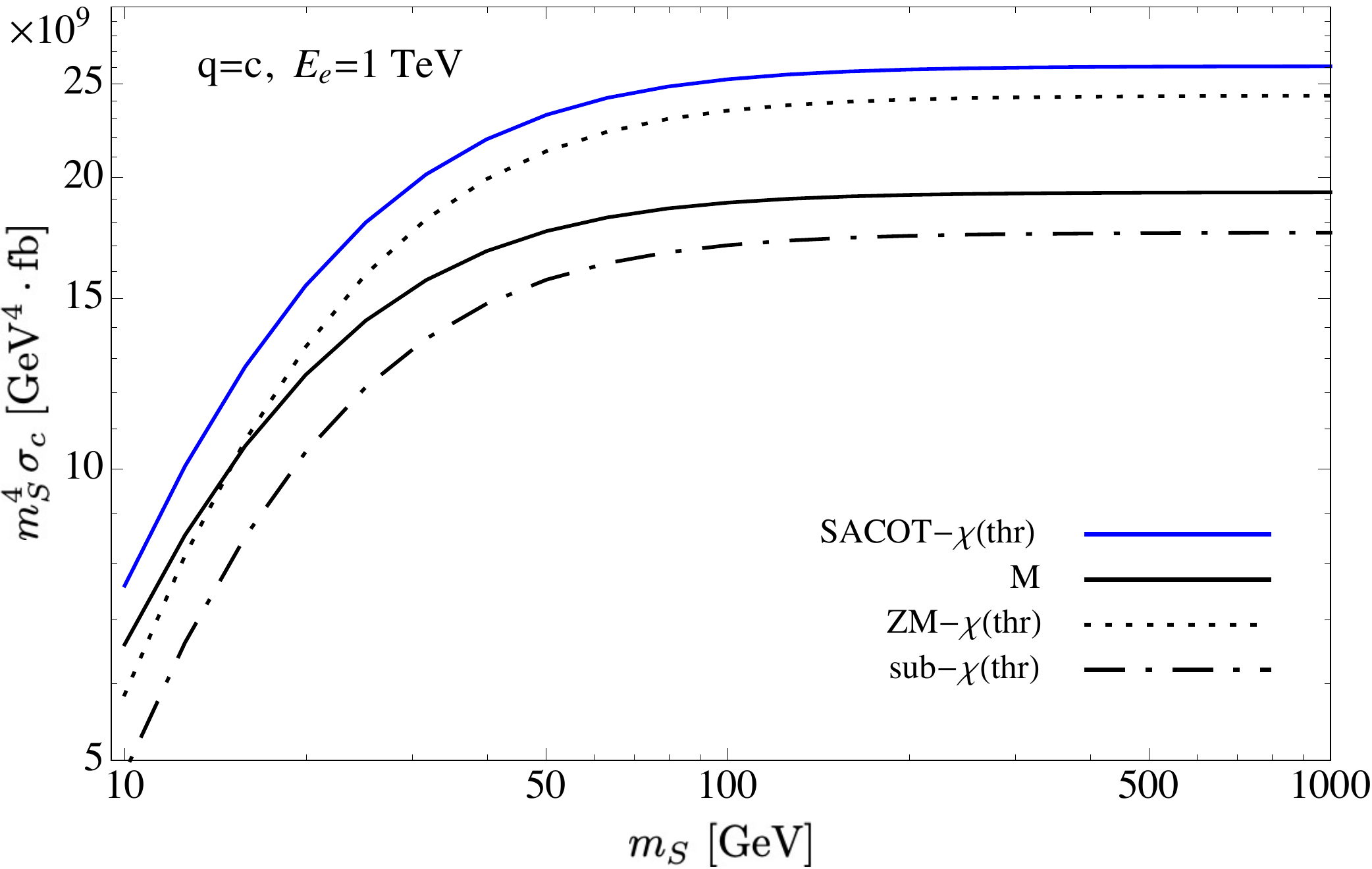}
\caption{
The total cross sections $m_S^4 \, \sigma_q$ are plotted as functions of $m_S$. 
The results for bottom quark ($q=b$)
 and charm quark ($q=c$) are shown respectively 
 in the upper and lower rows, and 
the  beam energies are $E_e=200\, {\rm GeV}$ 
and $E_e=1\, {\rm TeV}$ in the left and right columns.
}
\label{Fig:ms4XS_vs_ms}
\end{figure}

Here we discuss the dependences of the total cross sections 
on the collision energy $\sqrt{s}$ and on the scalar mass $m_S$
 in the \AC~(thr) scheme.
In Fig.~\ref{Fig:ms4XS_vs_ms} the total cross sections $m_S^4\, \sigma_q$  for CLFV DIS associated with heavy quark productions
are plotted as functions of scalar mass $m_S$. 
 The results for bottom and 
charm quark productions are presented in the upper and lower rows,
respectively, and the beam energies are 
$E_e=200\, {\rm GeV}$ and $E_e=1\, {\rm TeV}$
respectively in the left and right columns. 
The following behaviors concerning to the 
$m_S$-dependence can be read from the plots:
\begin{itemize}
 
\item  For the bottom quark production, 
all the curves for \AC~(thr), M, \ZMC~(thr), and \SubC~(thr) 
converge to their asymptotic (constant) values in large-$m_S$ limit.
The \AC~(thr) cross section fitted at the large-$m_S$ limit is 
$\sigma_{{\rm thr}, b}^{\mAC}\approx (6.6+1.0-0.9)\cdot 10^6\,  {\rm fb} 
\times (m_S/1\,{\rm GeV})^{-4}$, decomposing 
the components for M, \ZMC~(thr), \SubC~(thr), for $E_e=200\, {\rm GeV}$ 
and $\sigma_{{\rm thr}, b}^{\mAC}\approx (2.2+1.5-1.2)\cdot 10^9\,  {\rm fb} 
\times (m_S/1\,{\rm GeV})^{-4}$  for $E_e=1\, {\rm TeV}$, which agree with the values for $m_S=10^5\, {\rm GeV}$ in Eq.~\eqref{eq:SACOT_vs_components_b}.
The \AC~(thr) cross section evaluated at $m_S=50\, {\rm GeV}$ deviates 
from this asymptotic form by $4\%$ ($13\%$) for 
 $E_e=200\,{\rm GeV}$ ($E_e=1\,{\rm TeV}$). 
This bears out 
 that the value of $m_S=50\, {\rm GeV}$ can be regarded in 
 a good approximation as the large-$m_S$ limit for beam energies
  $E_e\lesssim 1\, {\rm TeV}$.
The plots for the bottom cross section also show 
that  the M scheme cross section is a good approximation to the \AC~(thr),
 regardless of the value of $m_S$ for $E_e\lesssim1\, {\rm TeV}$.

\item
For the charm quark production,  
all the curves for $E_e=200\, {\rm GeV}$ and $E_e=1\, {\rm TeV}$ 
converge to their asymptotic values similarly to the case of 
bottom quark production. The \AC~(thr) cross section fitted as
the large-$m_S$ limit is 
$\sigma_{{\rm thr}, c}^{\mAC}\approx (1.6+1.8-1.3)\cdot 10^9\,  {\rm fb} 
\times (m_S/1\,{\rm GeV})^{-4}$
for $E_e=200\, {\rm GeV}$ and 
$\sigma_{{\rm thr}, c}^{\mAC}\approx (1.9+2.4-1.8)\cdot 10^{10}\,   {\rm fb} 
\times (m_S/1\,{\rm GeV})^{-4}$
for $E_e=1\, {\rm TeV}$,
 which agree with the values for $m_S=10^5\, {\rm GeV}$ in Eq.~\eqref{eq:SACOT_vs_components_c}.
Comparing the asymptotic large-$m_S$ limit with
the values at $m_S=50\, {\rm GeV}$ bears out that
$m_S=50\,{\rm GeV}$ can be regarded in 
 a good approximation as the asymptotic large $m_S$-limit for beam energies $E_e\lesssim1\, {\rm TeV}$.
 For the charm case, 
the \ZMC~(thr) cross section is the dominant component for 
$m_S \gtrsim 15\, {\rm GeV}$ and $E_e\lesssim 1\, {\rm TeV}$,
but there is still sizable corrections from the components
M and \SubC~(thr) to match with the  value of the 
\AC~(thr) scheme cross section.
 
\end{itemize}

\begin{figure}[htbp]
\centering
\includegraphics[width=8cm]{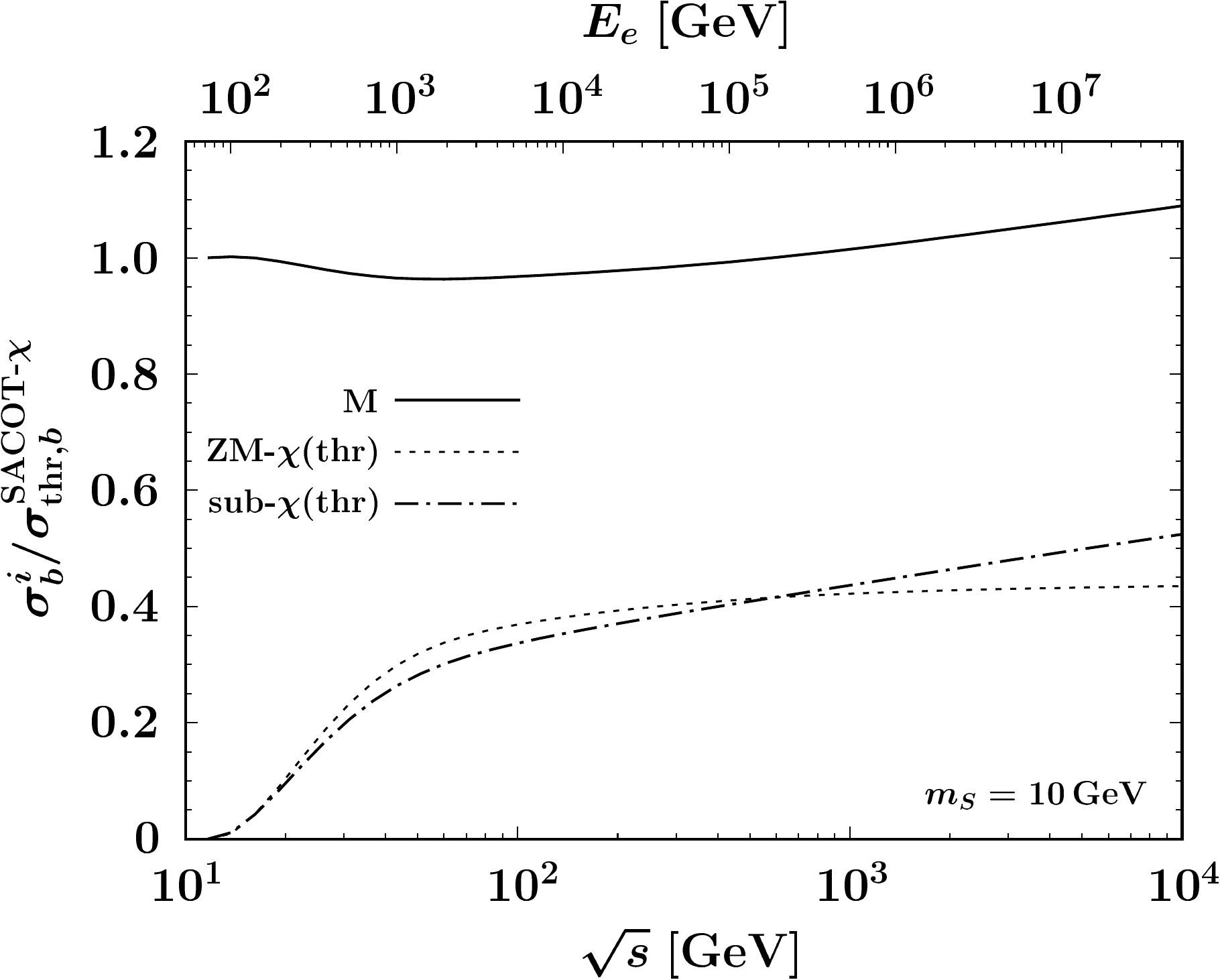}
\includegraphics[width=8cm]{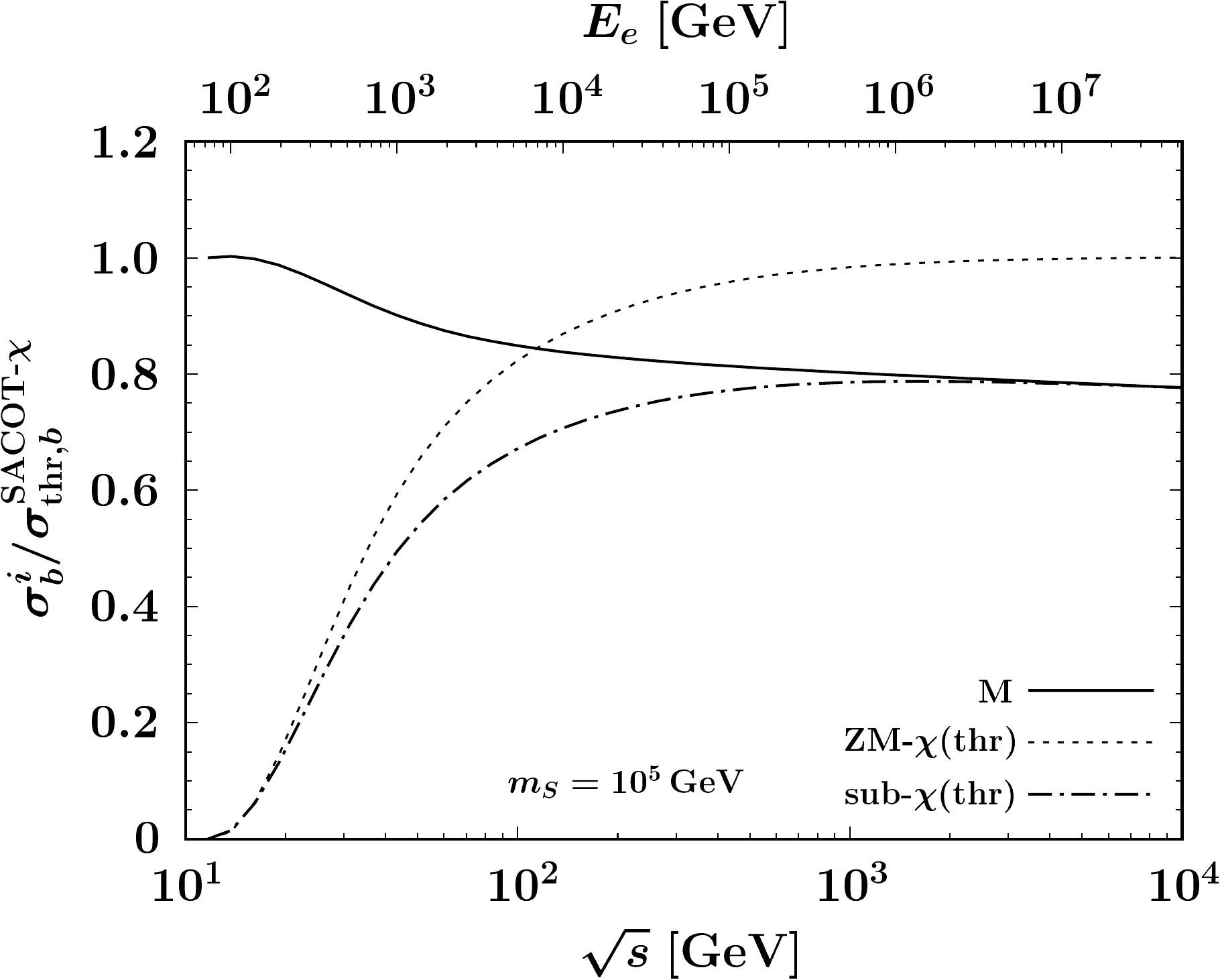}
\\
\vspace{5mm}
\includegraphics[width=8cm]{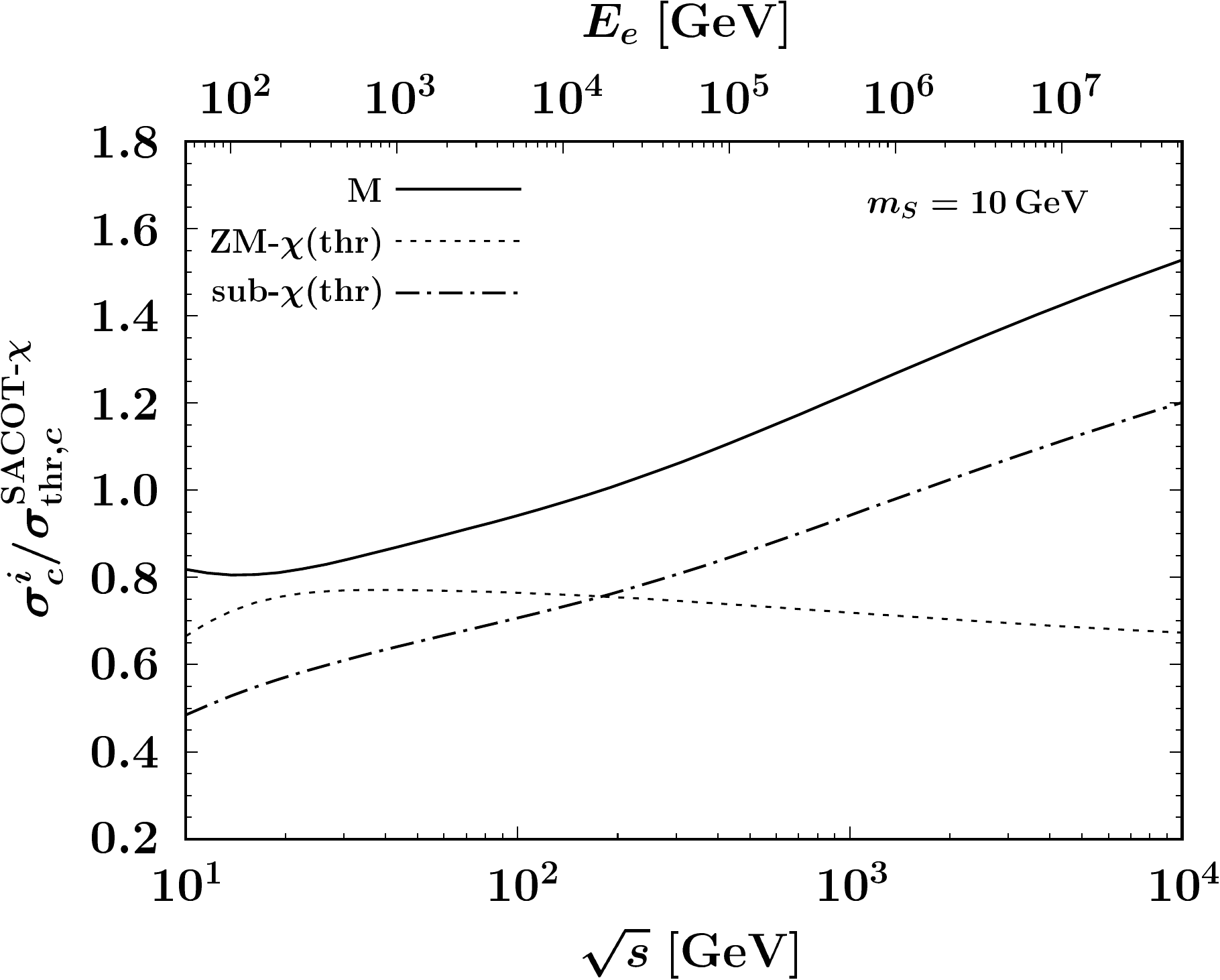}
\includegraphics[width=8cm]{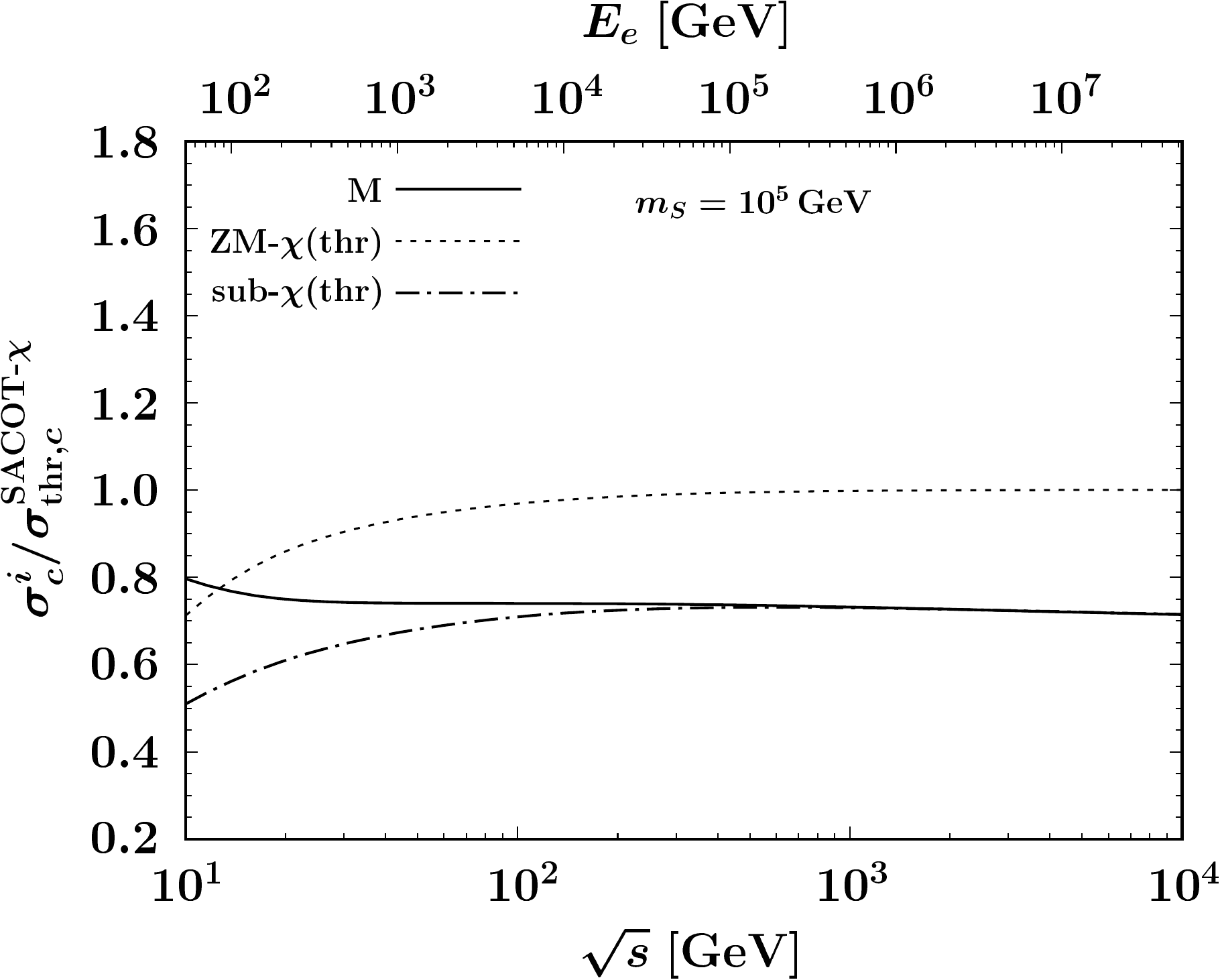}
\caption{
The components M, \ZMC~, and \MZC~ normalized to the 
\AC ~ as functions of the collision energy $\sqrt{s}$: 
the ticks of upper horizontal axis indicate the corresponding beam energy $E_e$.
The cross sections for the bottom and charm quark production
are shown in the upper and lower rows.
The scalar masses are $m_S=10\, {\rm GeV}$ and 
$m_S=10^5\, {\rm GeV}$ respectively in the left and right columns.
} 
\label{Fig:acot_bc}
\end{figure}

For comparison among the components of \AC~(thr) cross section
 in a wide range of the collision energy, the total cross sections
are plotted in Fig.~\ref{Fig:acot_bc} 
as functions of $\sqrt{s}$ for bottom and charm quark productions. 
Numerical values for the cross sections in the \AC~(thr) scheme and its components are listed in Table~\ref{tab:XS_ACOT}. Concerning to the
collision energy dependences following features can be observed from the 
plots:
\begin{itemize}
\item
For the bottom quark production the dominance of 
the M scheme contribution holds for the small scalar mass 
$m_S=10\, {\rm GeV}$ in all the range of $\sqrt{s}$. But 
for the large scalar mass $m_S=10^5\, {\rm GeV}$ the \ZMC~(thr) contribution 
becomes the dominant component for $\sqrt{s}\gtrsim100\, {\rm GeV}$
 ($E_e\gtrsim 5\, {\rm TeV}$). In large $\sqrt{s}$ limit for the 
 case of $m_S=10^5\, {\rm GeV}$ the cross sections in 
 M and \ZMC~(thr) schemes coincide each other above 
 $\sqrt{s} \approx 1 \, {\rm TeV}$. 
  Thus the M and \SubC~(thr) terms cancels each other
  and the \AC~(thr)   scheme cross section is approximated
  by the \ZMC~(thr) cross section for $\sqrt{s}\gtrsim 1\, {\rm TeV}$,
   i.e. $\sigma_{{\rm thr}, b}^{\mAC}\simeq \sigma_{{\rm thr}, b}^{\mZMC}$ 
 for $m_S=10^5\, {\rm GeV}$ and $\sqrt{s} \gtrsim 1 \, {\rm TeV}$.
  
\item 
For the charm quark production for small scalar mass $m_S=10\, {\rm GeV}$, there is no unique component which 
 dominates the cross section in all the energy range, and all the components M, \ZMC~(thr), and \SubC~(thr) are equally important for the total cross section $\sigma_{{\rm thr}, c}^{\mAC}$.
On the other hand, for the large scalar mass $m_S=10^5\, {\rm GeV}$,   the ZM-$\chi$ cross section describes the \AC~(thr) cross section well  for all the energy range except $\sqrt{s} \lesssim 100\, {\rm GeV}$. 
  Starting at $\sqrt{s} \sim 100\, {\rm GeV}$ the M and \SubC~(thr) cross sections coincide so that they cancel each other in the \AC~(thr) cross section, i.e. 
 $\sigma_{{\rm thr}, c}^{\mAC}\simeq \sigma_{{\rm thr}, c}^{\mZMC}$ 
 for $m_S=10^5\, {\rm GeV}$.
 \end{itemize}

\begin{table}[htbp]
\centering
\caption{The total cross sections for CLFV DIS associated with bottom quark production in \AC~-thr and its components. 
For the \ZMC~,  \MZC ~ and \AC~ schemes, the threshold factor $S_{\rm thr}$ is taken into account.}
\vspace{0.5cm}
\label{tab:XS_ACOT}
\begin{tabular}{|c|c|c|c|c|c|}
\hline
$m_S~[\rm {GeV}]$
& $E_e~{\rm [GeV]}$ 
& $\sigma_{b}^{\rm M} ~{\rm [fb]}$ 
& $\sigma_{{\rm thr}, b}^{\mZMC} ~{\rm [fb]}$   
& $\sigma_{{\rm thr}, b}^{\mMZC} ~{\rm [fb]}$  
& $\sigma_{{\rm thr}, b}^{\mAC}~{\rm [fb]}$    \\
\hline
& $10^2$ & ~$2.08\times 10^{-1}$~ & ~$1.95\times 10^{-3}$~ & ~$2.32\times 10^{-3}$~ & ~$2.07\times 10^{-1}$~
\\
& $10^3$ & $4.93\times 10^{4}$ & $1.53\times 10^{4}$ & $1.35\times 10^4$  & $5.11\times 10^{4}$
\\
$10$ 
& $10^4$ & $7.32\times 10^{5}$ & $2.87\times 10^{5}$ & $2.66\times 10^5$ & $7.53\times 10^{5}$
\\
& $10^5$ &  $3.72\times 10^{6}$ & $1.54\times 10^{6}$ & $1.51\times 10^6$ & $3.75\times 10^{6}$
\\
& $10^6$ &  $1.18\times 10^{7}$ & $4.90\times 10^{6}$ & $5.18\times 10^6$ & $1.15\times 10^{7}$
\\
& $10^7$ & $2.99\times 10^{7}$ & $1.22\times 10^{7}$ & $1.39\times 10^7$ & $2.81\times 10^{7}$
\\
\hline
& $10^2$ & $3.17\times 10^{-17}$ & $4.06\times 10^{-19}$ &$4.74\times 10^{-19}$ &  $3.17\times 10^{-17}$
\\
& $10^3$ &  $2.23\times 10^{-11}$ & $1.47\times 10^{-11}$ & $1.22\times 10^{-11}$&  $2.47\times 10^{-11}$
\\
$10^5$ 
& $10^4$ & $1.27\times 10^{-9}$ &  $1.32\times 10^{-9}$ & $1.07\times 10^{-9}$ & $1.51\times 10^{-9}$
\\
& $10^5$ & $2.57\times 10^{-8}$ & $3.03\times 10^{-8}$ &  $2.44\times 10^{-8}$ & $3.16\times 10^{-8}$
\\
& $10^6$ & $3.66\times 10^{-7}$ & $4.54\times 10^{-7}$ & $3.61\times 10^{-7}$  & $4.59\times 10^{-7}$
\\
& $10^7$ & $4.50\times 10^{-6}$ & $5.72\times 10^{-6}$ & $4.49\times 10^{-6}$ & $5.73\times 10^{-6}$
\\
\hline\hline
$m_S~[\rm {GeV}]$
& $E_e~{\rm [GeV]}$ 
& $\sigma_{c}^{\rm M} ~{\rm [fb]}$ 
& $\sigma_{{\rm thr}, c}^{\mZMC} ~{\rm [fb]}$   
& $\sigma_{{\rm thr}, c}^{\mMZC} ~{\rm [fb]}$  
& $\sigma_{{\rm thr}, c}^{\mAC}~{\rm [fb]}$    \\
\hline
& $10^2$ & ~$2.74\times 10^{4}$~ & ~$2.45\times 10^{4}$~ & ~$1.79\times 10^4$~ & ~$3.40\times 10^{4}$~
\\
& $10^3$ & $6.59\times 10^{5}$ &  $5.85\times 10^{5}$ & $4.86\times 10^5$  & $7.58\times 10^{5}$
\\
$10$ 
& $10^4$ &  $4.15\times 10^{6}$ & $3.25\times 10^{6}$ & $3.13\times 10^6$ & $4.27\times 10^{6}$
\\
& $10^5$ & $1.46\times 10^{7}$ & $9.73\times 10^{6}$ & $1.11\times 10^7$ & $1.32\times 10^{7}$
\\
& $10^6$ & $3.83\times 10^{7}$ &  $2.15\times 10^{7}$ & $2.96\times 10^7$ & $3.02\times 10^{7}$
\\
& $10^7$ & $8.68\times 10^{7}$ & $4.19\times 10^{7}$ & $6.78\times 10^7$ & $6.09\times 10^{7}$
\\
\hline
& $10^2$ &  $3.86\times 10^{-12}$ & $3.99\times 10^{-12}$ & $2.83\times 10^{-12}$ & $5.03\times 10^{-12}$
\\
& $10^3$ &  $1.93\times 10^{-10}$ & $2.43\times 10^{-10}$ &  $1.75\times 10^{-10}$ & $2.61\times 10^{-10}$
\\
$10^5$ 
& $10^4$ &  $3.51\times 10^{-9}$ & $4.64\times 10^{-9}$ & $3.41\times 10^{-9}$ & $4.75\times 10^{-9}$
\\
& $10^5$ & $4.66\times 10^{-8}$ & $6.29\times 10^{-8}$ & $4.62\times 10^{-8} $ & $6.33\times 10^{-8}$
\\
& $10^6$ & $5.48\times 10^{-7}$ & $7.50\times 10^{-7}$ & $5.47\times 10^{-7}$ & $7.50\times 10^{-7}$
\\
& $10^7$ &  $6.09\times 10^{-6}$ & $8.45\times 10^{-6}$ & $6.09\times 10^{-6}$ & $8.44\times 10^{-6}$ 
\\
\hline
\end{tabular}
\end{table}

\clearpage
\section{Summary}
In this article, we have considered a scenario 
where a CLFV (pseudo-)scalar mediator couples 
dominantly to heavy quarks and studied the CLFV 
DIS associated with heavy quark productions 
$\ell_i N\to \ell_j q\bar{q}X$ ($q=b,c$).
The CLFV DIS cross sections are written in terms of the leptonic and the hadronic parts.
We have computed the heavy quark contributions to the structure function 
in the \AC~(thr), M, \ZMC~(thr), and \MZC~(thr) schemes and presented 
their results. We have examined three improvements for the threshold 
 behavior of the structure function: the $\chi$-rescaling, choice of 
  factorization scale $\mu_f$, and inclusion of the threshold factor $S_{\rm thr}$. 
To our best knowledge the present article is the first 
application of the SACOT-$\chi$(thr) scheme for the CLFV DIS associated
with heavy quark productions. We have shown that only the SACOT-$\chi$(thr) scheme provides a reliable theory prediction for the CLFV DIS cross section in the wide kinematical region and in the full parameter space. Thus the systematic study for the CLFV signal 
search including the full DIS kinematical information will be available in the SACOT-$\chi$(thr) scheme. 

For the structure function we have made a comparison
 among the different computational schemes 
 and observed that the $\chi$-rescaling is effective in 
 $Q \lesssim 100 \, (50)$\,GeV 
for the bottom (charm) quark production. 
 It is mandatory to incorporate the $\chi$-rescaling 
 to predict the structure function especially on the 
 threshold behavior of  the heavy quark productions.

We also made a detailed analysis on the collision energy
 dependence and the mediator mass dependence of the 
 heavy quark production cross sections focusing on the beam 
energies up to $E_e=1\, {\rm TeV}$. 
For the bottom quark production, we showed that the M scheme cross section 
approximates that of the \AC~(thr) irrespective of the size of the scalar mass.
For the charm quark production, we showed that 
the \ZMC~(thr) cross section 
approximates that of the \AC~(thr) for $m_S\gtrsim 50\, {\rm GeV}$,
but for the small scalar mass $\sim 10\, {\rm GeV}$
 the M scheme cross section becomes the dominant component 
 for \AC~(thr).
We found that the ratio of the contributions in the massive and 
the zero-mass schemes  strongly depends on the mediator mass.
This is because when the mediator mass is small, the contributions from 
the low-$Q^2$ region are enhanced, for which the M scheme contribution 
is superior. 
%
%
%
%
%
We conclude that the SACOT-$\chi$(thr) prescription is 
indispensable to obtain the reliable sensitivity to the CLFV interactions in the next generation  experiments of the energy range  $E_e\lesssim 1\, {\rm TeV}$.

To utilize the processes for the CLFV signal searches,
we propose the measurements of the total cross sections 
and the momentum distributions
$d^2\sigma/d\bm{p}$ of the final lepton. 
The normalizations of the total cross sections depend on 
the combination of CLFV mediator couplings 
$|\rho_{qq}^\phi|^2 (|\rho_{ij}^\phi|^2+|\rho_{ji}^\phi|^2)$ and $m_S$, 
while the (normalized) momentum distributions depend on $m_S$.
We have shown for the case of $e N\to \tau q\bar{q}X$ that 
the momentum distributions can have a sensitivity on the size 
of scalar mass. To make a definite conclusion on the feasibility 
of the simultaneous determination of the CLFV couplings and the scalar mass,
more efforts have to be devoted to improve the precision of the 
theory computations. 
The issues of the scale dependence, 
PDF uncertainties, QCD radiative corrections, etc., 
have to be addressed together with more detailed simulation study 
taking the experimental uncertainties into account, which are beyond the scope 
of the present article, but we render them for a future work.

For the present model Lagrangian, it is known that 
the other types of subprocesses also contribute to the 
CLFV DIS via the gluonic operator ($\sim \phi \, G^a_{\mu\nu} G^{a\, \mu\nu}$) 
and the photonic dipole operator ($\sim \bar{\ell}_{i}\sigma_{\mu\nu} \ell_j F^{\mu\nu}$)  which generate the processes 
$\ell_i g\to \ell_j g$ and 
$\ell_i q_l \to \ell_j q_l$  ($q_l$ being the light quarks).
 The comprehensive analysis of the DIS observables taking these 
subprocesses into account is required to disentangle the 
type of the CLFV operators and eventually to unravel the 
properties of CLFV mediators. We leave these issues for 
our future works.

\section*{Acknowledgements}

This work was partly supported by MEXT Joint Usage/Research Center on Mathematics and Theoretical Physics JPMXP0619217849 (M.Y.) and JSPS KAKENHI Grant Numbers JP18K03611, JP16H06492, JP20H00160 (M.T.); JP18H01210, JP21H00081 (Y.U.); JP20H05852 (M.Y.).

\appendix
\section{Physical region of DIS kinematics}

\subsection{Range of $(x, y)$ in generic case}
We derive the physical region of kinematical variable $x$ and $y$ for CLFV DIS process
$\ell_i(k_i) p(k_p) \to \ell_j(k_j) \hat{X}(k_X)$, where $\hat{X}$ can be one particle or a 
multi-particle system. 
 We assume that the initial QCD parton $p$ is massless with 
 momentum $k_p=\xi P$, and other particles are massive: $k_i^2=m_i^2, k_p^2=0, k_j^2=m_j^2$, 
 and $k_X^2=w^2$. Similar analyses can be found in literatures \cite{Albright:1974ts, Hagiwara:2003di}.

First we express the kinematical variables 
in the center-of-mass frame of lepton and parton, where lepton energies
are given by
\begin{eqnarray}
E_i ~=~\frac{\hat{s}+m_i^2}{2\sqrt{\hat{s}}},
\hspace{4mm}
E_j ~=~\frac{\hat{s}+m_j^2-w^2}{2\sqrt{\hat{s}}},
\label{eq:Ei_and_Ej}
\end{eqnarray}
where $\hat{s}=(k_p+k_i)^2$ is related to the lepton-nucleon collision energy 
squared $s=(P+k_i)^2$ by $\hat{s}-m_i^2=\xi\, (s-m_i^2)$.
The momentum transfer $Q^2=-q^2=-(k_i-k_j)^2$ 
is given by 
$Q^{2} = -(m_i^2+m_j^2) + 2E_i E_j 
-2 |\boldsymbol{k}_i| 
|\boldsymbol{k}_j| \cos\theta_{ij}$ in the center-of-mass frame,
 where the product of spacial momenta in the last term
can be  replaced by $|\boldsymbol{k}_i||\boldsymbol{k}_j|
=\sqrt{E_i^2-m_i^2}\sqrt{E_j^2-m_j^2}$.
Now the angle $\theta_{ij}$ can be written as
\begin{equation}
\begin{split}
	\cos\theta_{ij}
	= 
	\frac{
	 2E_i E_j - xy(s-m_i^2) - (m_i^2+m_j^2) }
	{2 \sqrt{E_i^2-m_i^2}\sqrt{E_j^2-m_j^2}},
\label{Eq:costheta}	
\end{split}
\end{equation}
where $Q^2 = xy (s-m_i^2)$ is used.
The constraint $|\cos\theta_{ij}| \leq 1$ leads to  
an inequality
\begin{equation}
\begin{split}
	\bigl[2E_i E_j
        -xy \left( s-m_i^2 \right) 
        -\left( m_i^2+m_j^2 \right) 
	\bigr]^2
	\leq  4(E_i^2-m_i^2)(E_j^2-m_j^2).
\label{Eq:cos<1}	
\end{split}
\end{equation}
Substituting the lepton energies by Eq.~\eqref{eq:Ei_and_Ej}, we 
obtain 
\begin{eqnarray}
y\,
\big\{
(1-y) 
\big[x(s-m_i^2) +m_i^2\big]
-m_j^2
\big\}\geq 0,
\end{eqnarray}
where we used
$w^2=(k_p+q)^2=y(\hat{s}-m_i^2)(1-x/\xi)$.
This condition determines the upper bound of the inelasticity $y$
\begin{equation}
\begin{split}
	y \leq 1 - \frac{r_j}{x+r_i}.
\label{Eq:ymax}	
\end{split}
\end{equation}
Here we introduced dimensionless masses $r_a~(a=i, j)$
\begin{eqnarray}
r_i &=& \frac{m_i^2}{s-m_i^2},
\hspace{5mm}
r_j~=~ \frac{m_j^2}{s-m_i^2}.
\end{eqnarray}

The lower one is bound by the partonic phase space. 
We express the inelasticity parameter $y$ in terms of 
$x$ and $w^2$
\begin{eqnarray}
y&=&\frac{1}{\xi-x}\left(\frac{w^2}{s-m_i^2}\right).
\end{eqnarray}
The momentum fraction $\xi$ is expressed in terms of Lorentz invariance as 
$\xi = x\, \left( Q^2 + \omega^2 \right)/Q^2$, 
and its minimum needs to be less than $x$; 
\begin{equation}
\begin{split}
	\xi 
	= x \left(\frac{Q^2 + w_\text{min}^2}{Q^2} \right)
	\geq x. 
\end{split}
\end{equation}
Thus the lower bound of $y$ is given by 
the production threshold of $w^2_{\rm min}$ with 
maximum momentum fraction $\xi=1$:
\begin{eqnarray}
y\geq 
\frac{1}{1-x} \left(\frac{w^2_{\rm min}}{s-m_i^2}\right)
\end{eqnarray}
Thus the physical range of $y$ for fixed $x$ 
is given by $y_{\rm min}\leq y\leq y_{\rm max}$ with  
\begin{eqnarray}
y_{\rm min} (x) ~=~ \frac{r_{\rm min} }{1-x},
\hspace{1cm}
y_{\rm max} (x) ~=~ 1-\frac{r_j}{x+r_i},
\label{eq:y_min_max}
\end{eqnarray} 
where $r_{\rm min}\equiv w_{\rm min}^2/(s-m_i^2)$.
 
The physical range of $x$ can be obtained 
by requiring that the lower bound on $y$ should be smaller than the upper one, 
namely $y_{\rm min}(x) \leq y_{\rm max}(x)$ should hold.
This yields a relation
\begin{equation}
\begin{split}
x^2 -(1-r_i+r_j-r_{\rm min})x - (r_i-r_j-r_i \, r_{\rm min}) \leq 0.
\end{split}
\end{equation}
Solving the quadratic inequality for $x$, we find the range 
$x_-\leq x \leq x_+$ 
with
\begin{eqnarray}
x_\pm &=&
\frac{1}{2} \bigg[\left(1-r_i+r_j-r_{\rm min}\right) \pm 
\sqrt{(1-r_i+r_j-r_{\rm min})^2+4(r_i-r_j-r_i r_{\rm min})}
\bigg].
\label{eq:x_pm}
\end{eqnarray}

\subsection{Heavy quark pair production in terms of $(x, y)$}
Here we give concrete expression for the physical region 
by taking electron and tau leptons as initial and final leptons,
 and a heavy quark pair of mass $m_q$ in the hadronic part, i.e.
 $\hat{X}=q\bar{q}$. 
 We ignore the electron mass, but tau and heavy quark masses are kept.
 In this case Eq.~\eqref{eq:y_min_max} reduces to 
\begin{eqnarray}
y_{\rm min}(x)&=&\frac{4m_q^2}{(1-x)s}, 
\hspace{1cm}
y_{\rm max}(x) ~=~1-\frac{m_\tau^2}{x s},
\end{eqnarray}
and Eq.~\eqref{eq:x_pm} reduces to
\begin{eqnarray}
x_{\pm} &=& \frac{1}{2} \bigg[
1+\frac{m_\tau^2}{s} - \frac{4m_q^2}{s} 
\pm \sqrt{\left(1+\frac{m_\tau^2}{s}- \frac{4m_q^2}{s}\right)^2 -4 \frac{m_\tau^2}{s}}~
\bigg],
\label{eq:x_pm_concrete}
\end{eqnarray}
where the minimum of $w^2$ has been substituted by $4m_q^2$.

\subsection{Heavy quark pair production in terms of $(x,Q^2)$}

When one takes $(x, Q^2)$ as independent variables instead of $(x, y)$, 
the relation $xy=Q^2/s$ can be used. The physical region for $x$ is given by
$x_{\rm min}(s, Q^2) \leq  x \leq x_{\rm max}(Q^2)$ with 
\begin{eqnarray}
x_{\rm min}(s, Q^2) ~=~\frac{Q^2+m_\tau^2}{s},
\hspace{1cm}
x_{\rm max}(Q^2)~=~\frac{Q^2}{Q^2+4m_q^2},
\label{eq:xmin}
\end{eqnarray}
and for $Q^2$ the physical region is $Q_-^2\leq Q^2 \leq Q_+^2$ with 
\begin{eqnarray}
Q^2_\pm &=& sx_\pm -m_\tau^2,
\end{eqnarray}
where $x_\pm$ is defined in Eq.~\eqref{eq:x_pm_concrete}.

\bibliography{ref_ep2tx}

\end{document}